\documentclass{aa}  

\usepackage{graphicx}
\usepackage{txfonts}
\usepackage{natbib}
\usepackage{comment}
\usepackage{hyperref}
\hypersetup{
 colorlinks=true,
 citecolor=blue,
 linkcolor=blue,
 pdftitle={ Multi-messenger studies with PTA},
 }
\usepackage{adjustbox}
\usepackage{multirow}
\usepackage{colortbl}

\def \lgalaxies{\texttt{L-Galaxies}\,}

\usepackage{color}
\usepackage[dvipsnames]{xcolor}
\definecolor{myorange}{rgb}{0.8, 0.3, 0.0}

\begin{document}

   \title{Lighting up the nano-hertz gravitational wave sky: opportunities and challenges of multimessenger astronomy with PTA experiments}
   \subtitle{}
    \titlerunning{The hosts and AGNs of the MBHBs detected by PTA experiments}

   \author{Riccardo J. Truant\inst{1}\fnmsep\thanks{r.truant@campus.unimib.it} \and David Izquierdo-Villalba\inst{1,2} \and Alberto Sesana\inst{1,2,3} \and Golam Mohiuddin Shaifullah\inst{1,2,4} \and Matteo Bonetti\inst{1,2,3} \and Daniele Spinoso\inst{5} \and Silvia Bonoli\inst{6,7} \\  
          }

            \institute{Dipartimento di Fisica ``G. Occhialini'', Universit\`{a} degli Studi di Milano-Bicocca, Piazza della Scienza 3, I-20126 Milano, Italy
              \and
               INFN, Sezione di Milano-Bicocca, Piazza della Scienza 3, 20126 Milano, Italy
               \and
               INAF - Osservatorio Astronomico di Brera, via Brera 20, I-20121 Milano, Italy
               \and
               INAF - Osservatorio Astronomico di Cagliari, via della Scienza 5, 09047 Selargius (CA), Italy
               \and
               Como Lake Center for AstroPhysics,  University of Insubria, 22100, Como, Italy
               \and
                Donostia International Physics Centre (DIPC), Paseo Manuel de Lardizabal 4, 20018 Donostia-San Sebastian, Spain
                \and 
                IKERBASQUE, Basque Foundation for Science, E-48013, Bilbao, Spain\\ \\
             }

   \date{Received ---; accepted ---}
   
   \abstract{Pulsar Timing Array (PTA) experiments have the potential to unveil continuous gravitational wave (CGW) signals from individual, low redshift massive black hole binaries (MBHBs). Detecting these objects in both gravitational waves (GW) and the electromagnetic (EM) spectrum will open a new chapter in multimessenger astronomy. We investigate the feasibility of conducting multimessenger studies by combining the CGW detections from an idealized 30-year SKA PTA and the optical data from the forthcoming Legacy Survey of Space and Time (LSST). To this end, we employed the \texttt{L-Galaxies} semi-analytical model applied to the \texttt{Millennium} simulation. We generated 200 different all-sky lightcones that include galaxies, massive black holes, and MBHBs whose emission is consistently modeled based on their star formation histories and gas accretion physics. Our results predict an average of $\approx$33 CGW detections, with signal-to-noise ratios $\rm S/N \,{>}\,5$. The MBHBs associated with the detections are typically at $z\, {<}\,0.5$, with masses of $\rm {\sim}\,3\,{\times}\,10^{9}\, M_{\odot}$, mass ratios ${>}\,0.6$ and eccentricities ${\lesssim}\,0.2$. In terms of EM counterparts, we find less than 15\% of these systems to be connected with an AGN detectable by LSST, while their host galaxies are easily detectable (${<}\,23\, \rm mag$) massive ($\rm M_{\star}\,{>}\,10^{11}\, \rm M_{\odot}$) ellipticals with typical star formation rates ($10^{-15}\, {\rm yr}^{-1} \,{<}\,\rm sSRF\,{<}\,10^{-10}\, {\rm yr}^{-1}$). Although the CGW - EM counterpart association is complicated by poor sky localization (only 35\% of these CGWs are localized within $\rm 100\, deg^2$), the number of galaxy host candidates can be considerably reduced (thousands to tens, depending on the CGW S/N) by applying priors based on the galaxy-MBH correlations. However, picking the actual host among these candidates is highly non-trivial, as they occupy a similar region in any optical color-color diagram. Our findings highlight the considerable challenges entailed in opening the low-frequency multimessenger GW sky.}
   
   

   \keywords{ gravitational waves – methods: numerical – galaxies: nuclei }

   \maketitle

\section{Introduction}

The hierarchical assembly of structures and the presence of massive black holes (MBHs, mass $M\,\rm {>}\,10^6\,M_{\odot}$) lurking at the galactic centers hints at the existence of massive black hole binaries (MBHBs). As argued in \cite{Begelman1980}, following a galaxy merger, the dynamical friction caused by the stellar and gaseous components drags the MBHs of the interacting galaxies towards the nucleus of the newly formed system, enabling the formation of a gravitational bound MBHB at parsec separations \citep{Milosavljevic2001,Yu2002,Mayer2007,Fiacconi2013,Bortolas2020,Bortolas2022,Kunyang2022}. Beyond this point, the system keeps shrinking through the interaction with individual stars at the galactic center \citep{Quinlan1997,Sesana2006,Vasiliev2014,Sesana2015}  and/or torques extracted from a circumbinary gaseous disc \citep{Escala2004,Dotti2007,Cuadra2009,Bonetti2020,Franchini2021,Franchini2022}. These processes are relevant down to milli-parsec separations when the emission of gravitational waves (GWs) becomes the main process of extracting energy and angular momentum from the system, driving the MBHB to coalescence on timescales between Myrs and several Gyrs.\\

MBHBs with $M\,\rm {>}\,10^8\, M_{\odot}$ placed at low redshifts are loud GWs sources at nano-Hz frequencies ($10^{-9}\,{-}\,10^{-7}$ Hz). This window is currently probed by Pulsar Timing Array (PTA) experiments through monitoring the spatially correlated fluctuations in the arrival time of radio pulses from a network of galactic millisecond pulsars \citep{Foster1990}. Currently, there are  6 different PTA collaborations taking data: the \textit{European Pulsar Timing Array} \citep[EPTA,][]{Kramer2013,Desvignes2016}, the \textit{North American Nanohertz Observatory for Gravitational Waves} \citep[NANOGrav,][]{McLaughlin2013,Arzoumanian2015}, the \textit{Parkes Pulsar Timing Array} \citep[PPTA,][]{Manchester2013,Reardon2016}, the \textit{Indian PTA} \citep[InPTA,][]{Susobhanan2021}, the \textit{Chinese PTA} \citep[CPTA,][]{Lee2016} and the \textit{MeerKAT PTA} \citep[MPTA,][]{Miles2023}. All of them have recently reported evidence of a stochastic GW background (sGWB) with amplitude $[1.7\,{-}\,7] \,{\times}\,10^{-15}$ at $\rm 1\, yr^{-1}$ frequency, compatible with the existence of a cosmic MBHB population \citep{Antoniadis2023,Agazie2023,Reardon2023,Xu2023,Miles2024}.\\

In addition to the eventual detection of the nano-Hz sGWB, PTA collaborations also aim to uncover \textit{continous} GW (CGW) sources, i.e GWs from individual MBHBs whose deterministic signals can be separated from the sGWB \citep[][]{Sesana2009,Babak2012,Ravi2015,Kelley2018,Falxa2023}. 
Systematic CGW searches in simulated MBHB populations have shown that PTAs will predominantly select systems of ${>}\,10^9\, \rm M_{\odot}$ at low-$z$ ($z\,{<}\,0.5$) emitting GWs at ${\sim}\,10^8\, \rm Hz$ \citep{Rosado2015,Kelley2018,Truant2024}. The abundance of these detections seems to be moderately sensitive to binary evolution parameters such as environmental coupling or MBHB eccentricity \citep[][]{Kelley2018,Gardiner2023}. For instance, \cite{Truant2024} demonstrated that the number of sources detectable by PTAs with long observing baselines increases as the eccentricity of the underlying MBHB population rises. The noise characteristics of the PTA also appear to impact CGW detections. As indicated by \cite{Cella2024}, PTAs with large white noise (${>}\,300\, \rm ns$)
will hamper the detection of light MBHBs ($\rm {\lesssim}\,10^9\, M_{\odot}$).\\


The detection of single MBHBs can be complemented with multimessenger studies. These require the electromagnetic (EM) identification of the active galactic nucleus (AGN) or galaxy associated with the CGW. This task is especially challenging for PTAs given their poor sky-localization capabilities (few to thousands of $\rm deg^2$, depending on S/N, \citealt{SesanaVecchio2010}) and the impossibility, in most cases, of directly constraining the source redshift due to the  lack of MBHB frequency evolution within the observing time. There are typically several millions potential hosts within the localization errobox, making the identification of peculiar features associated with the presence of the binary vital for narrowing down the pool of candidates. For example, different studies have shown that CGWs tend to be hosted in massive galaxies ($\rm {\sim}\,10^{11} \, M_{\odot}$), brighter and redder than the overall massive galaxy population \citep[][]{Rosado2014,Simon2014,Mingarelli2017,Cella2024}. 
Besides studying peculiar features, the physical dependence of the CGW amplitude on the MBHB chirp mass and distance to the observer can be coupled with  MBH mass-host galaxy correlations to assign to each galaxy a probability of being the actual host. This approach was first introduced by \cite{Goldstein2019}, who 
demonstrated that there is a 90\% probability that the true host is among the 0.01\% more massive, nearby galaxies within the localization errorbox (usually few tens to few thousands massive ellipticals, see also \citealt{Petrov2024}).\\


While multimessenger studies may be feasible with PTAs, there are still important caveats to address. First, most of the pipelines used to detect the hosts of CGWs have been calibrated using galaxy catalogues complete up to $z\,{\sim}\,0.1$ \citep{Arzoumanian2021,Petrov2024}. However, different studies indicate that resolvable CGWs can also be found at higher redshifts ($z\,{\sim}\,0.5$, \citealt{Rosado2015}) dramatically incresasing the number of galaxies within the errorbox. 
Second, it is unclear whether the MBHBs detected by PTAs will be associated with an AGN emission \citep{BurkeSpolaor2013,Charisi2016,Charisi2018}. This latter point is particularly important, as the lower abundance of AGNs with respect to galaxies would help to narrow down the number of candidates \citep{Tanaka2012}. In this paper we systematically address these points by means of realistic simulations of the GW and EM sky obtained with the \lgalaxies semi-analytical model \citep[SAM,][]{IzquierdoVillalba2022}. The simulations are used to construct synthetic full-sky lightcones populated with realistic galaxies, MBHs and MBHBs. From those, we compute the nano-Hz GW signal and identify the MBHBs sourcing CGWs that can be resolved by an SKA-like PTA. We characterize their AGN emission and study the properties of their host galaxies. We then make use of the Fisher matrix calculations developed by \cite{Truant2024} to compute the sky localization error of each CGW and employ the methodology of \cite{Goldstein2019} to narrow down the number of galaxy host candidates inside this area. We contrast the properties of these candidates with those of the true hosts in search of distinctive signatures that might facilitate the CGW-EM counterpart association. This allows us to make a realistic assessment about the feasibility and challenges entailed in the pursue of multimessenger astornomy in the nano-Hz GW band.\\


The paper is structured as follows. Section~\ref{sec::SAM desct} presents the \lgalaxies SAM and the methodology used to generate the used lightcones. In Section~\ref{sec:EM_Modelling} we describe the approach used to generate synthetic photometry of galaxies, MBHs and MBHBs. Section~\ref{sec:Detection_CW} outlines the characteristics of the PTA used and describes the methodology for detecting CGWs, including how to identify their parameters. Section~\ref{sec:sky_localization_Methodology} shows the methodology used to define the sky-localization area and reduce the number of potential host candidates. We present our results in Section~\ref{sec:Results}, focusing on the properties of the CGW hosts, their AGN counterpart and the possibility of identifying them in the sky. We summarize our main findings in Section~\ref{sec:Conclusions}. A Lambda Cold Dark Matter $(\Lambda$CDM) cosmology with parameters $\Omega_{\rm m} \,{=}\,0.315$, $\Omega_{\rm \Lambda}\,{=}\,0.685$, $\Omega_{\rm b}\,{=}\,0.045$, $\sigma_{8}\,{=}\,0.9$ and $h \, {=} \, \rm H_0/100\,{=}\,67.3/100\, \rm km\,s^{-1}\,Mpc^{-1}$ is adopted throughout the paper \citep{PlanckCollaboration2014}.

\section{The virtual Universe: a nano-Hz lightcone} \label{sec::SAM desct}

To create the population of galaxies, MBHs and MBHBs used in this work, we employ \lgalaxies{}, a state-of-the-art SAM calibrated to reproduce a vast array of observational properties such as the galaxy stellar mass function, the cosmic star formation rate density, galaxy colors and the fraction of passive galaxies \citep{Henriques2015,Henriques2020,Yates2021}. In particular, we make use of \lgalaxies in the version presented in \cite{IzquierdoVillalba2022} whose modifications improve the physics of MBHs (formation and growth), introduce the formation and dynamical evolution of MBHBs and allow the creation of lightcones. In the following, we briefly summarize the main features of the model.

\subsection{Dark matter and baryons} \label{sec:DM_and_Baryons}
The \lgalaxies SAM is built on the subhalo (hereafter just halo) merger trees extracted from N-body dark matter simulations. In particular, \lgalaxies has been tested on the \texttt{Millennium} and \texttt{TNG-DARK} suit of simulations \citep{Guo2011,Henriques2015,Ayromlou2021}. Here, we make use of the merger trees from the \texttt{Millennium-I} (MS, \citealt{Springel2005}), whose minimum particle mass and large cosmological volume allow us to trace the build-up of halos of a wide range of masses. Specifically, MS follows the cosmological evolution of $2160^3$ DM particles of mass $\rm 8.6\,{\times}\,10^8 \, M_{\odot}/\mathit{h}$ inside a periodic box of $500\, \rm Mpc/\mathit{h}$ on a side, from $z\,{=}\,127$ to the present. Particle data of MS was stored at 63 different times or snapshots. Structures formed in these outputs were identified by applying the friend-of-fried and \texttt{SUBFIND} algorithms and linked across time as progenitors and descendants in the so-called \textit{merger tree} via the \texttt{L-HALOTREE} code \citep{Springel2001}. Finally, given the coarse resolution of the DM outputs, \lgalaxies performs an internal time interpolation (${\sim}\,5\,{-}\,50 \, \rm Myr$, depending on redshift) to improve the tracing of the baryonics physics involved in the assembly of galaxies, MBHs, and MBHBs. Concerning the cosmology, originally MS was run with the WMAP1 \& 2dFGRS concordance cosmological parameters. However, the version of the merger trees used here is re-scaled with the procedure of \cite{AnguloandWhite2010} to match the cosmological parameters provided by Planck first-year data \citep{PlanckCollaboration2014}.\\

The evolution of galaxies is traced by \lgalaxies using a sophisticated galaxy formation model able to trace the evolution of the gas and stellar component of structures formed along the mass assembly of DM merger trees. As soon as a DM halo is resolved in the merger tree, \lgalaxies assigns a brayon fraction to it, consistent with the cosmological baryon fraction \citep{WhiteFrenk1991}. These baryons are initially distributed in a quasi-static hot gas atmosphere which progressively cools down at a rate given by the halo dynamical time. This cooled gas inherits the specific angular momentum from its host DM halo and settles at its centre in the form of a disc-like structure. The continuous fuel of cold gas raises the mass of the cold disc triggering star formation events and leading to the assembly of a stellar component distributed in a disc \citep{Croton2006a}. The results of this star formation process cause short-lived and massive stars to explode as supernovae, injecting energy and metals into the cold gas disc, heating it and eventually pushing it beyond the virial radius of the DM halo \citep{Guo2011}. The cold gas component is also regulated by the radio-mode feedback of the central MBH, which efficiently reduces or even suppresses cooling flows in massive halos \citep{Croton2006a,Croton2006}. Thanks to the continuous star formation, galaxies can develop massive stellar discs that can be prone to non-axisymmetric instabilities (the so-called \textit{disc instabilities}) which lead to the formation of a central ellipsoidal component called \textit{pseudobulge}.\\

Besides internal processes, galaxies interact among themselves, as a result of the hierarchical growth of DM halos. The time scale of these interactions is given by the dynamical friction experienced by the merging galaxies, accounted for by using the \cite{BinneyTremaine2008} formalism. The outcome of the interaction depends on the baryonic mass ratio ($\rm m_R$) of the two interacting systems. For large mass ratios ($\rm m_R \,{>}\,0.2$, \textit{major mergers}), the interaction destroys the discs of the two galaxies and leads to the formation of pure spheroidal component (\textit{elliptical galaxy}) whose mass increases as a consequence of a collisional burst \citep{Henriques2015}. Conversely, small mass ratios ($\rm m_R \,{<}\,0.2$, \textit{minor mergers}) allow the survival of the most massive galaxy disc component (which undergoes a burst of star formation) and its bulge grows as a consequence of the integration for the entire stellar mass of the satellite galaxy (forming a \textit{classical bulge}). Besides these two merger types, the version of \lgalaxies used in this work includes a prescription for smooth accretion to deal with the physics of extreme minor mergers (see for more details \citealt{IzquierdoVillalba2019}). Finally, \lgalaxies also models large-scale effects like ram pressure stripping or galaxy tidal disruption \citep{Henriques2015}.

\subsection{Massive black holes} \label{sec:MassiveBlackHoles}

\lgalaxies includes a comprehensive physical model to follow the formation of the first MBHs \citep{Spinoso2023}. It tracks the spatial variation of the Lyman Werner radiation and metals to track the formation of massive (direct collapse) and intermediate-mass (collapse of dense, nuclear stellar clusters) seeds. A sub-grid approach also accounts for the formation of light seeds. This model can only be used when the resolution of the DM merger trees reaches the small halos (${<}\, 10^{8.5}\, \rm M_{\odot}$)\footnote{These masses are covered by the merger trees extracted from the \texttt{Millennium-II}, \texttt{TNG-DARK00} and \texttt{TNG-DARK50} simulations.} where the genesis of the first MBHs take place. Given the characteristics of the MS simulation, newly resolved DM halos are found as soon as their masses reach $\rm 10^{10} M_{\odot} /\mathit{h}$. This hinders the possibility of using the self-consistent seeding model included in \lgalaxies. Since we are interested here in massive, nearby MBHBs, this is not a limitation, because the early growth of MBHs tends to erase the signatures of the seed population. \\

Taking into account the above, when any DM halo is firstly detected in the MS merger trees, \lgalaxies places an MBH seed of initial mass $10^4 \, \rm M_{\odot}$ and spin, $\chi$, extracted randomly form a uniform distribution ranging between $0\,{<}\, \chi \,{<}\, 0.98$. After the seeding, the MBH can grow through three different mechanisms: cold gas accretion, hot gas accretion and mergers with other MBHs. In the following paragraphs, we outline the main characteristics and the default values of free parameters employed in the so-called \textit{Fiducial} model \citep[see also][]{IzquierdoVillalba2020,IzquierdoVillalba2022}.

\begin{enumerate}
    \item \textbf{Cold gas accretion}:  This mechanism is the main growth channel and it is activated by two processes (see Section~\ref{sec:DM_and_Baryons}): \\
    
    \begin{enumerate}
        \item[-] \textit{Galaxy mergers}: MBHs can accrete a part of the cold gas component of the galaxy  given by:
            \begin{equation} \label{merg_acc}
                \Delta \mathrm{M}_{\mathrm{BH}}^{\rm gas }\,{=}\,f_{\mathrm{BH}}^{\text {merger }}\left(1+z_{\text {merger }}\right)^{5 / 2} \frac{m_{\mathrm{R}} \, \mathrm{M}_{\rm gas }} {1+\left(V_{\mathrm{BH}} / V_{200}\right)^2},
            \end{equation}
            where  $m_{\mathrm{R}}\,{<}\,1$ is the ratio between the baryonic mass content of the two interacting galaxies, $V_{200}$ is the viral velocity of the dark matter halo hosting the interacting galaxies, $z_{\text {merger}} $ is the redshift at with the merger take place, $M_{\text {gas}}$ is the cold gas content of the host galaxy. $V_{\mathrm{BH}}$ and $f_{\mathrm{BH}}^{\text{merger}}$ are two adjustable parameters set to $V_{\mathrm{BH}}\,{=}\,280 \, \rm {km/s}$ and $0.025$, respectively.\\ 
            
        \item[-] \textit{Disc instabilities}: The gas accreted by any MBH after any instability is given by:
            \begin{equation} \label{Diks_acc}
                \Delta \mathrm{M}_{\mathrm{BH}}^{\text {gas }}\,{=}\,f_{\mathrm{BH}}^{\mathrm{DI}}\left(1+z_{\mathrm{DI}}\right)^{5 / 2} \frac{\Delta \mathrm{M}_{\mathrm{stars}}^{\mathrm{DI}}}{1+\left(\mathrm{V}_{\mathrm{BH}} / \mathrm{V}_{200}\right)^2},
            \end{equation}
        where $f_{\mathrm{BH}}^{\mathrm{DI}}$ similarly to a $f_{\mathrm{BH}}^{\text {merger }}$ is a free parameter that describes the efficiency of the accretion process and it set $f_{\mathrm{BH}}^{\mathrm{DI}}\,{=}\,0.0015$ and $z_{\mathrm{DI}}$ is the redshift at which the disc instability occurs. Finally, $\Delta \mathrm{M}_{\mathrm{stars}}^{\mathrm{DI}}$ is the amount of stars triggering the disc instability.\\
    \end{enumerate}
    
    \lgalaxies assumes that the cold gas mass available for accretion after mergers or disc instabilities is not instantaneously accreted. Instead, it is stored in a gas reservoir around the MBH and progressively consumed according to a first phase of Eddington-limited growth followed by a decaying sub-Eddington phase \citep[see the results of][]{Bonoli2009,HopkinsQuataert2010}. While this work does not utilize them, we emphasize that the latest modifications of \lgalaxies allow for the possibility that some MBHs undergo episodic super-Eddington accretion episodes \citep{IzquierdoVillalba2024}. In future works, we will investigate how incorporating these events could impact the detection prospects of CGWs.\\  

    \item \textbf{Hot gas accretion}: MBHs are fueled by the constant inflow of hot gas that surrounds the galaxy. This hot accretion mode is characterized by a small accretion rate given by \citep{Henriques2015}:
    \begin{equation} \label{eq:AGN accretion}
        \dot{\text{M}}_{\text{BH}} = k_{\text{AGN}} \Biggl( \dfrac{\text{M}_{\text{hot}}}{10^{11}\text{M}_\odot}\Biggr) \Biggl( \dfrac{\text{M}_{\text{BH}}}{10^{8}\text{M}_\odot} \Biggr), 
    \end{equation}
    where $\text{M}_{\text{hot}}$ is the total mass of hot gas surrounding the galaxy and $k_{\text{AGN}}$ is a free parameter chosen to reproduce the galaxy luminosity function set to $3.5\,{\times}\,10^{-3} \, \rm M_{\odot}/yr $. Despite being a constant process, the accretion flow in this regime is orders of magnitude smaller than the MBH Eddington limit and the growth of the MBH is minimal. 

\end{enumerate}

\subsection{Massive black hole binaries} \label{sec:MassiveBlackHoleBinaries}

On top of the evolution of single MBH, \lgalaxies includes analytical prescriptions to deal with the formation and dynamical evolution of MBHBs \citep{IzquierdoVillalba2022}. Following the seminal work by \cite{Begelman1980}, \lgalaxies follows the evolution of these systems in different stages: \textit{pairing} and \textit{hardening \& gravitational wave phase}. The first one starts
after a galaxy merger. During this phase, the dynamical friction caused by the stellar component of the remnant galaxy reduces the initial separation (${\sim}$ kpc) between the nuclear and the satellite MBHs, moving the latter towards the galactic centre. The time spent by the satellite MBH in this phase is set according to the expression of \cite{BinneyTremaine2008}. Once the satellite MBH reaches the galactic nucleus, it binds gravitationally with the central MBH ($\rm {\sim}\,pc$ separation) and the \textit{hardening \& GW phase} starts. From hereafter, it is assumed that the most massive black hole in the system is the \textit{primary} one (with mass $\rm M_{BH,1}$)  whereas the less massive one is tagged as the \textit{secondary} (with mass $\rm M_{BH,2}$). The initial eccentricity of the binary system, $e_0$, is chosen randomly between $[0\,{-}\,1]$ and the initial separation, $a_0$, is selected to be the scale at which $\rm M_{Bulge}({<}\mathit{a_0})\,{=}\,2\,M_{BH,2}$, where $\rm M_{Bulge}({<}\mathit{a_0})$ corresponds to the mass in stars of the hosting bulge within $a_0$. To compute the value of $a_0$, \lgalaxies assumes that the bulge mass density profile follows a Sérsic model \citep{Sersic1968}. During this hardening \& GW phase, the separation ($a_{\rm BH}$) and eccentricity ($e_{\rm BH}$) of the binary system are evolved depending on the environment in which the MBHB is embedded. If the gas reservoir around the binary is larger than its total mass, the evolution of the system is driven by the interaction with a circumbinary gaseous disc and then GWs emission \citep{PetersAndMathews1963,Dotti2015,Bonetti2018a}. Otherwise, the system evolves via the interaction with single stars embedded in a Sérsic profile and the emission of GWs \citep{PetersAndMathews1963,Quinlan1997,Sesana2006}. In some cases, the lifetime of an MBHB can be long enough that a third MBH can reach the galaxy nucleus and interact with the binary system. In this scenario, the final stage of the MBHB (prompt merger of the MBHB or the ejection of the lightest object) is determined using the probability distributions presented in \cite{Bonetti2018ModelGrid}. \\

Finally, \texttt{L-Galaxies} follows the growth of MBHBs in a different way than the one of single MBHs. In particular, the code assumes that the accretion rates of the two SMBHs of the binary are correlated, as proposed by \cite{Duffell2020}: 
\begin{equation} \label{eq:Relation_accretion_hard_binary_blac_hole}
\dot{\rm M}_{\rm BH_1} \, {=}\,  \dot{\rm M}_{\rm BH_2} (0.1+0.9\mathit{q}),
\end{equation} 
where $\dot{\rm M}_{\rm BH_1}$ and  $\dot{\rm M}_{\rm BH_2}$ are respectively the accretion rate of the primary and secondary MBHs and $q \,{=}\, \rm M_{BH,2}/M_{BH,1}$ the binary mass ratio. For simplicity, the latter is set by \lgalaxies to the Eddington limit. Notice that if the cold gas reservoir is depleted, the secondary MBH is not able to accrete anymore. However, the primary MBH is still able to consume matter from the hot gas mode of Eq.~\eqref{eq:AGN accretion}. 

\subsubsection{The \textit{Fiducial} and \textit{Boosted} models}

The work of \cite{IzquierdoVillalba2022} showed that the \textit{Fiducial} model described above ($f_{\rm BH}^{\rm merger} \,{=}\,0.025$  and $f_{\rm BH}^{\rm DI} \,{=}\,0.0015$) give rise to a population of MBHs in agreement with current observational constraints (quasar luminosity functions and black hole mass functions) but the resulting MBHB population produces an sGWB of $A_{yr^{-1}}\,{\sim}\,1.2\,{\times}\,10^{-15}$, in mild tension with recent PTA observations, although in agreement with most of the theoretical predictions in literature. To address this discrepancy, the authors explored a \textit{Boosted} model in which the two phases of growth during cold gas accretion were untouched, but the gas accretion efficiency during mergers was raised up to $f_{\rm BH}^{\rm merger} \,{=}\, 0.075$. The results showed that this model was able to generate an sGWB amplitude $A_{yr^{-1}}\,{\in}\,\left[1.9\,{-}\, 2.6 \right]\,{\times}\,10^{-15}$, in better agreement with the latest PTA results \citep{Agazie2023,Antoniadis2023,Reardon2023,Xu2023}. In this work, we make use of both models (\textit{Fiducial} and \textit{Boosted}).

\begin{figure}    
\centering  
\includegraphics[width=1\columnwidth]{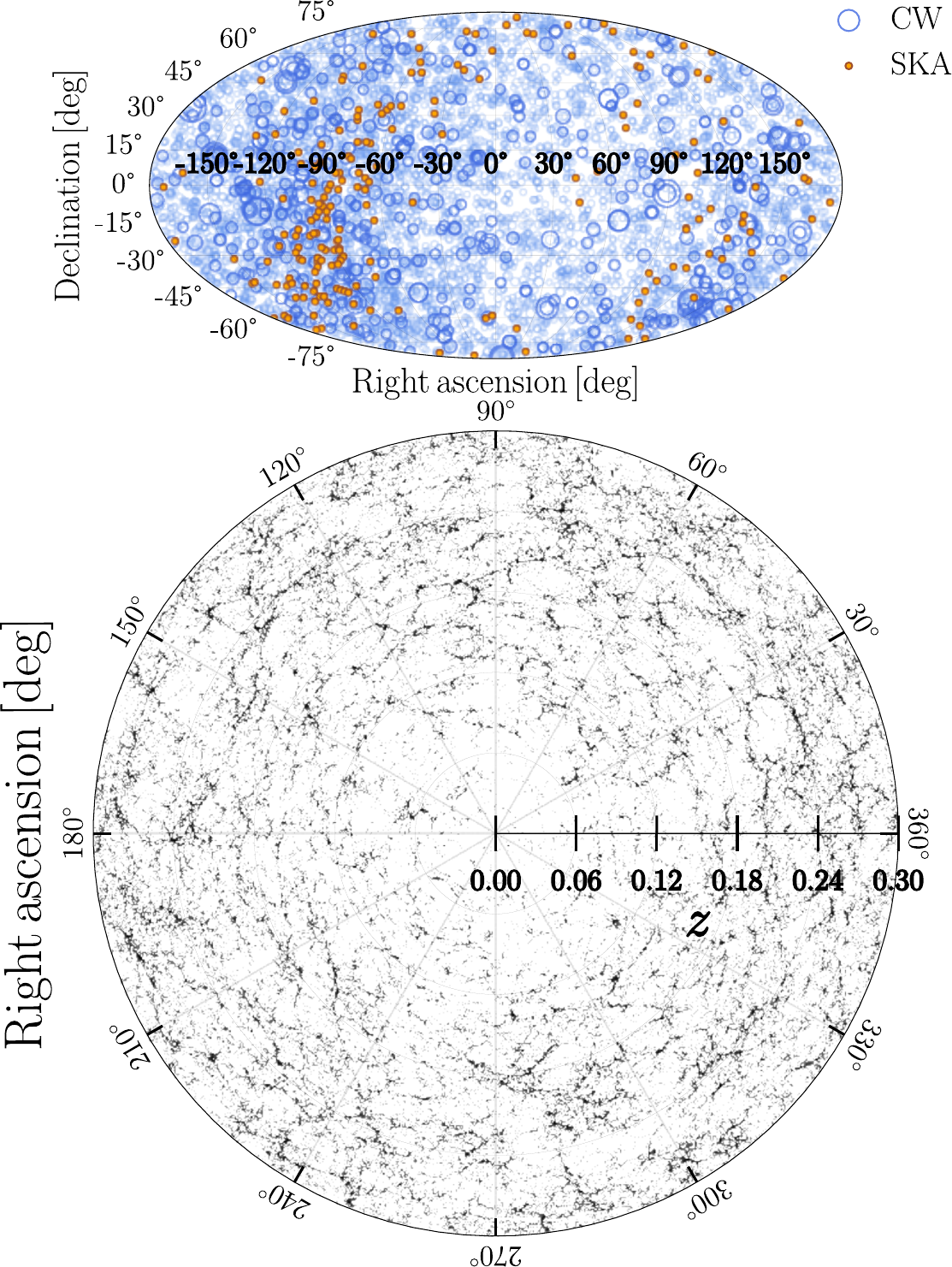}
\caption[]{\textbf{Upper panel}: Sky distribution of the SKA PTA pulsars (orange) and the MBHBs detected as CGW (blue) within the 200 MBHB realizations. The size of the blue points corresponds to the $\rm S/N$ of the CGW. \textbf{Lower panel}: Right ascension versus redshift for all the galaxies placed in the lightcone within a thin slide of $0.1 \, \rm  deg$. For the sake of clarity, we only showed the $z\,{<}\,0.3$ redshift range.}
\label{fig:Lightcone}
\end{figure}

\subsection{The simulated lightcones: MBHB population selection} \label{sec:Lightcone_And_MBHB_populations}


To make a more direct comparison with observations and build sky-localization areas, we need to create full-sky lightcones from the \lgalaxies outputs. To transform the discrete number of comoving boxes generated by \texttt{L-Galaxies} into a full-sky lightcone we use the method presented in \cite{IzquierdoVillalba2019LC}. In brief, the methodology takes the comoving boxes of galaxies created at different cosmological times and replicate them $\mathcal{N}$ times in each Cartesian coordinate. The value of $\mathcal{N}$ is set in such a way that the last replica displays a comoving distance from the observer as large as the desired redshift depth. The observer is placed in a position inside the first replica of the box (typically the first corner), and each galaxy is placed inside the lightcone by making use of the box replica which satisfies the condition that the galaxy crosses the observer past lightcone. For this work, we only keep the galaxies within the lightcone whose stellar mass $\rm M_{*}\,{>}\,10^{10.3} \, M_{\odot}$. We have checked that this cut does not affect the predictions about the sGWB but it improves the handling of the \lgalaxies outputs. To show the outcome of this procedure, the bottom panel of Fig.~\ref{fig:Lightcone} presents a slice of one of the full-sky lightcone used in this work.\\



We generate a total of 10 different lightcones: five  using the \textit{Fidicual} model, and five employing the \textit{Boosted} one. Since the observer plays an important role in determining which galaxies enter inside the lightcone, each lightcone is constructed assuming a different position of the observer in the simulation box.
Specifically, the observer has been placed in the following random positions: $\{ x_{\rm Obs},y_{\rm Obs},z_{\rm Obs}\} =$ $\{ 0,0,0\}\rm L_{box}$, $\{0.25,0.96,0.40\}\rm L_{box}$, $\{0.36,0.30,0.76\}\rm L_{box}$, $\{0.52,0.46,0.82\}\rm L_{box}$ and $\{0.74,0.45,0.10\}\rm L_{box}$ where $\rm L_{box}$ corresponds to the box size of the MS simulation. For each of these 10 lightcones, we have created 20 realizations, slightly changing the population of MBHBs. The first 10 were generated drawing 10 random polarization and inclination of each merging MBHB inside the lightcone. The other 10 were built by re-drawing each time the total mass of the MBHB according to the $\rm M_{stellar}\,{-}\, M_{BH}$ relation predicted by the lightcone at the redshift in which the binary is placed. In particular, the total mass was assigned randomly between the 10-90 percentile of the scaling relation and the mass of the primary and secondary MBH is assigned by keeping the mass ratio predicted by \lgalaxies.  Considering all the ariants, we create a total of 200 population of merging MBHBs.\\ 

\begin{figure}
    \centering
    \includegraphics[width=1\columnwidth]{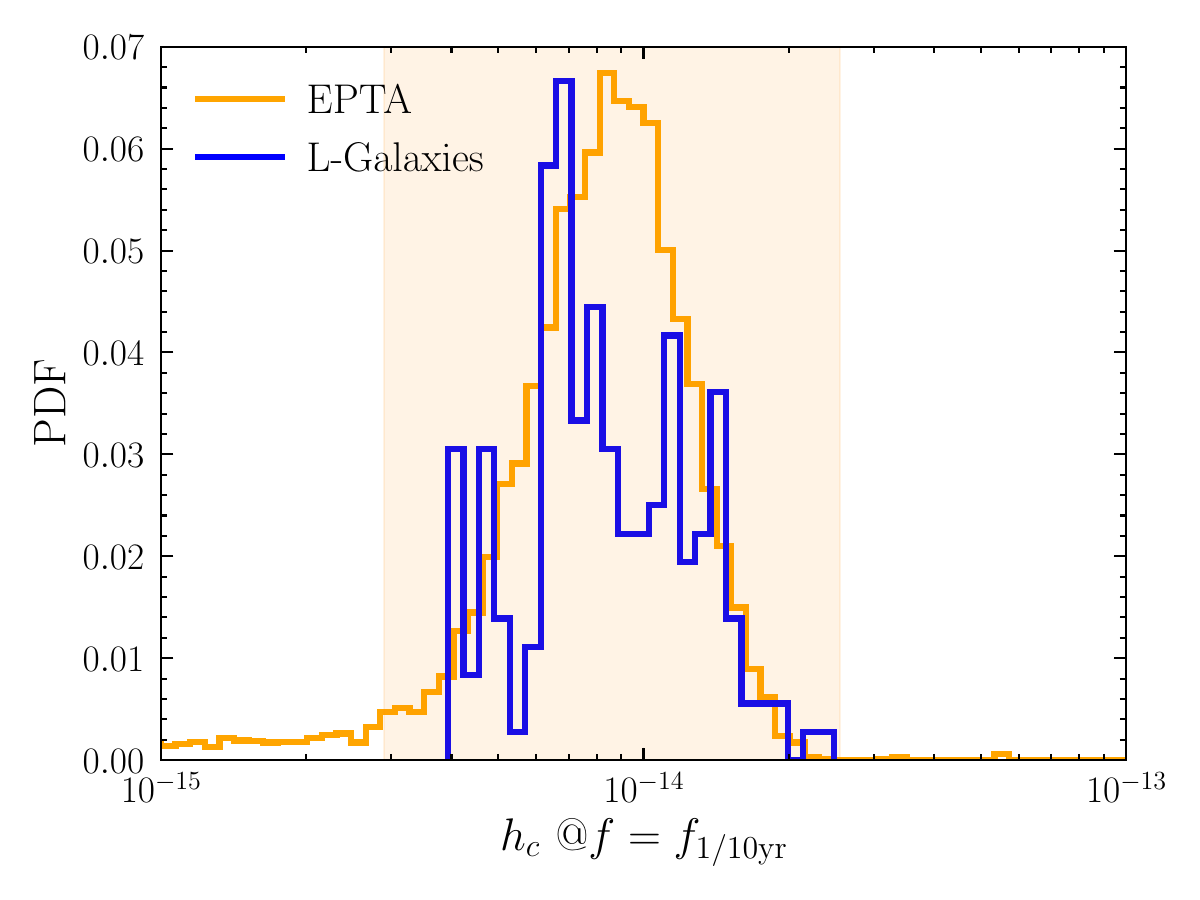}
   \caption{Normalized distribution of the sGWB characteristic strain amplitude at $f_{1/10yr}$. The orange histogram represents the posterior distribution inferred by the EPTA collaboration {\protect \citep{EPTA_GW_2023}}, while the orange shaded region represents the ${\pm}\,3 \, \sigma$ interval. The blue histogram corresponds to the strains produced by the MBHB population generated by \lgalaxies.}
    \label{fig:Amplitude_200_models}
\end{figure}

To pitch the most realistic scenario, among the 200 MBHB populations, we selected those consistent with the latest results presented by the European Pulsar Timing Array (EPTA) collaboration \citep{EPTA_GW_2023}. Given that the most constraining evidence for nanoHz gravitational waves comes from the lowest frequency bin probed by the array, $f \,{=}\, 1/10 \, yr^{-1}$ ($f_{1/10yr}$), we retained only the MBHB populations predicting an sGWB amplitude that fall within $3\sigma$ of the observed posterior distribution. 
The distribution of the sGWB amplitude at $f_{1/10yr}$ reported by EPTA and the \lgalaxies MBHB population is presented in Fig.~\ref{fig:Amplitude_200_models}. As shown, the generated populations are consistent with observational constraints within $3\sigma$ confidence level. Despite this, we stress that the characteristic strain produced by the generated MBHB population peaks at slightly lower values.\\ 

\section{The optical counterpart of galaxies and MBH(B)s} \label{sec:EM_Modelling}

Since we are interested in the optical identification of the CGW hosts, in this section we outline the procedure for computing the photometry of the galaxy stellar component and accreting MBHs/MBHBs. Notice that the final photometry of the galaxy will be the sum of these two contributions. Among all the optical surveys, we focus on the Vera C. Rubin Observatory’s Legacy Survey of Space and Time \citep[LSST,][]{Ivezic2019} whose observing strategy will enable the detection of very dim and high-$z$ objects across half of the sky. Specifically, the deep-observing mode of LSST will perform observations in the $u,g,r,i,z$ filters\footnote{We stress that in this work we do not consider the $y$ filter since we will only use the common ones with the Sloan Digital Sky Survey.} with limiting magnitudes corresponds to $u\,{\sim}\,26.1$, $g\,{\sim}\,27.4$, $r\,{\sim}\,27.5$, $i\,{\sim}\,26.8$ and $z\,{\sim}\,26.1$  \citep{LSSTScienceBook2009}.\\


\subsection{The galaxy photometry: Stellar population synthesis}

The galaxy magnitude in the $u,g,r,i,z$ filters is computed self-consistently by \lgalaxies using a simple stellar population synthesis model (SSP) combined with dust-reddening. In short, within a given filter system \lgalaxies creates tabulated grids of the observed luminosity generated by an SSP of fixed mass but different stellar ages and metallicities. Combining these grids with the star formation history of the galaxy and dust attenuation (due to the interstellar medium and molecular clouds) enables the computation of the observed photometry of a galaxy at any redshift. For further information, we refer the reader to \cite{Guo2011} and \cite{Henriques2015}. The typical photometry predicted by \lgalaxies for massive galaxies ($\rm M_{*}\,{>}\,10^{11}\, M_{\odot}
$) at low-$z$ is shown in the upper panels of Fig.~\ref{fig:SED}, which make clear that these galaxies can be up to two times brighter in the reddest filters. Finally, we stress that the version of the model used here is tuned to reproduce the fraction of passive galaxies at $z\,{<}\,3$ and the galaxy colors in the Sloan Digital Sky Survey optical bands \cite[see][]{Henriques2015}. \\

\subsection{The AGN photometry: Accreting MBHs and MBHBs}

The version of \lgalaxies used in this work does not include any model to compute the photometry of the AGN associated with an accreting MBH. Thus, to determine the magnitude of these objects in a given generic filter ``$j$'' we need to determine its spectral energy distribution. In particular, the magnitude ($m_j$) is determined by:
\begin{equation}\label{eq:magnitude}
    m_j \,{=}\, - 2.5 \log_{10} \left(\frac{\int_{\lambda_{\rm min}}^{\lambda_{\rm max}} d\lambda \, T_j(\lambda) \, \lambda^{-1} \, L_{\nu}^{\rm obs}(\lambda)}{ 4 \pi D_L^2 \int_{\lambda_{\rm min}}^{\lambda_{\rm max}}{d \lambda \, T_j(\lambda) \, \lambda^{-1}}}  \right) - 48.6,  
\end{equation}
where $L_{\nu}^{\rm obs}(\lambda) \, [\rm erg \,\, s^{-1} \, Hz^{-1}]$ is the observed spectral energy distribution of the source\footnote{Notice that $L_{\nu}^{\rm obs}(\lambda) = (1+z) L_{\nu}(\lambda)$, being $z$ the redshift of the source and $L_{\nu}(\lambda)$ its rest-frame SED.} at the observed wavelength $\lambda$, $D_L$ the source luminosity distance, $T_j(\lambda)$ the transition curve of the ``$j$'' filter and $\lambda_{\rm min}$ and $\lambda_{\rm max}$ the minimum and maximum wavelength covered by it. Taking into account this, in the upcoming sections we describe the approach used to associate an SED to each accreting MBH/MBHB taking into account the accretion properties of the system.   

\subsubsection{The spectral energy distribution of MBHBs} \label{sec:SED_MBHs_MBHBs}

\begin{figure*} 
   \centering
   \includegraphics[width=2\columnwidth]{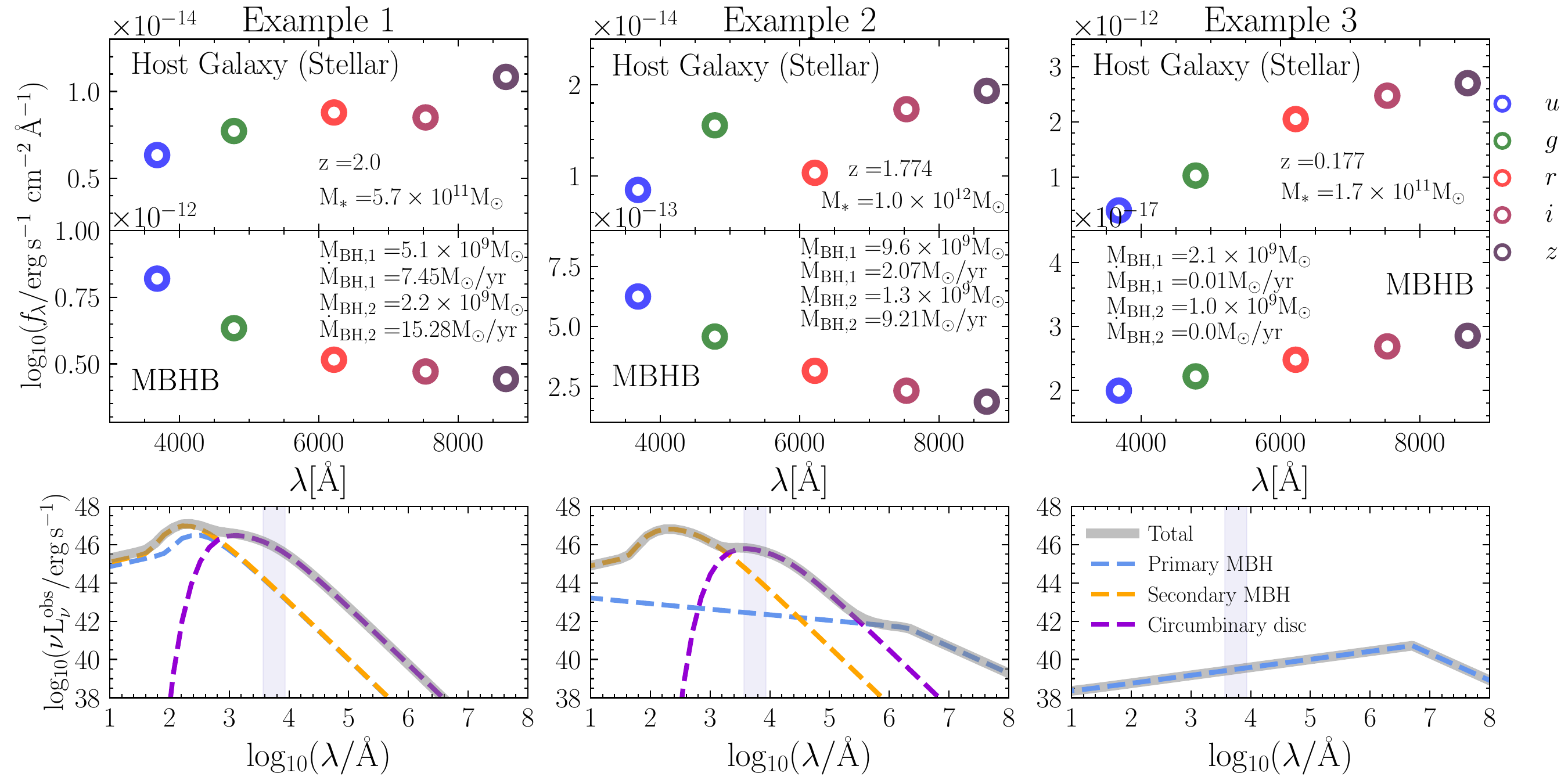}
   \caption{\textbf{Upper and middle panels}: Observed photometry in LSST for three detected MBHB (upper panels) and the stellar component of their host galaxies (middle panels). Blue, green, red, purple and brown dots correspond to the flux inside the $u,g,r,i,z$ filter, respectively. The properties of the binary and the host galaxy are indicated in each panel. \textbf{Lower panels}: Observed spectral energy distribution ($\nu\,L_{\nu}^{obs}$) of the MBHB used to compute the photometry ($\nu$ and $\lambda$ are the observed frequency and wavelength, respectively). The blue, orange and purple lines correspond to the emission of the primary MBH, secondary MBH and circumbinary disc, respectively; while the gray thick line is the sum of the three components. The vertical shaded area corresponds to the optical range covered by LSST.}
   \label{fig:SED}
\end{figure*}

Different hydrodynamical simulations have demonstrated that gas around MBHBs settles in the form of a \textit{circumbinary disc} \citep[see e.g][]{Ivanov1999,Haiman2009,Lodato2009}. Besides this, it has been shown that the gravitational torque exerted by MBHB is capable of opening a large cavity around the MBHB, with gas streams flowing inside and leading to the formation of \textit{mini-discs} structures around each MBH \citep{Ragusa2016,Fontecilla2017,Franchini2023}. 
Considering this gas structure, the total rest-frame spectral energy distribution of an MBHB, $L_{\nu}$, can be characterized by:
\begin{equation}\label{eq:EmissionTotalMBHB}
    L_{\nu} \,{=}\, L_{\nu}^{\rm CBD} \,{+}\, \sum_{i\,{=}\,1}^{i\,{=}\,2} L_{\nu,i}^{\rm Md} \, {+} \, L_{\nu}^{\rm Corona}
\end{equation}
where $L_{\nu}^{\rm Md}$ is the emission generated by the mini-disc of the primary ($i \,{=}\,1$) and secondary ($i \,{=}\,2$) MBH, $L_{\nu}^{\rm CBD}$ the one originated by the circumbinary disc and $L_{\nu}^{\rm Corona}$ the radiation coming from a non-thermal corona. In the following paragraphs, we outline the procedure used to compute each component of Eq.~\eqref{eq:EmissionTotalMBHB}. We stress that we do not model any infrared emission coming from a torus structure surrounding the MBHB. Furthermore, we do not account for any obscuration in the emission of AGNs caused by gas and dust in the galaxy's interstellar medium. Therefore, the SEDs and photometry of AGNs derived in this work should be regarded as the optimal observing scenario.\\ 

\noindent - \textbf{Circumbinary disc}:  The spectral emission radiated by the structure is computed according to the thin disc model \citep{Shakura&Sunyaev1973}:
    \begin{equation}\label{eq:EmissionSS}
        L_{\nu} = \frac{2 \pi h r_{\rm Schw}}{c^{2}}\nu^{3}
        \ \int_{x_{\rm in}}^{x_{\rm out}}
        \ \frac{4 \pi x }{{\rm exp}\left(\frac{h \nu}{k_{\rm B} \, T(x)}\right)-1} dx 
        \end{equation}
    where $k_{\text{\tiny B}}$ is the Boltzmann constant, $h$ the Planck constant, $c$ the speed of light and $\nu$ the rest-frame frequency. The variable $x$ is defined as $x \, {\equiv}\, r/r_{\rm Schw}$ being $r$ the radial distance to the MBH and $r_{\rm Schw}$ the Schwarzshild radius. The temperature of the circumbinary disc at a given distance $x$, $T(x)$, is determined by:
    \begin{equation} \label{eq:TemperatureSS}
        T(x) \,{=}\, \left[ \frac{3\dot{M}_{\rm BH}c^6}{64\pi \sigma_{\rm B} G^2 M^{2}_{\rm BH}} \right]^{\frac{1}{4}} \left[ \frac{1}{x^{3}} \left(1 - \sqrt{\frac{3}{x}}\right)\right]^{\frac{1}{4}}
    \end{equation}
    where $G$ the gravitational constant and $\sigma_{\rm B}$ is the Stefan-Boltzmann constant. In this case, $\rm M_{BH} \,{=}\, M_{BH,1} \,{+}\,M_{BH,2}$ being $\rm M_{BH,1}$ ($\rm M_{BH,2}$) the mass of the primary (secondary) MBH. Similarly, $\rm \dot{M}_{\rm BH}\,{=}\,\dot{M}_{BH,1} \,{+}\, \dot{M}_{BH,2}$ where $\rm \dot{M}_{\rm BH,1}$ ($\rm \dot{M}_{\rm BH,2}$) corresponds to the accretion rate onto the primary (secondary) MBH. Finally, $\left(x_{\rm in}, x_{\rm out} \right)\,{=}\,\left(2\,a_{\rm BHB}/r_{\rm Schw},10\,a_{\rm BHB}/r_{\rm Schw} \right)$, being $a_{\rm BHB}$ the binary semi-major axis \citep[see similar integration approach in][]{Cocchiararo2024}. \\
    
\noindent - \textbf{Mini-disc}: The emission raised from the mini-disc structures is going to be modeled differently according to the value of $f_{\rm Edd}\,{=}\,\rm L_{bol}/L_{Edd}$, predicted by \lgalaxies. Specifically, if $f_{\rm Edd}\,{>}\,0.03$ the accretion flow around the MBH can be considered as a thin disc (see Eq.~\ref{eq:EmissionSS}). On the contrary, the accretion is modeled according to an advection-dominated accretion flow where the gas is stored in an optically thin and geometrically thick disk. In the following, we describe the equations used to determine the emission of these structures. 
        \begin{itemize}
            \item[*] \textit{Thin disc accretion flow}: We follow the modeling of Eq.~\eqref{eq:EmissionSS} where the integration is performed from $x_{\rm in}\,{=}\,3$ to $x_{\rm out}\,{=}\,\mathrm{min} \left( 3000, r_{\rm Hill}/r_{\rm Schw}\right)$. $r_{\rm Hill}$ corresponds to the Hill radius and accounts for the fact that the presence of a massive companion can truncate the extension of an MBH mini-disc \citep[see a similar approach in][]{Kelley2019}. Here we define $r_{\rm Hill}$ as:
                \begin{equation} \label{eq:HillRadius}
                    r_{\rm Hill} \,{=}\, \dfrac{a_{\rm BHB}}{2} \left(\dfrac{\rm M_{BH,i}}{\rm M_{BHB}}\right)^{1/3},
                \end{equation}
                $\rm M_{BH,i}$ is the mass of the primary ($i\,{=}\,1$) or secondary ($i\,{=}\,2$) MBH and $\rm M_{BHB}$ is the total mass of the binary. We stress that for the sake of simplicity, Eq.~\ref{eq:HillRadius} ignores the eccentricity of the binary system. As we will see in Section~\ref{sec:Results}, we do not expect that this simplification affects our results since the targeted sample features a low eccentricity.\\
    
            \item[*] \textit{Advection dominated accretion flow} (ADAF): We closely follow the work of \cite{Mahadevan1996} which showed that the spectral energy distribution of an ADAF process is the result of three physical processes: synchrotron radiation ($L_{\nu}^{\text{\tiny Sync}}$), inverse Compton scattering ($L_{\nu}^{\text{\tiny Comp}}$) and bremsstrahlung radiation ($L_{\nu}^{\text{\tiny Brem}}$):
                \begin{equation} \label{eq:ADAF_model}
                  L_{\nu} \,{=}\, L_{\nu}^{\rm  Sync} + L_{\nu}^{\rm Comp} + L_{\nu}^{\rm Brem}.
                \end{equation}
            The expression of each component is formulated as:
            \begin{equation*}
                L_{\nu}^{ \rm Sync}\,{=}\, (s_1 \ s_2)^{8/5} \  s_3 \ \mathrm{M_{BH,i}}^{6/5} \ \dot{m}_{i}^{4/5} \ T_e^{21/5}\nu^{2/5},
            \end{equation*}
            \begin{equation*}
                L_{\nu}^{\rm Comp}\,{=}\, L_{\nu}^{\rm Sync} \left(\frac{\nu}{\nu_p}\right)^{\alpha_c},
            \end{equation*}
            \begin{equation*}
            \begin{split}
                L_{\nu}^{\rm Brem} = \mathcal{A}\, \alpha^{-2} c_1^{-2} \ln \left(\frac{r_{\rm  max}}{r_{\rm min}}\right)   \frac{\mathcal{F}(\theta_e)}{T_e} \mathrm{M_{BH,i}} \, \dot{m}_{i}^2 \exp\left(-\frac{h \nu}{ k_{\rm  B} T_e}\right),
            \end{split}
            \end{equation*}
            where $m_i\,{=}\,\rm \dot{M}_{BH,i}/\dot{M}_{Edd,i}$ is the accretion rate of the $i$ MBH in units of Eddington accretion, we set $\alpha \,{=}\,0.3$, $\beta\,{=}\,0.5$, $\delta\,{=}\,2000^{-1}$, $\nu_p$ is the frequency at which the synchrotron radiation is maximum, $T_e$ is the electron temperature, $r_{\rm min}\,{=}\,3\,r_{\rm Schw}$, $r_{\rm max}\,{=}\,10^3\,r_{\rm Schw}$ and $s_1$, $s_2$, $s_3$, $\mathcal{A}$ are constants. For the sake of brevity, we refer to \cite{Mahadevan1996}
            for the specific values and computations of the parameters.             
        \end{itemize}
        
\noindent - \textbf{Non-thermal corona}: We assume that the non-thermal emission produces a spectrum according to \citep{Regan2019,Cocchiararo2024}:
\begin{equation}
    L_{\nu}^{\rm Corona} \,{=}\, \mathcal{C} \nu^{\tau} \exp\left({-\nu/\nu_c}\right),
\end{equation}
where $\tau \,{=}\,-1.7$, $\nu_c\,{=}\,7.2\,{\times}\,10^{19} \, \rm Hz$ and
\begin{equation}
    \mathcal{C}\,{=}\,\frac{(1\,{+}\,\tau)\,L_{\rm X-ray} }{\nu_{f}^{\tau+1} - \nu_{0}^{\tau+1}}, 
\end{equation}
where $\nu_{0}\,{=}\, 4.8\,{\times}\,10^{16} \, \rm Hz$, $\nu_{f}\,{=}\, \rm 2.4\,{\times}\,10^{18} \, Hz$. The value of $L_{\rm X-ray}$ is the sum of the emission produced by each MBH in the hard X-rays (2-10 KeV), computed after applying the corrections of \cite{Shen2020} to the total bolometric luminosity of the MBHB.\\

To show the potential of our methodology, in Fig.~\ref{fig:SED} we present the SED and LSST photometry of three MBHBs inside our lightcones. As expected, the ones accreting during the thin disc regimen exhibit a predominantly blue SED, resulting in brighter photometry in the $u$ and $g$ filters. On the contrary, the MBHB accreting in the ADAF mode features a dominant red SED and photometry. Taking into account the above model and the MBH/MBHB accretion physics included in \lgalaxies (Section~\ref{sec:MassiveBlackHoles} and Section~\ref{sec:MassiveBlackHoleBinaries}), there are three possible SED combinations assigned to our MBHBs. The first case occurs when the two MBHs in the binary accrete in the thin disc regime, being very common in binaries with similar mass ratios (see the middle/lower left panel of Fig.~\ref{fig:SED}). The second case happens when the secondary MBH lies in the thin disc regime (always set by default at the Eddington limit) but the primary undergoes ADAF accretion (see the middle/lower center panel of Fig.~\ref{fig:SED}). This case is common for very unequal mass systems. Finally, in the third scenario the secondary is not accreting while the primary is in the ADAF accretion mode due to hot gas accretion (see the middle/lower right panel of Fig.~\ref{fig:SED}).\\

\subsubsection{The spectral energy distribution of MBH} \label{sec:SED_MBHs}

In the case of single MBHs, we employ the same formalism presented in the previous section. Specifically, Eq.~\eqref{eq:EmissionSS} (thin disc) and Eq.~\eqref{eq:ADAF_model} (ADAF). The main difference is the absence of a circumbinary disc and the transformation of the mini-disc around the MBH into a regular accretion disc with no truncation at the Hill's radius.


\section{Detecting continuous gravitational wave sources} \label{sec:Detection_CW}

In this section, we summarize the methodology used to detect eccentric MBHBs using an idealized SKA PTA experiment with 30 years of observations. For the sake of brevity, we will only refer to the main equations employed in this work. For a detailed computation of the procedure, we refer the reader to \cite{Truant2024}. 

\subsection{Signal-to-noise computation, parameter estimation and noise modelling} \label{sec:SNR computation}

To determine the possibility of detecting the signal of an MBHB in the nHz regime, it is necessary to compute the so-called \textit{signal-to-noise ratio} ($\rm S/N$). Assuming that a CGW is present in the timing residual of a pulsar, the $\rm S/N$ can be computed as:
\begin{equation} \label{eq:SNR_Geneal}
    \left(\dfrac{S}{N}\right)^2=4 \int_{0}^{\infty} df \dfrac{\lvert {\tilde{s}}(f) \rvert ^2}{S_k(f)}.
\end{equation} 
where $S_K(f)$ is the noise power spectral density (NPSD) and $\tilde{s}(f)$ the time residuals 
in the frequency domain. Following \cite{Truant2024}, the $\rm S/N$ for a generic eccentric MBHB detected with an array of pulsars ($N_{\rm Pulsars}$) can be expressed as:
\begin{equation} \label{eq:SNR_ecc}
    \left(\dfrac{S}{N}\right)^2_{tot}\,{=}\, \sum_{p=1}^{N_{\rm Pulsar}} \left[\sum_{n=1}^{\infty} \dfrac{2}{S_{k,p}(f_n)} \int_{t_0}^t  dt'\, s_{n,p}^2(t')\right],
\end{equation}
where $S_{k,p}(f_n)$ and $s_{n,p}(t^{'})$ correspond to the NPSD and the time residuals of the $p$ pulsar.\\

In the case of high $\rm S/N$, the MBHB parameters can be estimated through the computation of the \textit{Fisher Information Matrix} ($\Gamma$)\footnote{The inverse of the Fisher matrix provides a lower limit to the error covariance of unbiased estimators. For further details about the Fisher Information Matrix, we refer the reader to \cite{Husa2009} and \cite{SesanaVecchio2010}.}. Specifically, here we consider 9 free parameters:
\begin{equation} \label{Eq:free_par}
    \vec{\lambda}=(\zeta,f_k,e,i,\psi,l_0,\gamma,\phi,\theta).
\end{equation}
where $\zeta$ is the GW amplitude given $\zeta \,{=}\, (G\mathcal{M}_z)^{5/3}/c^4D_L$ being $\mathcal{M}_z$ the redshifted chirp mass, $D_L$ the luminosity distance, $G$ the gravitational constant and $c$ the light speed.  $f_k$ represents the orbital angular frequency, $e$ the binary eccentricity, $\iota$ the inclination angle, $\psi$ the polarization angle, $l_0$ the initial phase of orbit and $\gamma$ the direction of the pericenter with respect to $(\hat{\Omega} + \hat{L}\cos{\iota})/\sqrt{1-\cos^2{\iota}}$ where $\hat{L}$ and $\hat{\Omega}$ correspond to the orbital angular momentum and GW propagation direction, respectively. Finally, $(\theta, \phi) = (\pi/2 -DEC, RA)$, being DEC the declination and RA the right ascension. Following \cite{Truant2024}, the Fisher Information Matrix of a generic eccentric binary can be re-written as:
\begin{equation} \label{eq:FisherMatrixEccentric}
    \Gamma_{ij} \,{\simeq}\,  \sum_{p=1}^{N_{\rm Pulsar}} \left[\sum_{n=1}^{\infty} \dfrac{2}{S_{k,p}(f_n)} \int_{t_0}^{t}  dt' \, \partial_i s_{n,p}(t') \partial_j s_{n,p}(t')\right],
\end{equation}
where $\partial_i$ and $\partial_j$ are the partial derivatives of time residual in the frequency domain, $s(f)$, with respect to the $\lambda_i$ and $\lambda_j$ parameters, respectively. Notice that the covariance matrix is simply the inverse of the Fisher information matrix ($\Gamma^{-1}$). Therefore, the elements on the diagonal correspond to the parameter variance ($\sigma_{ii}^2\,{=}\,\Gamma^{-1}_{ii}$), while the off-diagonal terms represent the correlation coefficients between parameters ($\sigma_{ij}^2\,{=}\,\Gamma^{-1}_{ij}/\sqrt{\sigma_{i}^2\sigma_{j}^2}$).\\

Finally, the pulsar NPSD can be characterized by splitting it into two separate terms:
\begin{equation} \label{eq:NPSD}
    S_k(f) \,{=}\, S_p(f) \,{+}\, S_h(f).
\end{equation}
where term $S_p(f)$ encapsulates all noise sources related to the telescope sensitivity or intrinsic noise in the pulsar emission mechanism (among many others). It is usually described as a combination of three terms:
\begin{equation} \label{Eq:tot_NPSD}
    S_p(f) \,{=}\, S_{w} \,{+}\, S_{\rm DM}(f) \,{+}\, S_{\rm red}(f).
\end{equation}
$S_{w}$ accounts for stochastic errors in the measurement of a pulsar arrival time and is modeled as $S_{w} \,{=}\, 2\Delta t_{\rm cad} \sigma^2_{w}$. The value of $\Delta t_{\rm cad}$ corresponds to the observation cadence and $\sigma_{w}$ is determined in such a way that the pulse irregularity is a random Gaussian process described by the root mean square value $\sigma_{w}$. $S_{\rm red}(f)$ and $S_{\rm DM}(f)$ are the achromatic and chromatic red noise contributions, respectively. These two red noises are usually modeled as a stationary stochastic process, described as a power law and fully characterized by an amplitude and a spectral index.\\

The term $S_h(f)$ represents the red noise contributed at each given frequency by the sGWB \citep{Rosado2015}: 
\begin{equation} \label{eq:NPSD_blackhole}
S_h(f) \,{=}\,  \dfrac{h_c^2(f)}{12\pi^2f^3},
\end{equation}
given that PTA estimate the noise at each resolution frequency bin of the array (i.e $\Delta{f_i}=[i/T_{\rm obs},(i+1)/T_{\rm obs}]$, with $i=1,...,N$), the value $h_c^2$ of Eq.~\eqref{eq:NPSD_blackhole} associate to each frequency resolution element can be written as\footnote{We stress that when evaluating the detectability of a given MBHB we will not take into account the contribution of its $h_{c,j}^2(f)$ when computing the value of $S_h(f)$.}: 
\begin{equation}\label{eq:hc_disc}
    h_c^2(f_i) \,{=}\, \sum_{j=1}^{N_S} h_{c,j}^2(f)\delta(\Delta{f_i}\,{-}\,f).
\end{equation}
where the sum is over all sources, $N_S$, and $\delta(\Delta{f_i}\,{-}\,f)$ is a delta function that selects only MBHBs emitting within the considered bin. $h_{c,j}^2(f)$ is the squared characteristic strain of the  $j{-}\rm th$ source. Since we follow \cite{Truant2024} and consider eccentric MBHBs, Eq.~\eqref{eq:hc_disc} can be rewritten as:
\begin{equation}
    h_c^2(f_i) \,{=}\, \sum_{j=1}^{N_S} \sum_{n=1}^{\infty}  h_{c,n,j}^2(nf_k) \, \delta(\Delta{f_i}\,{-}\,nf_k),
\end{equation}
where the value of $h^2_{c,n}$ is given by \cite{Amaro_Seoane2010}. 

\subsection{Pulsar Timing Array: Square Kilometer Array Mid telescope} \label{sec::SKA}

We use an idealized PTA by employing the Square Kilometer Array Mid telescope (SKA, \citealt{Dewdney2009}) planned to be operative in 2027. SKA will be a large radio interferometer telescope whose sensitivity and survey speed will be an order of magnitude greater than any current radio telescope. For this work, we simulate the same array as the one presented in \cite{Truant2024}, i.e a 30-year SKA PTA with 200 pulsars featuring a white noise of $\sigma_w\,{=}\,100\, ns$ and an observing cadence of 14 days. We also include red noise to the total noise power spectral density in Eq. \eqref{Eq:tot_NPSD}, characterized as a power law \citep[see e.g][]{Lentati2015}:
\begin{equation}
S_{\rm red}(f) \,{=}\, \dfrac{A_{\rm red}^2}{12 \pi^2} \bigg( \dfrac{f}{f_{\rm yr}}\bigg)^{-\gamma_{\rm red}} \rm yr^{3},
\end{equation}
where $A_{\rm red}$ corresponds to the amplitude at one year and $\gamma_{\rm red}$ is the spectral index. The red noise properties are chosen to be consistent with those measured in the EPTA DR2Full. To this end, we fit a linear ${\rm log}\,A_{\rm red}-\gamma$ relation to the measured red noises in Table 4 of \cite{Custom_noise_EPTA}. Then, we assign $A_{\rm red}$ and $\gamma$ parameters according with this relation to 30\% of the pulsars in the SKA array, extracting $A_{\rm red}$ randomly from a uniform log-distribution in the interval $-15\,{<}\,{\rm log_{10}}(A_{\rm red})\,{<}\,-14$, for which the corresponding $\gamma$ is ${>}\,3$. In this manner, we mimic the fraction and properties of EPTA DR2Full pulsars in the SKA array with a robust red noise contribution. Note that while the remaining 70\% pulsars will likely display some lower level of red noise, this is unlikely to affect the properties of the detected CGWs. In fact, for those pulsars the main stochastic red noise component is going to be the sGWB itself, which is already included in our calculation. We refer the reader to \cite{Truant2024} for the sensitivity curve associated with this PTA experiment. We assume that an MBHB is detected by SKA PTA as a CGW when its $\rm S/N\,{\geq}\,5$.\\

Finally, the sky position of the 200 pulsars has been set similarly to \cite{Truant2024}. In brief, we draw a realistic sky distribution by using the pulsar population synthesis code \texttt{PsrPopPY}\footnote{\hyperlink{https://github.com/samb8s/PsrPopPy}{https://github.com/samb8s/PsrPopPy}.} \citep{Bates2014}. This code generates and evolves realistic pulsar populations drawn from physically motivated models of stellar evolution, calibrated against observational constraints on pulse periods, luminosities, and spatial distributions. The final population of pulsars ($10^5$) is selected such that they would be observable ($\rm S/N\,{>}\,9$) by a SKA survey with an antenna gain of 140 K/Jy and integration time of 35 minutes. To avoid a particularly (un)fortunate pulsar sky disposition from this distribution, we select a set of 200 pulsars randomly chosen according to a weight given by their radio flux. This procedure is done for each one of the 200 MBHB populations presented in Section~\ref{sec:Lightcone_And_MBHB_populations}. The sky distribution of one realization of SKA pulsars is presented in the upper panel of Fig.~\ref{fig:Lightcone}. Since \texttt{PsrPopPy} simulates realistic distributions of pulsars generated using theoretical considerations and observational constraints, the bulk of the generated full population of pulsars lies close to the Galactic plane. 

\section{CGW sky-localization and potential hosts} \label{sec:sky_localization_Methodology}

The success of multimessenger astronomy depends on unequivocally identifying in the sky the AGN or galaxy associated with the detected CGW. This requires having some priors in the source right ascension (RA) and declination (DEC) together with its luminosity distance (or its equivalent redshift). In particular, PTA experiments can give constraints in sky localization $(\theta,\phi) \,{=}\,(\rm \pi/2\,{-}\,DEC, RA)$ but generally not in luminosity distance. In fact, chirp mass and luminosity distance can only be disentangled by measuring the GW frequency evolution, $\dot{f}$, which is typically not possible for the bulk of PTA targets \citep{SesanaVecchio2010}. Under these circumstances, it is only possible to constrain the GW amplitude of the source, $\zeta$, i.e a combination of the chirp mass $\mathcal{M}_z$ and the luminosity distance $D_L$. We acknowledge that $\dot{f}$ might be detected for some high-frequency, high S/N CGWs, especially for long observation times like the 30 years considered here. However, in this work, we make the conservative assumption that $\dot{f}$ can not be measured. 
In the next sections, we describe the selection of the sky-localization area and how we select the final host candidates to perform multimessenger follow-ups within that sky area. \\

\subsection{Defining the sky localization area} \label{sec:Sky_Localization}

\begin{figure} 
   \centering
   \includegraphics[width=1\columnwidth]{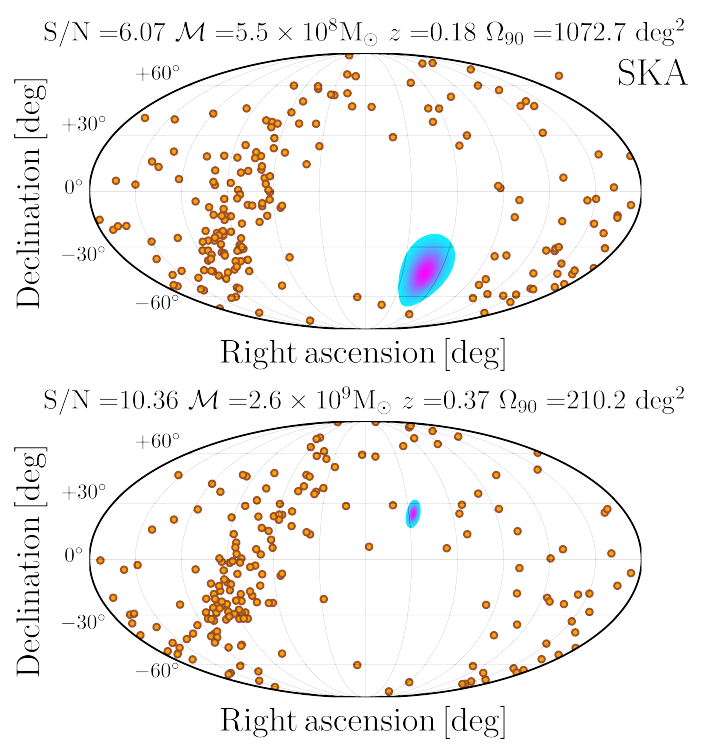}
   \caption{Sky-localization of two different random sources inside our lightcones (blue and purple areas). The upper one features a $\rm S/N\,{\sim}\,6$ while the lower one corresponds to $\rm S/N\,{\sim}\,10$. The orange points correspond to the position of the SKA PTA (different from sample to sample as explained in Section~\ref{sec::SKA}).}
   \label{fig:Sky_Localization_Examples}
\end{figure}

Given the lack of constraints on the luminosity distance, the sky error box associated with the CGW will consist of a region in the $(\theta,\phi)$ plane. To determine the area and shape of this region we follow \cite{Goldstein2019} and make use of a likelihood defined as:
\begin{equation} \label{eq::Likelihood_Sky_Area}
    \mathcal{L}(\theta,\phi) \,{=}\, \dfrac{1}{\sqrt{(2\pi)^{^3}\det\Gamma^{-1}}} \exp{\left[-\dfrac{1}{2} (\vec{\lambda_i} - \vec{\lambda_c})^{\rm T} \, \Gamma \, (\vec{\lambda_i} - \vec{\lambda_c}) \right]},
\end{equation}
where $\Gamma^{-1}$ is the $2\,{\times}\,2$ covariance matrix for the $\theta$ and $\phi$ parameters. The vector $\vec{\lambda_i}\,{=}\,\{\theta_i,\phi_i\}$ represents the position of the $i^{\rm th}$-galaxy in the sky and $\vec{\lambda_c}\,{=}\,\{\theta_c,\phi_c\}$ corresponds to a random point in the sky selected randomly from a 2-dimensional Gaussian probability density function with centers given by the true values of the hosts $(\theta_h,\phi_h)$ and variances the ones computed from the Fisher matrix\footnote{Notice that we select the random values of $\vec{\lambda_h}$ within the $ {\pm}\, 3\sigma$, where $\sigma$ is the variance associated to each parameter}. Taking into account Eq.\eqref{eq::Likelihood_Sky_Area}, the error box associated with a CGW (hereafter $\Omega_{90}$) will be defined as the sky region containing $90\%$ of the likelihood. Similarly, the total number of galaxies or AGNs candidates to be the true hosts (hereafter $N_{100}$) will be the ones included within $ \Omega_{90}$. To guide the reader, in Fig.~\ref{fig:Sky_Localization_Examples} we present two examples of the sky-localization areas associated with two detected CGWs at different $\rm S/N$. As expected, the source featuring the larger $\rm S/N$ is better localized.\\

\subsection{Reducing the number of candidates within the sky-localization area} \label{sec:Sky_Localization_Computation}

The absence of information concerning the luminosity distance of the CGW means that no redshift constraints can be applied to the galaxies within $\Omega_{90}$. This leads to a large number of potential host candidates, even in small areas, which complicates practical multimessenger follow-ups. To lower the value of $N_{100}$, we follow \cite{Goldstein2019} which showed that the number of potential candidates within $\Omega_{90}$ can be significantly decreased by exploiting the physical dependence of the CGW amplitude with the MBHB chirp mass and distance to the observer, together with empirical MBH mass-host galaxy correlations. Considering this, we assign to the $i^{\rm th}$-galaxy within $\Omega_{90}$ a probability, $\mathcal{P}$, of being the true host according to the three-dimensional likelihood function ($\mathcal{L}(\theta, \phi, \ln \zeta)$) and the prior probability ($P$) which encapsulates the correlation between the galaxy and MBH properties.\\


Following Section~\ref{sec:Sky_Localization} and under the assumption of high $\rm S/N$, the likelihood marginalized over $\theta, \phi$ and $\ln \zeta$ (hereafter, $\mathcal{L}(\theta, \phi, \ln \zeta)$) can be written as a three-dimensional multivariate Gaussian centered on the most probable parameter values:
\begin{equation} \label{eq:Likelihood_Amplitude}
    \mathcal{L}(\theta,\phi,\ln \zeta)  \,{=}\, \dfrac{1}{\sqrt{(2\pi)^{^3}\det\Gamma^{-1}}} \exp{\left[-\dfrac{1}{2} (\vec{\lambda_i} - \vec{\lambda_c})^{\rm T} \, \Gamma \, (\vec{\lambda_i} - \vec{\lambda_c}) \right]}.
\end{equation}
In this case $\Gamma^{-1}$ corresponds to the $3\,{\times}\,3$ covariance matrix for the $\theta$, $\phi$, $\ln \zeta$ values\footnote{Note that we construct our Fisher matrix by taking the logarithmic derivative of the S/N with respect to the amplitude $\zeta$. This ensures a physical range of the gravitational wave amplitude, avoiding negative values.}. The vector $\vec{\lambda_c}\,{=}\,\{\theta_c,\phi_c,\ln\zeta_c\}$ corresponds to a random point in the sky and GW amplitude selected randomly from a 3-dimensional Gaussian probability density function with centers given by the true values of the hosts ($\theta_h, \phi_h, \zeta_h$) and its variance the one computed from the Fisher matrix. For consistency, the values of $\theta_c,\phi_c$ will be fixed to the ones computed in the Section~\ref{sec:Sky_Localization_Computation}. The vector $\vec{\lambda_i}\,{=}\,\{\theta_i,\phi_i,\ln\zeta_i\}$ represents the sky position and GW amplitude of the possible MBHB placed in the $i^{\rm th}$-galaxy. The value of $\zeta_i$ 
is computed according to: 
\begin{equation}
    \zeta_i \,{=}\,  \left[\dfrac{G^{5/3}}{c^4 D_{L,i}}\right]  \left[\dfrac{(1 \,{+}\, z_i)\,\text{M}_{\rm{MBH},i} \,q_i^{3/5} }{(1\,{+}\,q_i)^{6/5}} \right]^{5/3},
\end{equation}
where $D_{L,i}$ is the $i^{\rm th}$-galaxy luminosity distance assumed to be known once its redshift ($z_i$) is measured, $q_i$ is the binary mass ratio selected randomly between $[0.01\,{-}\,1]$ and $\rm M_{MBH,i}$ the total mass of the binary assumed to be placed in the $i^{\rm th}$-galaxy and drawn from the well-established scaling relation between the MBH and galaxy bulge ($\rm M_{Bulge}$):
\begin{equation} \label{eq:MBH_MBuge_relation}
    \log_{10} \Biggl( \dfrac{\rm M_{MBH}}{M_\odot} \Biggr) = \alpha + \beta \log_{10}  \Biggl( \dfrac{\rm M_{Bulge}}{\rm 10^{11}M_\odot} \Biggr),
\end{equation}
where $\alpha$ and $\beta$ sets the slope and amplitude of the relation. Here we rely on the results presented in \cite{Chen2019} which pointed out $\alpha\,{=}\,8.40$. Concerning the normalization, we follow \cite{InterpretationPaperEPTA2023} and select $\beta\,{=}\,1.01$ which leads to an agreement with the recent result presented in \cite{EPTA_GW_2023}. On top of this, we assume that the relation presented in Eq.~\eqref{eq:MBH_MBuge_relation} displays an intrinsic scatter, $\epsilon$, set to a conservative number of $\epsilon\,{=}\,0.4$. We stress that we do not assume any redshift evolution of the scaling relation, given that this is still under debate \citep[see][and references therein]{Marshall2020}\\

On the other hand, the prior that we will use corresponds to the well-established relation between the MBH mass and the galaxy stellar bulge mass ($\rm M_{Bulge}$):
\begin{equation} \label{eq:Prior}
\begin{split}
      & P({\rm M_{MBH} | M_{Bulge}})  \,{=}\,  \\ = & \dfrac{1}{\sqrt{2\pi\epsilon^2}} \exp \Biggl\{ - \dfrac{  \biggl[ \log_{10} \dfrac{\rm M_{MBH}}{\rm M_\odot} - \biggl( \alpha + \beta \log_{10} \left( \dfrac{\rm M_{Bulge}}{\rm 10^{11} M_\odot } \right) \biggr)^2  \Biggr] }{2 \epsilon^2} \Biggr\}, 
\end{split}
\end{equation}
where $\epsilon$ corresponds to the intrinsic scatter of the relation. For simplicity, we will assume that the galaxy bulge mass is always know and the $\epsilon$ value is set to the conservative value of $0.4$.\\

By combining Eq.~\eqref{eq:Likelihood_Amplitude} and Eq.~\eqref{eq:Prior}, the probability $P$ of the $i^{\rm th}$-galaxy to be the true host can be written as:
\begin{equation} \label{Eq::Galaxy_host_prob}
P \,{=}\, \int^{\rm M_{MBH,max}}_{\rm M_{MBH,min}}\int^{q_{\rm max}}_{q_{\rm min}}\mathcal{L}(\zeta,\phi,\theta) \, P({\rm M_{MBH}} |{\rm M_{Bulge}}) \, d{\rm M_{MBH}} \, dq,
\end{equation}
where $q$ is assumed to have flat prior between the reasonable range $[q_{\rm min}, q_{\rm max}] \,{=}\,[0.01\,{-}\,1]$ and $\rm [M_{MBH,min}, M_{MBH,max}] \, {=} $ [Eq.\eqref{eq:MBH_MBuge_relation} - $3\epsilon$, Eq.\eqref{eq:MBH_MBuge_relation} + $3\epsilon$]. We refer the reader to \cite{Goldstein2019} for further details about the methodology.\\

After quantifying the values of $P$ for all galaxies within $\Omega_{90}$, we normalize the total probability across the sky-localization area and rank the galaxies from the most to the least probable. We then calculate the cumulative probability for the entire galaxy sample and select only those objects that contribute to 90\% of the total probability. These selected galaxies will be tagged as $N_{90}$ and will be the only ones used in our search for the true CGW host. By construction, there is a $90\%$ probability that the true host lies within the $N_{90}$ sample. \\

\section{Results} \label{sec:Results}

We now present our main results. We investigate the AGN counterparts of the detected MBHBs and what are the properties of the galaxies in which they are hosted. We then discuss challenges and strategies for unequivocally identifying the MBHB host within its sky-localization error box.

\begin{figure}
    \centering
    \includegraphics[width=1\columnwidth]{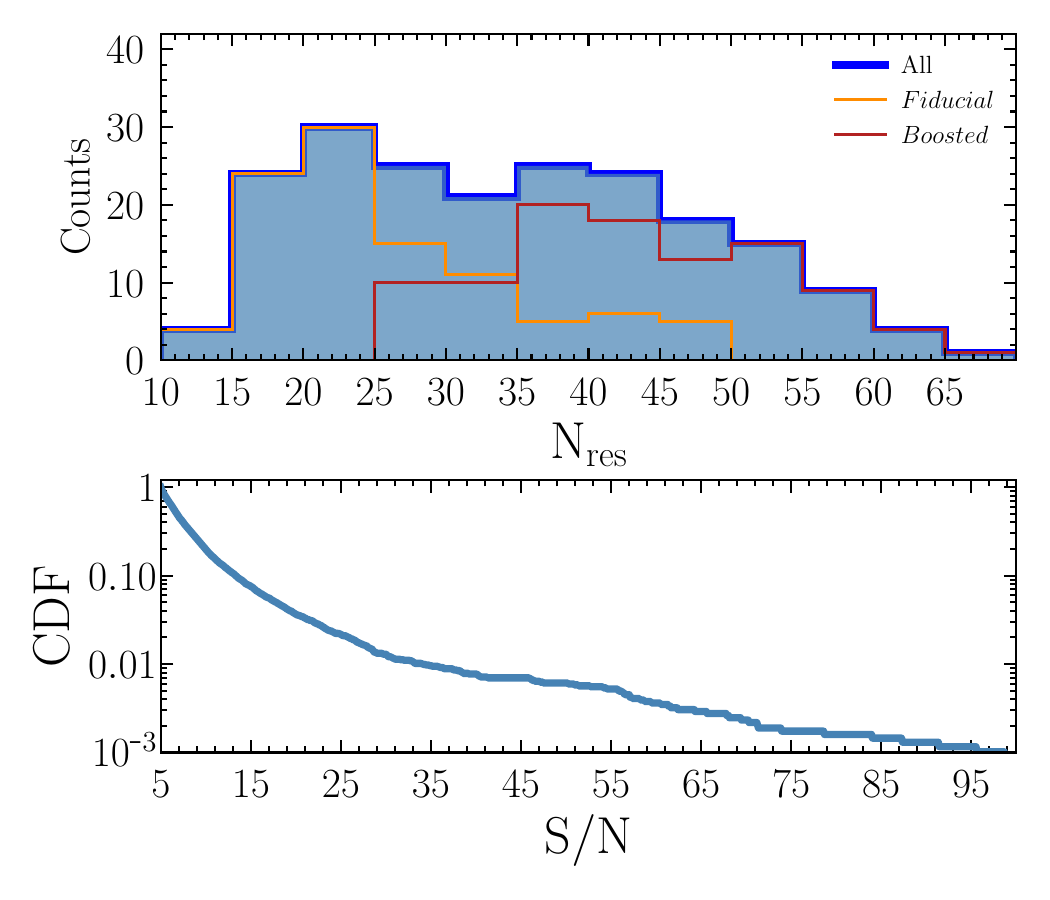}
    \caption{\textbf{Upper panel}: Distribution of the number of MBHBs resolved by SKA 30-yr ($\rm N_{res}$) in each lightcone generated by \texttt{L-Galaxies}. \textbf{Lower panel}: Cumulative distribution function (CDF) of the signal-to-noise ratio ($\rm S/N$) for the resolved MBHBs.}
    \label{fig:N_res}
\end{figure}

\subsection{Resolvable MBHBs and their properties} \label{sec:BinaryProperties}


In Fig.~\ref{fig:N_res} we show the distribution of the number of resolvable sources ($N_{\rm res}$) in our lightcones. The distribution is broad, with a plateau extending from 15 to 40 and a median value ${\sim}\,33$. 
In the same plot, we have divided the distribution between the \textit{Fiducial} and \textit{Boosted} models, showing that the cases with large $N_{\rm res}$ correspond to the latter models where the sGWB is larger than the average one of our populations (see Fig.~\ref{fig:Amplitude_200_models}). These results agree with the recent work of \cite{Truant2024} which, using populations of MBHBs in agreement with recent PTA results, showed that a 30-year SKA PTA would be able to resolve on average ${\approx}\,30$ CGWs, with a mild dependence on the typical eccentricity of the MBHB population. The lower panel of Fig.~\ref{fig:N_res} displays the cumulative distribution function of the $\rm S/N$ of the resolved sources: 90\% of the CGW feature $\rm S/N\,{\lesssim}\,15$, with a long tail extending to $\rm S/N\,\approx100$. Considering that, on average the array can detect $\approx 33$ CGWs, the typical highest $\rm S/N$ in a single universe realization is around 20-25.\\ 

\begin{figure}
    \centering
    \includegraphics[width=1\columnwidth]{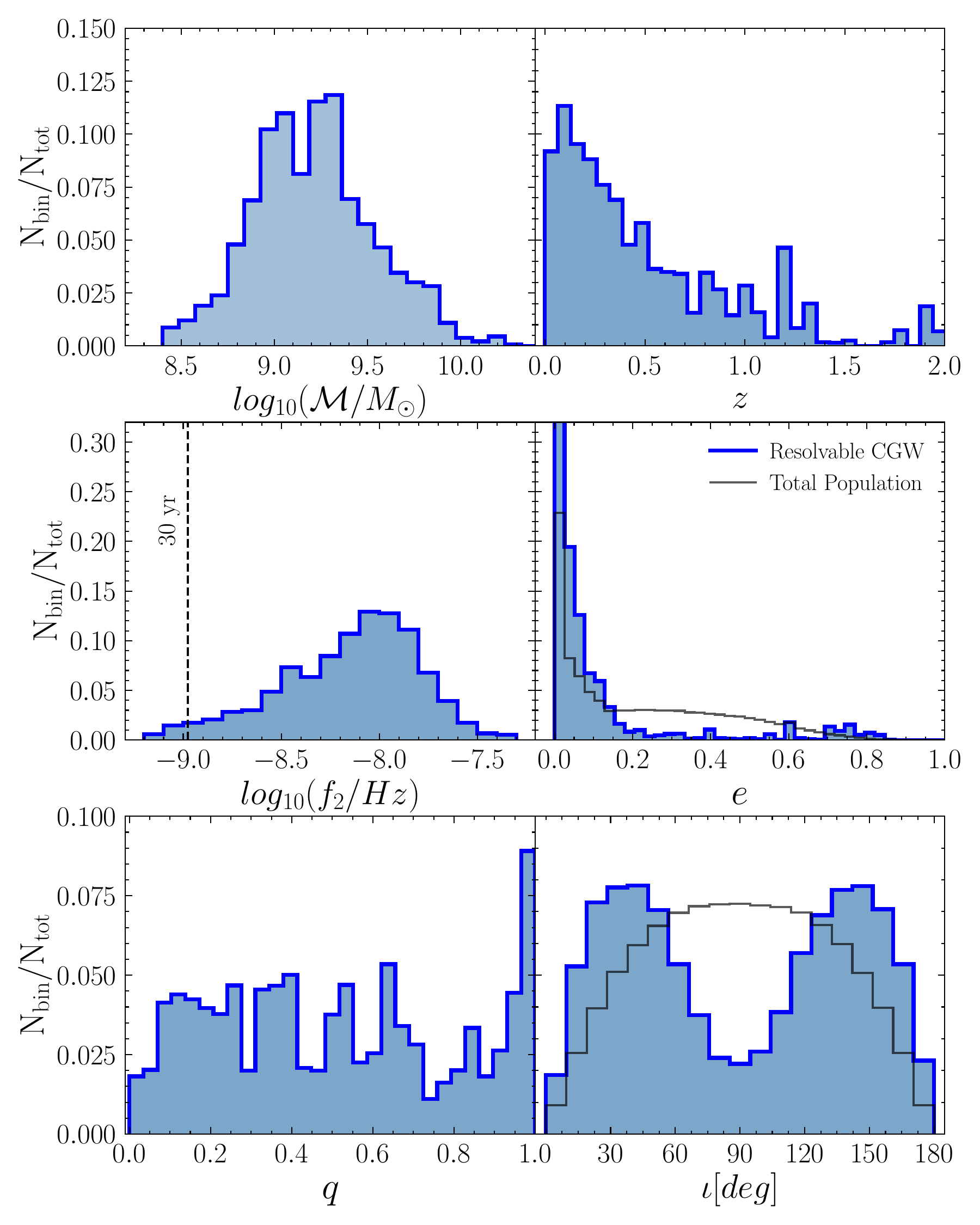}
    \caption{Properties of the detectable sources by 30-yr SKA: Chirp mass distribution ($\mathcal{M}$, upper left panel),  redshift ($z$, upper right panel), twice the observed Keplerian frequency ($f_2$, center left panel), eccentricity ($e$, center right panel), mass ration ($q$, lower left panel) and inclination angle ($i$, lower right panel) of the detected MBHBs. $N_{Bin}$ represents the number of objects in a given bin of the histogram, while $N_{tot}$ is the total number of objects analyzed. For reference, we present with black lines the distribution of eccentricity and inclination angle for all the MBHBs presented inside the lightcone.}
    \label{fig:Properties_Binaries}
\end{figure}

In Fig.~\ref{fig:Properties_Binaries} we present the properties of the detected MBHBs. As we can see, the detected sources can be found up to $z\,{\sim}\,2$ but the bulk of the CGWs are sourced by MBHBs at $z\,{<}\,0.5$. The  minimum chirp is $\rm \mathcal{M}\,{\sim}\,10^{8.5}\, M_{\odot}$, but typical values are in the range $\rm 10^{9}\,M_{\odot} \,{-}\, 10^{9.5}\,M_{\odot}$. The mass ratios are distributed in the range $0.1\,{<}\,q\,{<}\,1$, with a peak towards equal masses.
Eccentricities tend to be relatively small with a clear predominance of values $e\,{<}\,0.2$, although a long tail of eccentric binaries extends up to $e\,{\gtrsim}\,0.8$
The fundamental GW frequency (i.e twice the Keplerian frequency, $f_2$) of the MBHBs peaks at $10^{-8}\, \rm Hz$ and declines rapidly towards larger values. Interestingly, there are a few cases where the frequency is smaller than the one associated $1/T_{\rm obs}$. As described in \cite{Truant2024}, these sources are eccentric MBHBs that can enter inside the PTA band thanks to the power emission along many different harmonics. Finally, the inclination distribution features a bimodality, having a preference for face-on/face-off binaries. This can be simply explained by the angular pattern emission of GWs, which are stronger along the binary orbital angular momentum axis. These properties along with previous results indicate that future detection of nano-Hz MBHBs will unveil the most massive population of MBHBs in the relatively local Universe \citep[see similar results in e.g.][]{Rosado2015,Kelley2018,Cella2024}\\

\begin{figure}
    \centering
    \includegraphics[width=1\columnwidth]{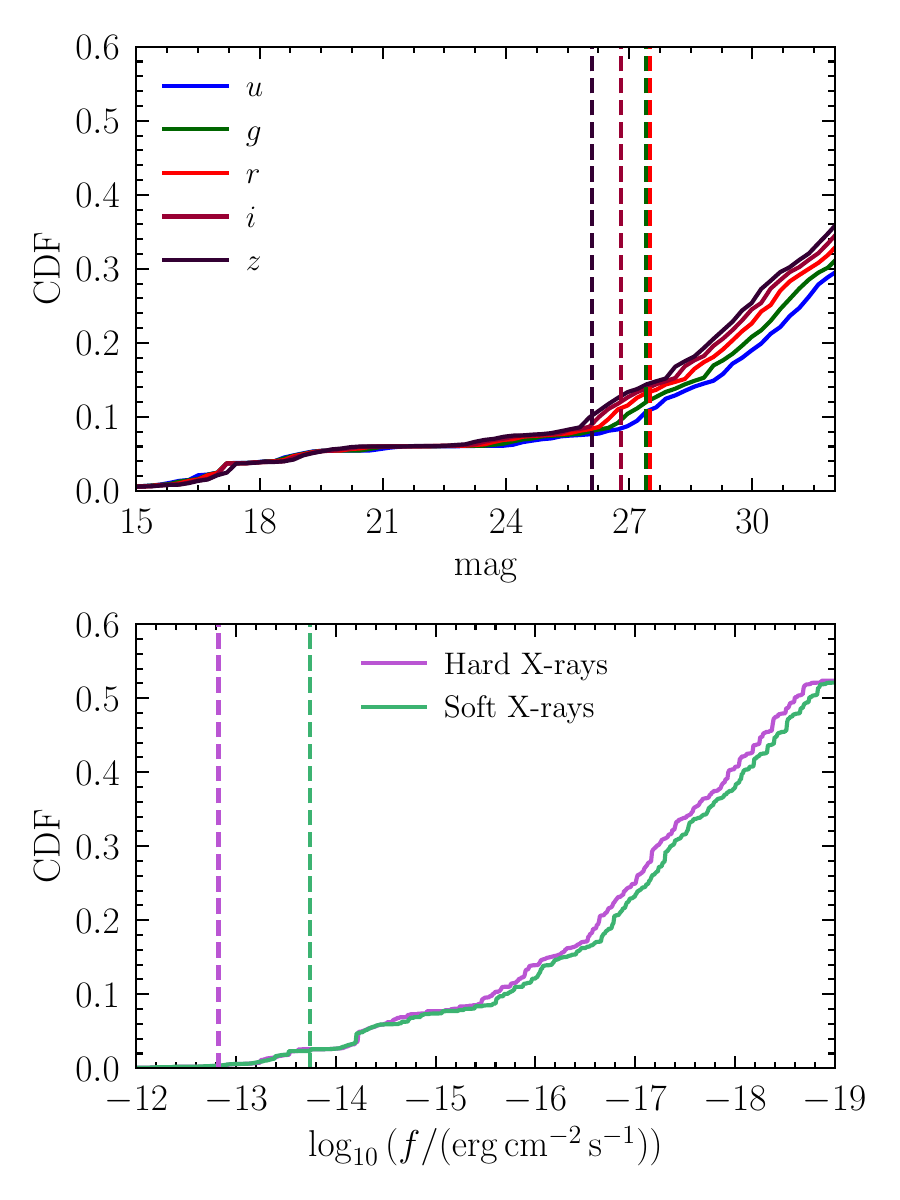}
    \caption{\textbf{Upper panel}: Cumulative distribution function (CDF) of the AGN magnitude associated with the MBHB sourcing the CGWs. Each color corresponds to a different LSST filter: $u,g,r,i,z$. Dashed vertical lines represent the limiting magnitudes of the LSST in each filter. \textbf{Lower panel}: Cumulative distribution function of the X-ray flux in the soft (0.5-2 KeV, green) and hard X-ray (2-10 KeV, purple) bands. Vertical dashed lines highlight the detection limit of eRosita.}
    \label{fig:OpticalAndAGNs_MBHBs}
\end{figure}

Once the properties of the detected MBHBs have been characterized, it is interesting to explore their emission properties.
The upper panel of Fig.~\ref{fig:OpticalAndAGNs_MBHBs} presents the cumulative distribution function of the associated AGN magnitudes in the optical $u,g,r,i,z$ filters. As we can see, less than 20\% of them feature an optical emission with magnitude $mag\,{<}\,30$ independently of the filter. This implies that a large fraction of our detected MBHBs will be essentially inactive, hindering the identification of the CGW EM counterpart based on AGN selection. This fraction decreases down to $\approx$10\% when we request that the AGN counterpart associated with the MBHB is optically detectable by LSST. In the lower panel of Fig.~\ref{fig:OpticalAndAGNs_MBHBs} we present the cumulative distribution function of the X-ray counterpart associated with the MBHB\footnote{To estimate the luminosity in the soft ($0.5\,{-}\,2\, \rm  keV$) and hard ($\rm 2\,{-}\,10 \, \rm keV$) X-ray band we have adopted the bolometric correction presented in \cite{Shen2020}. Specifically, $\log _{10}\left(\mathrm{~L}_{\mathrm{Hx}} / \mathrm{L}_{\text {bol}}\right)\,{=}\,-1.69-0.257 \mathcal{L} \,{-}\, 0.0078\mathcal{L}^2 \,{+}\, 0.0018\mathcal{L}^3$ and $\log _{10}\left(\mathrm{~L}_{\mathrm{Sx}} / \mathrm{L}_{\text {bol}}\right)\,{=}\,-1.84-0.260 \mathcal{L} \,{-}\, 0.0071\mathcal{L}^2 \,{+}\, 0.0020\mathcal{L}^3$, where $\mathcal{L} \,{=}\, \log_{10}(L_{\text{bol}}/\text{L}_\odot)\,{-}\,12 $ and  $L_{\text{bol}}$ the total SMBHB luminosity bolometric luminosity.}. As shown, only 50\% of the sources display hard $2\,{-}\,10\, \rm  keV$ and soft $0.5\,{-}\,2\, \rm  keV$ X-ray fluxes ${>}\,10^{-19}\,\rm erg\,s^{-1}\, cm^{-2}$. This value drops down to 5\% when fluxes ${>}\,10^{-14} \rm erg\,s^{-1}\, cm^{-2}$ are imposed. The X-ray emission is compared with the detection limit of eRosita \citep{Merloni2012}. As we can see, only ${\sim}\,3\%$ of the sources will be detected in the soft X-rays and none in the hard X-rays by eRosita. Considering next-generation X-ray observatories such as Athena \citep{2023MNRAS.521.2577P} yields more promising results. With flux limits of $2\times 10^{-17} \rm erg\,s^{-1}\, cm^{-2}$ and $10^{-16} \rm erg\,s^{-1}\, cm^{-2}$ in the soft and hard X-ray respectively, up to 20\% of the systems might be within reach for Athena. However, as we will see below, the tipical source localization in the sky is poor, and tailing the location error region with long Athena exposures might be unfeasible.
As discussed in Section~\ref{sec:EM_Modelling}, we stress that the X-ray results should be considered upper limits since attenuation due to neutral gas has not been included.

Based on the results presented above and considering an average number of 33 resolved CGWs, three-four of them might be associated with weak optical AGNs and only one might be associated with soft-X AGN activity within the eROSITA sensitivity limits. This suggests that multimessenger studies with PTA should also consider the search for the galaxy host rather than focusing exclusively on AGNs.


\subsection{Host galaxy properties}

\begin{figure}
    \centering
    \includegraphics[width=1\columnwidth]{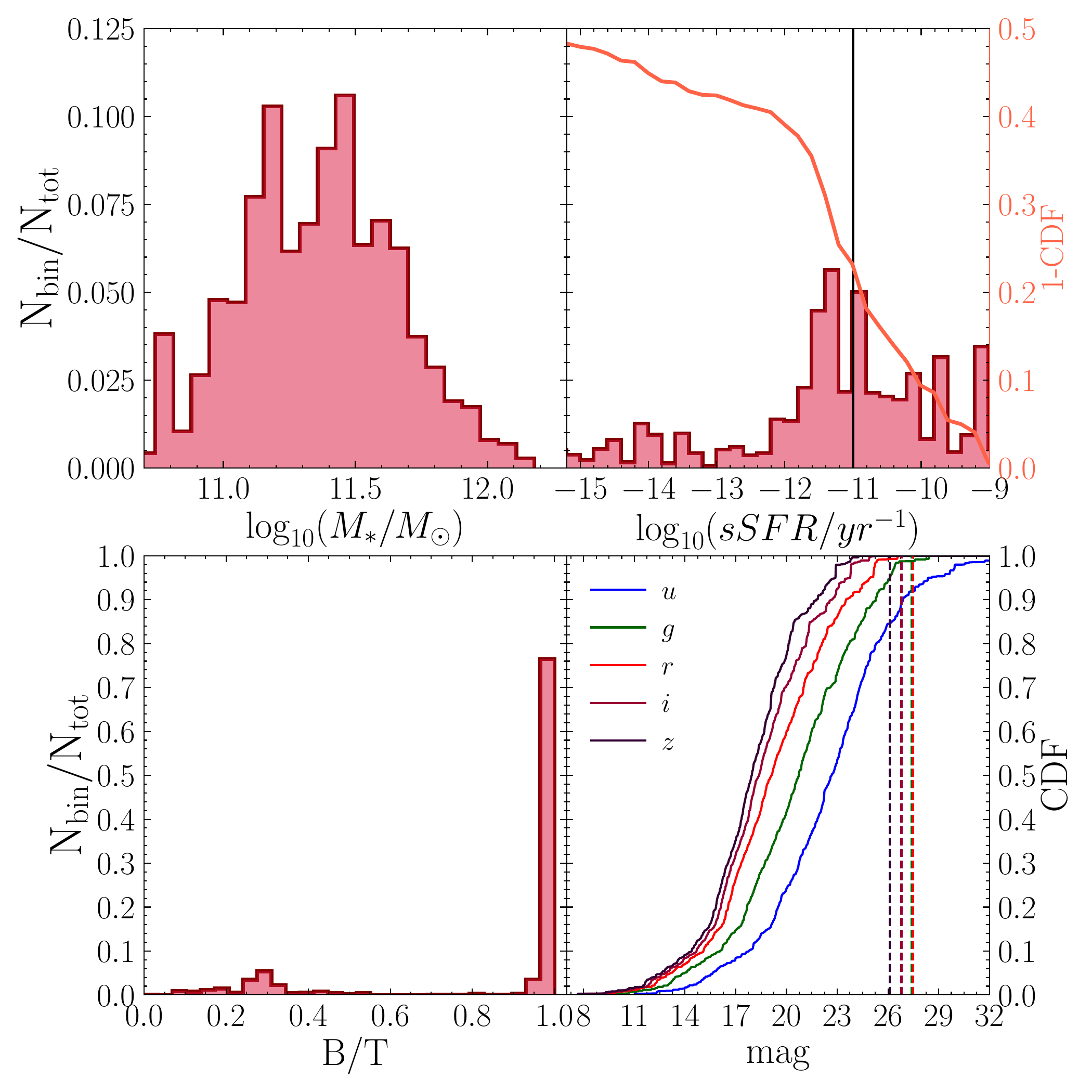}
    \caption{Properties of the galaxies hosting the MBHBs sourcing the CGWs detected by 30-yr SKA: Galaxy stellar mass ($\rm M_{*}$, upper left), specific star formation rate ($\rm sSFR$, upper right), bulge-to-total ratio ($\rm B/T$, lower left) and stellar magnitude in the ($u,g,r,i,z$) set of filters (lower right). In the panel with the sSFR the vertical line highlights the $\rm 10^{-11}\, \rm yr^{-1}$ value. On the other hand, the blue line corresponds to the cumulative distribution. Finally, the vertical lines correspond to the magnitude limits of LSST in the $u,g,r,i,z$ filters.}
    \label{fig:Properties_HostGalaxies}
\end{figure}


In Fig.~\ref{fig:Properties_HostGalaxies} we present the properties of the galaxies hosting the detected MBHBs. As shown, these systems are typically harbored in massive $\rm M_{\star}\,{>}\,10^{11} \, M_{\odot}$ elliptical (bulge-to-total ratio $\rm B/T\,{>}\,0.9$) galaxies, 
although there are cases where the galaxy can be disc-dominated ($\rm B/T\,{<}\,0.4$). Fig.~\ref{fig:Properties_HostGalaxies} also depicts the specific star formation rate of these galaxies. The values are typical for massive galaxies, with most hosts lying in the range 
$10^{-12}\, {\rm yr}^{-1}\,{<}\,{\rm sSRF}\,
\,{<}\,10^{-10}\, {\rm yr}^{-1}$. However, there is a long tail (${>}\,50\%$) of passive hosts extending down to ${\rm sSRF}\,{\sim}\,10^{-15}\, {\rm yr}^{-1}$.\\


The lower panel of Fig.~\ref{fig:Properties_HostGalaxies} presents the cumulative distribution function of the galaxy magnitude in the ($u,g,r,i,z$) filter set. As we can see, the detection of galaxies in the bluer filters ($u,g$) is disfavored with respect to the redder ones ($r,i,z$) given that their distributions are shifted towards larger magnitudes (i.e lower fluxes). For instance, ${\sim}\,90\%$ of the sources in the $r$ filter feature ${<}\,23 \, \rm mag$  while only ${\sim}\,50\%$ feature the same limit in the $u$ filter band. These results, together with the sSFR distributions, imply that the hosts of the detected MBHBs will be placed in red galaxies \citep[see similar results in][]{Cella2024}. Finally, when comparing the magnitude distributions with the LSST expected magnitude limits, we can see that the vast majority of the objects will be detected. This value reaches $\approx 100\%$ for ($g,r,i,z$) filters and $\approx 85\%$ for the $u$ filter.

\subsection{Host identification}
In this section, we explore the feasibility of unequivocally identifying the galaxy host of the MBHB sourcing the CGW detected by the SKA PTA. To this end, we will explore the typical sky localization area, the number of galaxies within it and different selection techniques to reduce the number of host candidates. We stress that we will only consider galaxies that are observable in the $r$-band, which is the deepest of the LSST filters\footnote{Similar results are found if we use different detection bands.}. We refer the reader to Appendix~\ref{appendix:Magnitude_N90} for the magnitude distribution of the host candidates.

\subsubsection{Potential hosts: number and properties}

\begin{figure}
    \centering
    \includegraphics[width=1\columnwidth]{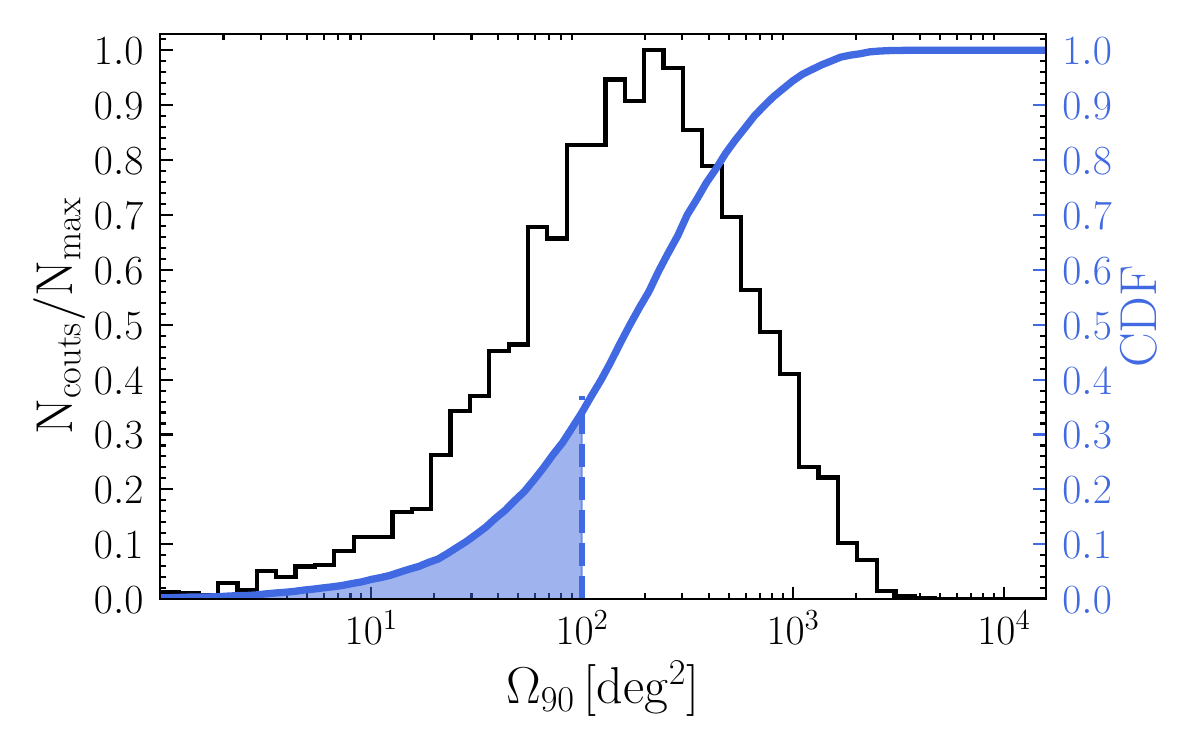}
    \caption{Histogram and cumulative density function (in blue)of $\Omega_{90}$ of the resolved sources. The light blue area represents the CGW with an $\Omega_{90}<100 \rm \,deg^2$ on which we will consider a doable electromagnetic follow-up.}
    \label{fig:Amplitude_And_Sky_Localization_Relation}
\end{figure}



\begin{figure}
    \centering
    \includegraphics[width=1\columnwidth]{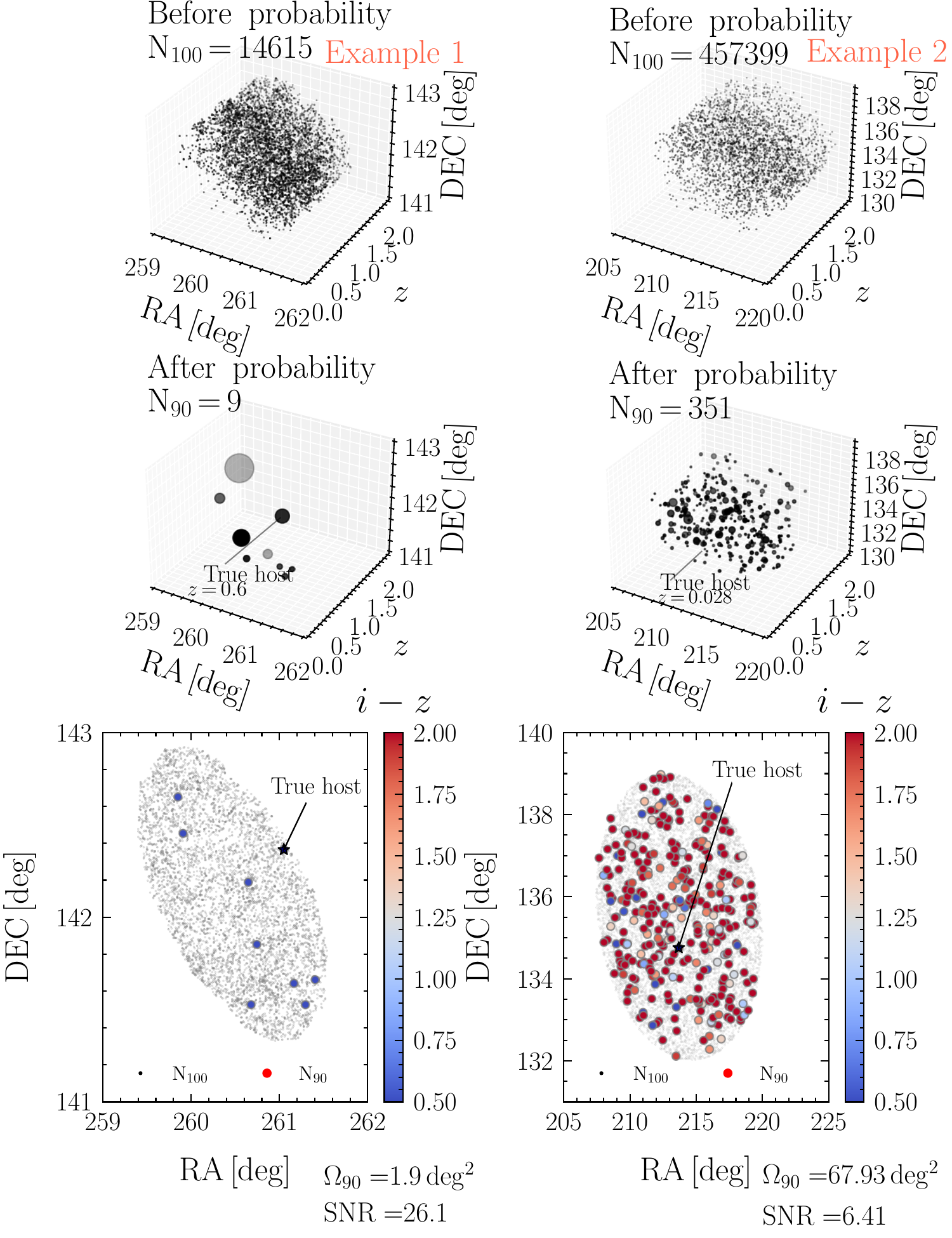}
    \caption{Distribution in the (RA, DEC, $z$) space of the galaxies with $\rm M_{*}\,{>}\,10^{10.3}\, M_{\odot}$ and $r$-band ${<}\,27.5\, \rm mag$ placed within the sky-localization area constrained by SKA PTA. \textbf{Top panel}: Distribution of the galaxies when no probability selection is performed. $\rm N_{100}$ indicates the total number of these galaxies. \textbf{Middle panel}: The same as before but when presenting only the galaxies which contribute with the 90\% probability of being the true host (see Section~\ref{sec:sky_localization_Methodology}). The size of the circle is proportional to the probability. \textbf{Lower panel}: Right ascension (RA) and declination (DEC) plane. The grey (colored) points correspond to all the galaxies in the $\rm N_{100}$ ($\rm N_{90}$) sample. The color encodes the photometric color $i\,{-}\,z$.}
    \label{fig:3D_Sky_Map}
\end{figure}

Fig.~\ref{fig:Amplitude_And_Sky_Localization_Relation} presents the distribution of $\Omega_{90}$ for all SKA PTA resolvable CGW sources. The distribution peaks at $\approx300\, \rm deg^2$ indicating that the search of the galaxy hosting the MBHB can be strongly disfavored. Despite that, ${\sim}\,35\%$ of the detected sources feature $\Omega_{90} \,{<}\,100 \, \rm deg^2$. From hereafter, we will examine the feasibility of unequivocally identifying the CGW hosts for those systems with $\Omega_{90}\,{<}\,100\, \rm deg^2$. To guide the reader, the left and right panels of Fig.~\ref{fig:3D_Sky_Map} depict two examples of galaxy distributions within the sky-localization regions associated to a detected CGW with $\Omega_{90} \,{<}\,100 \, \rm deg^2$. As shown, the lack of redshift constrains results in the true host being buried among hundreds of thousands of galaxies observable by LSST. However, as demonstrated in the middle panel of Fig.~\ref{fig:3D_Sky_Map}, by applying the methodology described in Section~\ref{sec:sky_localization_Methodology} we can significantly reduce the number of potential candidates.\\ 

To quantify the trend presented above, in Fig.~\ref{fig:N100_N90_N50} we present the number of candidates within the sky-localization area depending on the source $\rm S/N$ \citep[see similar results in][]{Tanaka2012}. Without any selection procedure, the total number of candidates ($\rm N_{100}$) varies between $3\,{\times}\,10^4\,{-}\,2\,{\times}\,10^5$, being the CGWs with larger $\rm S/N$ the ones featuring the lowest values. These large numbers would hinder a viable search for the true host. However, when we select only galaxies contributing 90\% ($\rm N_{90}$) and 50\% ($\rm N_{50}$) of the whole probability, the number of candidates drops by $\rm 2\,{-}\,5 \, dex$, depending on the $\rm S/N$. Despite this improvement, CGWs detected at $\rm S/N\,{\lesssim}\,25$ still have a large number of potential hosts (${>}\,100$). Conversely, the origin of the rarer CGWs detected with $\rm S/N\,{>}\,25$ can be traced to ${<}\,100$ potential candidates, making them promising targets for multimessenger studies. 

\subsubsection{Strategies for identifying hosts of CGWs with $\rm S/N\,{<}25$: Galaxy color cuts}
As shown in the previous section,  CGWs detected at $\rm S/N\,{>}\,25$ have  $\rm N_{90}\,{\lesssim}\,100$, making them the perfect targets for multimessenger follow-ups. However, these cases are rare, accounting for less than 5\% of the future detections  (i.e. ${\approx}\, 1$ CGW in a 30yr SKA PTA, see Fig.~\ref{fig:N_res}).  On the other hand, CGWs with $\rm S/N\,{<}\,25$ represent the most common scenarios, but they typically features $\rm 10^2{<}\,N_{90}\,{<}\,10^3$. Therefore, it is crucial to investigate whether candidates linked to CGWs with $\rm S/N\,{<}\,25$ show any differences in the optical properties compared to the true host, that can be easily identified by LSST\footnote{We refer to Appendix~\ref{appendix:Colors_High_SNR} for the same analysis but for sources with $\rm S/N{>}\,25$.}.\\

\begin{figure}
    \centering
    \includegraphics[width=1\columnwidth]{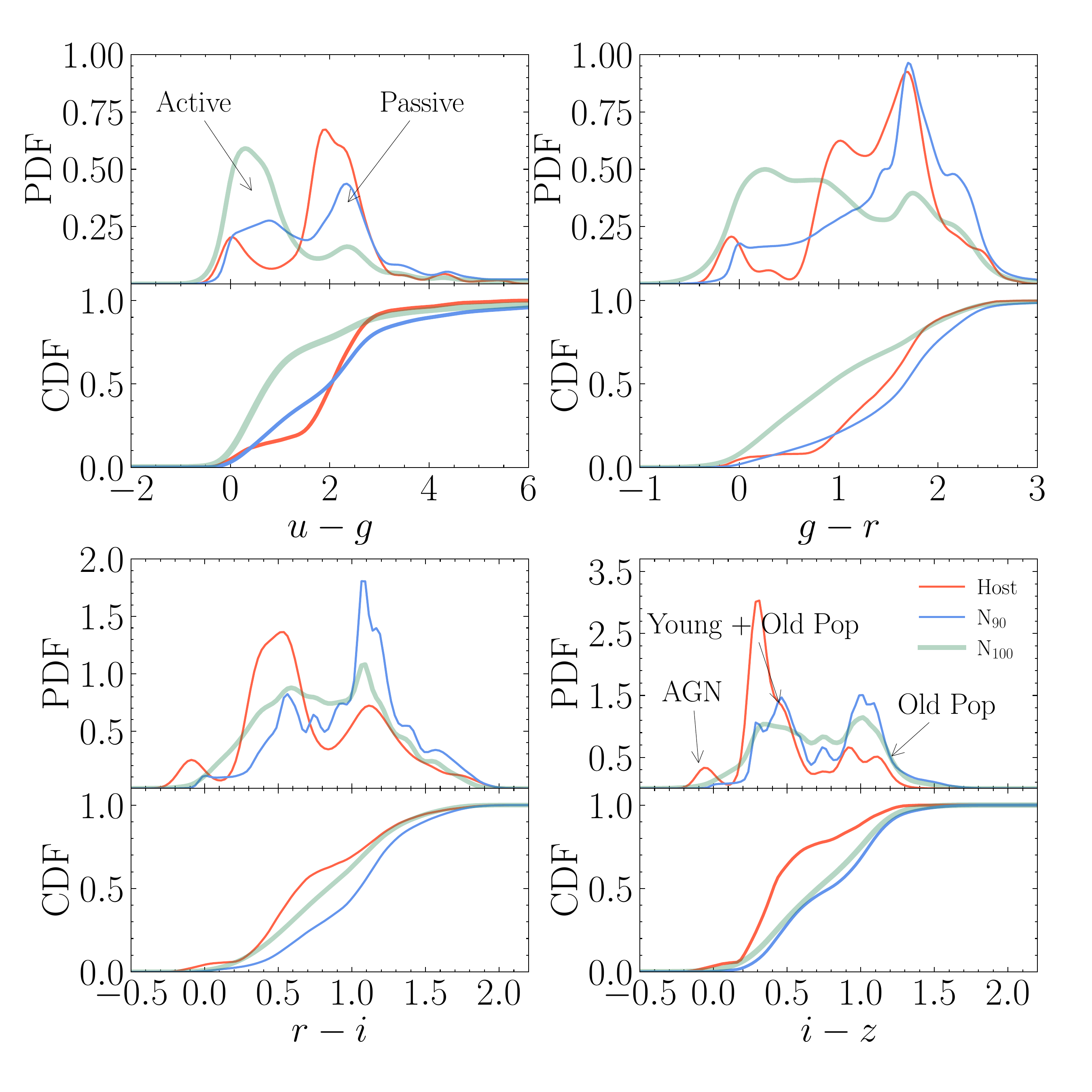}
    \caption{\textbf{Upper panels}: Probability distribution function (PDF) of the $u\,{-}\,g$, $g\,{-}\,r$, $r\,{-}\,i$ and $i\,{-}\,z$ colors for the galaxies within the sky-localization area ($\rm N_{100}$, green), galaxies within the sky-localization area contribution to the 90\% probability of being the true host ($N_{90}$, blue) and the true host (Host, red). \textbf{Lower panel}: Cumulative distribution function (CDF) of the $u\,{-}\,g$, $g\,{-}\,r$, $r\,{-}\,i$ and $i\,{-}\,z$ colors for $\rm N_{100}$, $\rm N_{90}$ and true host. In all the plots, the distributions correspond to the galaxies associated with CGWs whose $\rm S/N \,{<}\,25$.}
    \label{fig:Color_distribution}
\end{figure}

In Fig.~\ref{fig:Color_distribution}  we present the distribution of the optical color of the candidate galaxies ($N_{90}$) and true CGW hosts. Specifically, we have explored $u\,{-}\,g$, $g\,{-}\,r$, $r\,{-}\,i$ and $i\,{-}\,z$\footnote{We have checked that $\left(u\,{-}\,r, u\,{-}\,i, u\,{-}\,z\right)$ colors show similar trends of $u\,{-}\,g$. Similarly, $\left( g\,{-}\,i, g\,{-}\,z\right)$ are analogous to $g\,{-}\,r$ and $r\,{-}\,z$ to $r\,{-}\,i$.}. For instance, while the former color encodes the current star formation activity of the galaxy, the latter traces its past star formation history \citep[see e.g.][]{Strateva2001}. As presented in the figure, the distributions of the two samples overlap in all the presented colors and feature two predominant peaks at comparable values. For example, in the case of $u\,{-}\,g$, the true hosts exhibit a primary peak at $0.1$ and a secondary one at $2.5$. Host candidates also display a similar bi-modality, but their primary peak occurs at $u\,{-}\,g\,{=}\,2.5$. This behavior stems from the fact that galaxies composing the two samples are formed by star-forming (bluer colours) and passive (redder colours) systems. However, the relative abundance of these two classes of galaxies varies between the CGW hosts and $N_{90}$: while the predominant peak in the colors of the CGW  hosts rises due to the predominance of passive galaxies, the peak in the $\rm N_{90}$ sample is made of star-forming galaxies. A similar trend is observed in the $r\,{-}\,i$ and $i\,{-}\,z$ colors, although the two peaks convey slightly different information. The peaks at lower values correspond to galaxies that have undergone a relatively recent star formation event, which is now nearly complete or already finished, resulting in a mix of old and young stellar populations. For instance, in the case of CGW hosts, this last star formation process is likely associated with the galaxy merger that led to the formation of the MBHB. In contrast, the peaks at higher values represent galaxies with very evolved and old stellar populations that have not experienced any significant star formation activity in the Gyr.\\

\begin{figure}
    \centering
    \includegraphics[width=1\columnwidth]{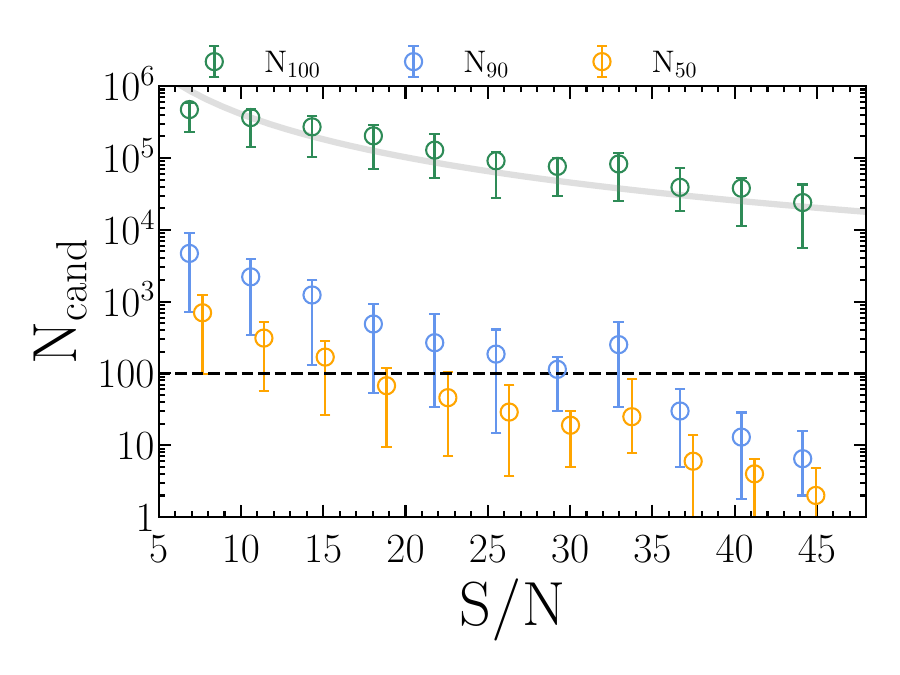}
    \caption{Number of candidates ($\rm N_{cand}$) within the sky-localization area depending on the $\rm S/N$ of the detected CGW source. Green circles correspond to the total number of galaxies with $\rm M_{*}\,{>}\,10^{10.3}\, M_{\odot}$ within the detection area ($\rm N_{100}$). Blue and orange circles represent the same thing, but only including the galaxies contributing 90\% and 50\% of the total probability of being the true host. To improve the plot clarity, orange points have been shifted by a factor of 0.5. The gray line corresponds to the expected $\rm 1/(S/N)^{2}$ relation.}
    \label{fig:N100_N90_N50}
\end{figure}

To determine if a simple color cut can reduce a significant fraction of the $\rm N_{90}$ galaxies, in the lower panels of  Fig.~\ref{fig:Color_distribution} we present the cumulative distribution function (CDF). As shown, the $g\,{-}\,r$ color is the one that provides the worst candidate reduction, since the CDF of candidates and true hosts are very similar. In contrast, the $i\,{-}\,z$ and $r\,{-}\,i$ colors are the most effective ones at distinguishing between candidates and true hosts. For instance, in the case of $i\,{-}\,z$ ($r\,{-}\,i$), we can reduce the number of sources by focusing on the 65\% of candidates that have colors ${\lesssim}\,0.80$ (${\lesssim}\,1.15$) which are compatible with those of the CGW hosts. 
In Table.~\ref{Table:Fraction_Lost}  we summarize the color cut required to minimize the risk of excluding the true host while reducing the number of potential candidates. As depicted in the table, the most promising scenario only envisions the reduction of ${\sim}\,37\%$ of the sources, which does not add decisive information for efficient multimessenger follow-up of low $\rm S/N$ CGWs (see Fig.~\ref{fig:N100_N90_N50}). Due to the absence of a single color cut that could effectively narrow down the candidate pool, in Fig.~\ref{fig:Color_color_Diagram}  we investigate whether the potential candidates and true host galaxies occupy distinct regions in color-color space.  As shown, no significant differences are seen in any of the color diagrams explored.\\

\renewcommand{\arraystretch}{1.05}
\begin{table}[]
\centering
\begin{adjustbox}{width=0.8\columnwidth,center}
\begin{tabular}{lcccc}
\hline 
\multicolumn{1}{c}{Color} & \multicolumn{1}{c}{$u-g$} & \multicolumn{1}{c}{$g-r$} & \multicolumn{1}{c}{$r-i$} & \multicolumn{1}{c}{$i-z$} \\ \hline \hline
 \cellcolor[HTML]{EFEFEF} Hosts lost                & \multicolumn{4}{c}{\cellcolor[HTML]{EFEFEF} \textbf{5\%}} \\
Color cut                 &  ${>}\,0.02$    &   ${>}\,-0.02$  &  ${<}\,1.48$    &  ${<}\,1.10$  \\
$\rm \, N_{90}$ reduction  &  1.78\%   &  0.67\%   &   8.20\%  &  13.64\%  \\ \hline \hline
\cellcolor[HTML]{EFEFEF} Hosts lost                & \multicolumn{4}{c}{\cellcolor[HTML]{EFEFEF} \textbf{10\%}} \\
Color cut                 &  ${>}\,0.20$    &  ${>}\,0.77$    &  ${<}\,1.32$    &  ${<}\,0.98$  \\
$\rm \, N_{90}$ reduction &  3.76\%   &  6.43\%    & 15.46\%    &  25.40\%  \\ \hline \hline
\cellcolor[HTML]{EFEFEF} Hosts lost                & \multicolumn{4}{c}{\cellcolor[HTML]{EFEFEF} \textbf{20\%}} \\
Color cut                 &  ${>}\,1.35$    & ${>}\,0.96$     &  ${<}\,1.15$    &  ${<}\,0.80$   \\
$\rm \, N_{90}$ reduction &  20.46\%   &   9.50\%   &   34.71\%  & 37.15\%  \\\hline 
\end{tabular}
\end{adjustbox}
\caption{Fraction of $\rm N_{90}$ galaxies removed after applying different optical color cuts with LSST filters. The host lost refers to the probability that the true host is removed after the color cut. These numbers only consider the galaxies associated with CGWs whose $\rm S/N\,{<}\,25$.}
\label{Table:Fraction_Lost}
\end{table}

Considering the results presented above, the absence of a definitive method to significantly narrow down the candidate pool underlines the complexities involved in identifying suitable PTA CGW sources for multimessenger follow-up observations.

\section{Conclusions} \label{sec:Conclusions}
In this paper, we have investigated the feasibility of performing multimessenger astronomy with continuous gravitational wave sources (CGWs) detected by an idealized 30-year SKA Pulsar Timing Array. To this end, we have employed the \texttt{L-Galaxies} semi-analytical model applied to the \texttt{Millennium} simulation. Specifically, we have generated 200 different all-sky lightcones that include galaxies, MHBs, and MBHBs whose emission has been modeled consistently based on their star formation histories and gas accretion physics. By applying the detection methodology presented in \cite{Truant2024} in our simulated Universes we find that:

\begin{figure}
    \centering
    \includegraphics[width=1\columnwidth]{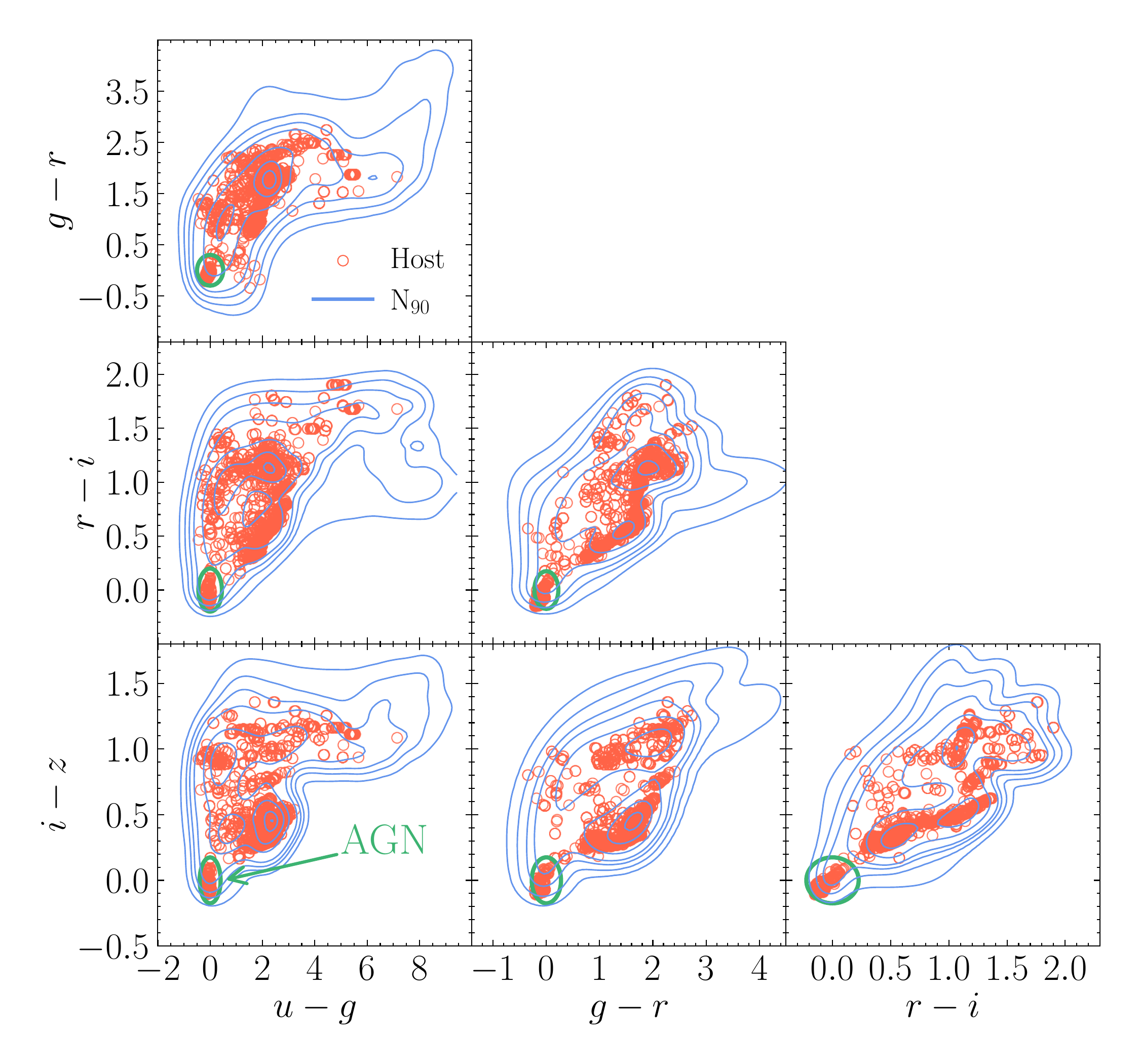}
    \caption{Color-color diagrams for the galaxies hosting the CGW (Hosts, red circles) and the galaxies inside the sky-localization area contributing to 90\% of the probability ($\rm N_{90}$, blue contours). In all the plots, the distributions correspond to the galaxies associated with CGWs with $\rm S/N \,{<}\,25$. Green circles represent the region where AGNs are expected to be placed in the color-color diagram.}
    \label{fig:Color_color_Diagram}
\end{figure}

\begin{itemize}
    \item SKA PTA is expected to detect an average of ${\sim}\,33$ CGWs. The majority of these detections (${>}\,90\%$) have $\rm S/N$ lower than 15. \\
    \item The detected CGW signals come from MBHBs placed at low-$z$ ($z\,{<}\,0.5$) with masses of $\rm {\sim}\,3\,{\times}\,10^9 \, M_{\odot}$, mass ratios ${>}\,0.1$ and eccentricities ${\lesssim}\,0.2$. \\
    \item Only 10\% of the detected CGWs are related to an AGN with optical magnitude ${\lesssim}\,27\, \rm mag$, i.e the typical detection limit of the deep-observing mode of LSST. Furthermore, only about 3\% of the CGWs reach the detection limit of the e-Rosita X-ray telescope.\\
    \item Detected CGWs are hosted in massive ($\rm M_{\star}\,{>}\, 10^{11}\, M_{\odot}$) 
    elliptical ($\rm B/T\,{>}\,0.9$) galaxies with typical sSFR, 
    easily detectable by LSST in the redder bands (${<}\,23\, \rm  mag$).\\
\end{itemize}

In light of these results, our study suggests that the unequivocal EM detection of PTA CGW counterparts should prioritize the search for galaxies over AGNs. With this in mind, we have explored the potential of conducting multimessenger astronomy using the legacy data of LSST which will allow us to create complete sky galaxy catalogues over about half of the sky. Our findings have shown that the typical sky-localization of our CGW sources is $\rm {\sim}\,300\, deg^2$ with about 35\% of them associated with areas $\rm {<}\,100\, deg^2$. Given the large sky-localization regions, we have only investigated the feasibility of unequivocally identifying the CGW hosts in the cases where the restricted area is lower than $\rm 100\, deg^2$.  The results can be summarized as follows:

\begin{itemize}

    \item The total number of candidates detectable by LSST ($\rm N_{100}$) varies between $3\,{\times}\,10^4\,{-}\,2\,{\times}\,10^5$. As expected, the CGWs with larger $\rm S/N$ ratios are the ones featuring the lowest $\rm N_{100}$ values because of their better sky-localization areas.\\  

    \item The number of potential candidates within the sky-localization area can be significantly reduced by using the methodology outlined in \cite{Goldstein2019}. This approach assigns a probability of being the true hosts by examining the physical relationship between the  CGW amplitude and the MBHB chirp mass, alongside empirical correlations between MBH mass and its host galaxy. By selecting only those candidates that contribute to 90\% of the total probability ($\rm N_{90}$), we find that for systems with $\rm S/N\,{<}\,25$ the values of $\rm N_{90}$ falls between $10^3\,{-}\,100$. For sources with an even higher S/N, $\rm N_{90}$ decreases to a range between $100\,{-}\,5$, making them the perfect sources for multimessenger follow-ups. \\

    \item  In cases where CGWs are detected with low $\rm S/N$ (${<}\,25$), applying certain optical color cuts can help reduce the number of $\rm N_{90}$. Specifically, the  $i\,{-}\,z$ color is the most effective for this purpose, as it distinguishes between galaxies with an old or mixed stellar population. For example, by selecting only galaxies with color of $i\,{-}\,z\,{<}\,0.80$, we can eliminate up to 37\% galaxies within $\rm N_{90}$ while having only a 20\% chance of mistakenly discarding the true host. Similar results have been found for the $r\,{-}\,i$ color. Despite this improvement, the number of candidates is still large. Color-color diagrams can also help in reducing  $\rm N_{90}$ but no strong improvement with respect to a single color cut has been noticed.\\ 

    \item In cases where CGWs are detected with large $\rm S/N$ (${>}\,25$), optical color cuts are more efficient in reducing the number of $\rm N_{90}$ candidates. In these cases, a cut in  $i\,{-}\,z\,{<}\,0.33$ can eliminate up to 70\% galaxies within $\rm N_{90}$ with a 20\% chance of mistakenly discarding the true host. This means the number of candidates can typically be reduced to ${<}\,50$.\\
    

\end{itemize}

We stress that these findings are based on a few conservative assumptions. One important caveat is that we assumed that PTAs cannot measure the frequency evolution ($\dot{f}$) of the CGW. Although this is a realistic assumption in most cases, it might be possible for a SKA array to measure $\dot{f}$ for some high S/N, high-frequency CGWs. This would allow to place a constraint on the source luminosity distance, possibly substantially narrowing down the number of candidate host galaxies within the (now 3-D) error volume. Furthermore, we ignored in our Fisher analysis the CGW pulsar terms, which might significantly improve source sky localization and distance measurement. It should be noticed, however, that identification of individual pulsar terms, especially in the presence of multiple overlapping CGWs in the data might be extremely challenging and has not been demonstrated.\\  

In summary, our investigation indicates that CGWs detected with high $\rm S/N$ will serve as ideal sources for conducting multimessenger studies with optical follow-ups. However, these studies will be more challenging for MBHBs detected with low $\rm S/N$, as there is no definitive methodology to significantly reduce the pool of candidates within their sky-localization areas. Since CGW detections will be dominated by low $\rm S/N$ systems, multi-messenger astronomy involving PTA sources is anticipated to be challenging.


\begin{acknowledgements}
        We thank the B-Massive group at Milano-Bicocca University for useful discussions and comments. R.T., D.I.V., A.S. and G.M.S acknowledge the financial support provided under the European Union’s H2020 ERC Consolidator Grant B Massive (Grant Agreement: 818691) and Advanced Grant PINGU (Grant Agreement: 101142097). M.B. acknowledges support provided by MUR under grant ``PNRR - Missione 4 Istruzione e Ricerca - Componente 2 Dalla Ricerca all'Impresa - Investimento 1.2 Finanziamento di progetti presentati da giovani ricercatori ID:SOE\_0163'' and by University of Milano-Bicocca under grant ``2022-NAZ-0482/B''. S.B.acknowledges support from the Spanish Ministerio de Ciencia e Innovación through project PID2021-124243NB-C21.
\end{acknowledgements}

%
%

\bibliographystyle{aa} 
\bibliography{references}

\begin{thebibliography}{107}
\expandafter\ifx\csname natexlab\endcsname\relax\def\natexlab#1{#1}\fi

\bibitem[{{Agazie} {et~al.}(2023){Agazie}, {Anumarlapudi}, {Archibald},
  {Arzoumanian}, {Baker}, {B{\'e}csy}, {Blecha}, {Brazier}, {Brook},
  {Burke-Spolaor}, {Burnette}, {Case}, {Charisi}, {Chatterjee},
  {Chatziioannou}, {Cheeseboro}, {Chen}, {Cohen}, {Cordes}, {Cornish},
  {Crawford}, {Cromartie}, {Crowter}, {Cutler}, {Decesar}, {Degan}, {Demorest},
  {Deng}, {Dolch}, {Drachler}, {Ellis}, {Ferrara}, {Fiore}, {Fonseca},
  {Freedman}, {Garver-Daniels}, {Gentile}, {Gersbach}, {Glaser}, {Good},
  {G{\"u}ltekin}, {Hazboun}, {Hourihane}, {Islo}, {Jennings}, {Johnson},
  {Jones}, {Kaiser}, {Kaplan}, {Kelley}, {Kerr}, {Key}, {Klein}, {Laal}, {Lam},
  {Lamb}, {Lazio}, {Lewandowska}, {Littenberg}, {Liu}, {Lommen}, {Lorimer},
  {Luo}, {Lynch}, {Ma}, {Madison}, {Mattson}, {McEwen}, {McKee}, {McLaughlin},
  {McMann}, {Meyers}, {Meyers}, {Mingarelli}, {Mitridate}, {Natarajan}, {Ng},
  {Nice}, {Ocker}, {Olum}, {Pennucci}, {Perera}, {Petrov}, {Pol}, {Radovan},
  {Ransom}, {Ray}, {Romano}, {Sardesai}, {Schmiedekamp}, {Schmiedekamp},
  {Schmitz}, {Schult}, {Shapiro-Albert}, {Siemens}, {Simon}, {Siwek}, {Stairs},
  {Stinebring}, {Stovall}, {Sun}, {Susobhanan}, {Swiggum}, {Taylor}, {Taylor},
  {Turner}, {Unal}, {Vallisneri}, {van Haasteren}, {Vigeland}, {Wahl}, {Wang},
  {Witt}, {Young}, \& {Nanograv Collaboration}}]{Agazie2023}
{Agazie}, G., {Anumarlapudi}, A., {Archibald}, A.~M., {et~al.} 2023, \apjl,
  951, L8

\bibitem[{{Amaro-Seoane} {et~al.}(2010){Amaro-Seoane}, {Sesana}, {Hoffman},
  {Benacquista}, {Eichhorn}, {Makino}, \& {Spurzem}}]{Amaro_Seoane2010}
{Amaro-Seoane}, P., {Sesana}, A., {Hoffman}, L., {et~al.} 2010, \mnras, 402,
  2308

\bibitem[{{Angulo} \& {White}(2010)}]{AnguloandWhite2010}
{Angulo}, R.~E. \& {White}, S.~D.~M. 2010, \mnras, 405, 143

\bibitem[{{Antoniadis} {et~al.}(2023){Antoniadis}, {Arumugam}, {Arumugam},
  {Babak}, {Bagchi}, {Bak Nielsen}, {Bassa}, {Bathula}, {Berthereau},
  {Bonetti}, {Bortolas}, {Brook}, {Burgay}, {Caballero}, {Chalumeau},
  {Champion}, {Chanlaridis}, {Chen}, {Cognard}, {Dandapat}, {Deb}, {Desai},
  {Desvignes}, {Dhanda-Batra}, {Dwivedi}, {Falxa}, {Ferdman}, {Franchini},
  {Gair}, {Goncharov}, {Gopakumar}, {Graikou}, {Grie{\ss}meier}, {Guillemot},
  {Guo}, {Gupta}, {Hisano}, {Hu}, {Iraci}, {Izquierdo-Villalba}, {Jang},
  {Jawor}, {Janssen}, {Jessner}, {Joshi}, {Kareem}, {Karuppusamy}, {Keane},
  {Keith}, {Kharbanda}, {Kikunaga}, {Kolhe}, {Kramer}, {Krishnakumar},
  {Lackeos}, {Lee}, {Liu}, {Liu}, {Lyne}, {McKee}, {Maan}, {Main},
  {Mickaliger}, {Nitu}, {Nobleson}, {Paladi}, {Parthasarathy}, {Perera},
  {Perrodin}, {Petiteau}, {Porayko}, {Possenti}, {Prabu}, {Quelquejay Leclere},
  {Rana}, {Samajdar}, {Sanidas}, {Sesana}, {Shaifullah}, {Singha}, {Speri},
  {Spiewak}, {Srivastava}, {Stappers}, {Surnis}, {Susarla}, {Susobhanan},
  {Takahashi}, {Tarafdar}, {Theureau}, {Tiburzi}, {van der Wateren}, {Vecchio},
  {Venkatraman Krishnan}, {Verbiest}, {Wang}, {Wang}, \& {Wu}}]{Antoniadis2023}
{Antoniadis}, J., {Arumugam}, P., {Arumugam}, S., {et~al.} 2023, arXiv
  e-prints, arXiv:2306.16214

\bibitem[{{Arzoumanian} {et~al.}(2021){Arzoumanian}, {Baker}, {Brazier},
  {Brook}, {Burke-Spolaor}, {Becsy}, {Charisi}, {Chatterjee}, {Cordes},
  {Cornish}, {Crawford}, {Cromartie}, {Decesar}, {Demorest}, {Dolch},
  {Elliott}, {Ellis}, {Ferrara}, {Fonseca}, {Garver-Daniels}, {Gentile},
  {Good}, {Hazboun}, {Islo}, {Jennings}, {Jones}, {Kaiser}, {Kaplan}, {Kelley},
  {Key}, {Lam}, {Lazio}, {Luo}, {Lynch}, {Ma}, {Madison}, {McLaughlin},
  {Mingarelli}, {Ng}, {Nice}, {Pennucci}, {Pol}, {Ransom}, {Ray},
  {Shapiro-Albert}, {Siemens}, {Simon}, {Spiewak}, {Stairs}, {Stinebring},
  {Stovall}, {Swiggum}, {Taylor}, {Vallisneri}, {Vigeland}, {Witt}, \&
  {Nanograv Collaboration}}]{Arzoumanian2021}
{Arzoumanian}, Z., {Baker}, P.~T., {Brazier}, A., {et~al.} 2021, \apj, 914, 121

\bibitem[{{Arzoumanian} {et~al.}(2015){Arzoumanian}, {Brazier},
  {Burke-Spolaor}, {Chamberlin}, {Chatterjee}, {Christy}, {Cordes}, {Cornish},
  {Demorest}, {Deng}, {Dolch}, {Ellis}, {Ferdman}, {Fonseca}, {Garver-Daniels},
  {Jenet}, {Jones}, {Kaspi}, {Koop}, {Lam}, {Lazio}, {Levin}, {Lommen},
  {Lorimer}, {Luo}, {Lynch}, {Madison}, {McLaughlin}, {McWilliams}, {Nice},
  {Palliyaguru}, {Pennucci}, {Ransom}, {Siemens}, {Stairs}, {Stinebring},
  {Stovall}, {Swiggum}, {Vallisneri}, {van Haasteren}, {Wang}, {Zhu}, \&
  {NANOGrav Collaboration}}]{Arzoumanian2015}
{Arzoumanian}, Z., {Brazier}, A., {Burke-Spolaor}, S., {et~al.} 2015, \apj,
  810, 150

\bibitem[{{Ayromlou} {et~al.}(2021){Ayromlou}, {Nelson}, {Yates}, {Kauffmann},
  {Renneby}, \& {White}}]{Ayromlou2021}
{Ayromlou}, M., {Nelson}, D., {Yates}, R.~M., {et~al.} 2021, \mnras, 502, 1051

\bibitem[{{Babak} \& {Sesana}(2012)}]{Babak2012}
{Babak}, S. \& {Sesana}, A. 2012, \prd, 85, 044034

\bibitem[{{Bates} {et~al.}(2014){Bates}, {Lorimer}, {Rane}, \&
  {Swiggum}}]{Bates2014}
{Bates}, S.~D., {Lorimer}, D.~R., {Rane}, A., \& {Swiggum}, J. 2014, \mnras,
  439, 2893

\bibitem[{{Begelman} {et~al.}(1980){Begelman}, {Blandford}, \&
  {Rees}}]{Begelman1980}
{Begelman}, M.~C., {Blandford}, R.~D., \& {Rees}, M.~J. 1980, \nat, 287, 307

\bibitem[{{Binney} \& {Tremaine}(2008)}]{BinneyTremaine2008}
{Binney}, J. \& {Tremaine}, S. 2008, {Galactic Dynamics: Second Edition}

\bibitem[{{Bonetti} {et~al.}(2018){Bonetti}, {Haardt}, {Sesana}, \&
  {Barausse}}]{Bonetti2018ModelGrid}
{Bonetti}, M., {Haardt}, F., {Sesana}, A., \& {Barausse}, E. 2018, \mnras, 477,
  3910

\bibitem[{{Bonetti} {et~al.}(2020){Bonetti}, {Rasskazov}, {Sesana}, {Dotti},
  {Haardt}, {Leigh}, {Arca Sedda}, {Fragione}, \& {Rossi}}]{Bonetti2020}
{Bonetti}, M., {Rasskazov}, A., {Sesana}, A., {et~al.} 2020, \mnras, 493, L114

\bibitem[{{Bonetti} {et~al.}(2019){Bonetti}, {Sesana}, {Haardt}, {Barausse}, \&
  {Colpi}}]{Bonetti2018a}
{Bonetti}, M., {Sesana}, A., {Haardt}, F., {Barausse}, E., \& {Colpi}, M. 2019,
  \mnras, 486, 4044

\bibitem[{{Bonoli} {et~al.}(2009){Bonoli}, {Marulli}, {Springel}, {White},
  {Branchini}, \& {Moscardini}}]{Bonoli2009}
{Bonoli}, S., {Marulli}, F., {Springel}, V., {et~al.} 2009, \mnras, 396, 423

\bibitem[{{Bortolas} {et~al.}(2022){Bortolas}, {Bonetti}, {Dotti}, {Lupi},
  {Capelo}, {Mayer}, \& {Sesana}}]{Bortolas2022}
{Bortolas}, E., {Bonetti}, M., {Dotti}, M., {et~al.} 2022, \mnras, 512, 3365

\bibitem[{{Bortolas} {et~al.}(2020){Bortolas}, {Capelo}, {Zana}, {Mayer},
  {Bonetti}, {Dotti}, {Davies}, \& {Madau}}]{Bortolas2020}
{Bortolas}, E., {Capelo}, P.~R., {Zana}, T., {et~al.} 2020, \mnras, 498, 3601

\bibitem[{{Burke-Spolaor}(2013)}]{BurkeSpolaor2013}
{Burke-Spolaor}, S. 2013, Classical and Quantum Gravity, 30, 224013

\bibitem[{{Cella} {et~al.}(2024){Cella}, {Taylor}, \& {Kelley}}]{Cella2024}
{Cella}, K., {Taylor}, S.~R., \& {Kelley}, L.~Z. 2024, arXiv e-prints,
  arXiv:2407.01659

\bibitem[{{Charisi} {et~al.}(2016){Charisi}, {Bartos}, {Haiman},
  {Price-Whelan}, {Graham}, {Bellm}, {Laher}, \& {M{\'a}rka}}]{Charisi2016}
{Charisi}, M., {Bartos}, I., {Haiman}, Z., {et~al.} 2016, \mnras, 463, 2145

\bibitem[{{Charisi} {et~al.}(2018){Charisi}, {Haiman}, {Schiminovich}, \&
  {D'Orazio}}]{Charisi2018}
{Charisi}, M., {Haiman}, Z., {Schiminovich}, D., \& {D'Orazio}, D.~J. 2018,
  \mnras, 476, 4617

\bibitem[{{Chen} {et~al.}(2019){Chen}, {Sesana}, \& {Conselice}}]{Chen2019}
{Chen}, S., {Sesana}, A., \& {Conselice}, C.~J. 2019, \mnras, 488, 401

\bibitem[{{Cocchiararo} {et~al.}(2024){Cocchiararo}, {Franchini}, {Lupi}, \&
  {Sesana}}]{Cocchiararo2024}
{Cocchiararo}, F., {Franchini}, A., {Lupi}, A., \& {Sesana}, A. 2024, \aap,
  691, A250

\bibitem[{{Croton}(2006)}]{Croton2006}
{Croton}, D.~J. 2006, \mnras, 369, 1808

\bibitem[{{Croton} {et~al.}(2006){Croton}, {Springel}, {White}, {De Lucia},
  {Frenk}, {Gao}, {Jenkins}, {Kauffmann}, {Navarro}, \&
  {Yoshida}}]{Croton2006a}
{Croton}, D.~J., {Springel}, V., {White}, S. D.~M., {et~al.} 2006, \mnras, 365,
  11

\bibitem[{{Cuadra} {et~al.}(2009){Cuadra}, {Armitage}, {Alexander}, \&
  {Begelman}}]{Cuadra2009}
{Cuadra}, J., {Armitage}, P.~J., {Alexander}, R.~D., \& {Begelman}, M.~C. 2009,
  \mnras, 393, 1423

\bibitem[{{Desvignes} {et~al.}(2016){Desvignes}, {Caballero}, {Lentati},
  {Verbiest}, {Champion}, {Stappers}, {Janssen}, {Lazarus}, {Os{\l}owski},
  {Babak}, {Bassa}, {Brem}, {Burgay}, {Cognard}, {Gair}, {Graikou},
  {Guillemot}, {Hessels}, {Jessner}, {Jordan}, {Karuppusamy}, {Kramer},
  {Lassus}, {Lazaridis}, {Lee}, {Liu}, {Lyne}, {McKee}, {Mingarelli},
  {Perrodin}, {Petiteau}, {Possenti}, {Purver}, {Rosado}, {Sanidas}, {Sesana},
  {Shaifullah}, {Smits}, {Taylor}, {Theureau}, {Tiburzi}, {van Haasteren}, \&
  {Vecchio}}]{Desvignes2016}
{Desvignes}, G., {Caballero}, R.~N., {Lentati}, L., {et~al.} 2016, \mnras, 458,
  3341

\bibitem[{{Dewdney} {et~al.}(2009){Dewdney}, {Hall}, {Schilizzi}, \&
  {Lazio}}]{Dewdney2009}
{Dewdney}, P.~E., {Hall}, P.~J., {Schilizzi}, R.~T., \& {Lazio}, T.~J.~L.~W.
  2009, IEEE Proceedings, 97, 1482

\bibitem[{{Dotti} {et~al.}(2007){Dotti}, {Colpi}, {Haardt}, \&
  {Mayer}}]{Dotti2007}
{Dotti}, M., {Colpi}, M., {Haardt}, F., \& {Mayer}, L. 2007, \mnras, 379, 956

\bibitem[{{Dotti} {et~al.}(2015){Dotti}, {Merloni}, \& {Montuori}}]{Dotti2015}
{Dotti}, M., {Merloni}, A., \& {Montuori}, C. 2015, \mnras, 448, 3603

\bibitem[{{Duffell} {et~al.}(2020){Duffell}, {D'Orazio}, {Derdzinski},
  {Haiman}, {MacFadyen}, {Rosen}, \& {Zrake}}]{Duffell2020}
{Duffell}, P.~C., {D'Orazio}, D., {Derdzinski}, A., {et~al.} 2020, \apj, 901,
  25

\bibitem[{{EPTA Collaboration} {et~al.}(2023{\natexlab{a}}){EPTA
  Collaboration}, {InPTA Collaboration}, {Antoniadis}, {Arumugam}, {Arumugam},
  {Babak}, {Bagchi}, {Bak Nielsen}, {Bassa}, {Bathula}, {Berthereau},
  {Bonetti}, {Bortolas}, {Brook}, {Burgay}, {Caballero}, {Chalumeau},
  {Champion}, {Chanlaridis}, {Chen}, {Cognard}, {Dandapat}, {Deb}, {Desai},
  {Desvignes}, {Dhanda-Batra}, {Dwivedi}, {Falxa}, {Ferdman}, {Franchini},
  {Gair}, {Goncharov}, {Gopakumar}, {Graikou}, {Grie{\ss}meier}, {Guillemot},
  {Guo}, {Gupta}, {Hisano}, {Hu}, {Iraci}, {Izquierdo-Villalba}, {Jang},
  {Jawor}, {Janssen}, {Jessner}, {Joshi}, {Kareem}, {Karuppusamy}, {Keane},
  {Keith}, {Kharbanda}, {Kikunaga}, {Kolhe}, {Kramer}, {Krishnakumar},
  {Lackeos}, {Lee}, {Liu}, {Liu}, {Lyne}, {McKee}, {Maan}, {Main},
  {Mickaliger}, {Ni{\c{t}}u}, {Nobleson}, {Paladi}, {Parthasarathy}, {Perera},
  {Perrodin}, {Petiteau}, {Porayko}, {Possenti}, {Prabu}, {Quelquejay Leclere},
  {Rana}, {Samajdar}, {Sanidas}, {Sesana}, {Shaifullah}, {Singha}, {Speri},
  {Spiewak}, {Srivastava}, {Stappers}, {Surnis}, {Susarla}, {Susobhanan},
  {Takahashi}, {Tarafdar}, {Theureau}, {Tiburzi}, {van der Wateren}, {Vecchio},
  {Venkatraman Krishnan}, {Verbiest}, {Wang}, {Wang}, \& {Wu}}]{EPTA_GW_2023}
{EPTA Collaboration}, {InPTA Collaboration}, {Antoniadis}, J., {et~al.}
  2023{\natexlab{a}}, \aap, 678, A50

\bibitem[{{EPTA Collaboration} {et~al.}(2024){EPTA Collaboration}, {InPTA
  Collaboration}, {Antoniadis}, {Arumugam}, {Arumugam}, {Babak}, {Bagchi}, {Bak
  Nielsen}, {Bassa}, {Bathula}, {Berthereau}, {Bonetti}, {Bortolas}, {Brook},
  {Burgay}, {Caballero}, {Chalumeau}, {Champion}, {Chanlaridis}, {Chen},
  {Cognard}, {Dandapat}, {Deb}, {Desai}, {Desvignes}, {Dhanda-Batra},
  {Dwivedi}, {Falxa}, {Ferdman}, {Franchini}, {Gair}, {Goncharov}, {Gopakumar},
  {Graikou}, {Grie{\ss}meier}, {Gualandris}, {Guillemot}, {Guo}, {Gupta},
  {Hisano}, {Hu}, {Iraci}, {Izquierdo-Villalba}, {Jang}, {Jawor}, {Janssen},
  {Jessner}, {Joshi}, {Kareem}, {Karuppusamy}, {Keane}, {Keith}, {Kharbanda},
  {Kikunaga}, {Kolhe}, {Kramer}, {Krishnakumar}, {Lackeos}, {Lee}, {Liu},
  {Liu}, {Lyne}, {McKee}, {Maan}, {Main}, {Mickaliger}, {Ni{\c{t}}u},
  {Nobleson}, {Paladi}, {Parthasarathy}, {Perera}, {Perrodin}, {Petiteau},
  {Porayko}, {Possenti}, {Prabu}, {Quelquejay Leclere}, {Rana}, {Samajdar},
  {Sanidas}, {Sesana}, {Shaifullah}, {Singha}, {Speri}, {Spiewak},
  {Srivastava}, {Stappers}, {Surnis}, {Susarla}, {Susobhanan}, {Takahashi},
  {Tarafdar}, {Theureau}, {Tiburzi}, {van der Wateren}, {Vecchio}, {Venkatraman
  Krishnan}, {Verbiest}, {Wang}, {Wang}, {Wu}, {Auclair}, {Barausse},
  {Caprini}, {Crisostomi}, {Fastidio}, {Khizriev}, {Middleton}, {Neronov},
  {Postnov}, {Roper Pol}, {Semikoz}, {Smarra}, {Steer}, {Truant}, \&
  {Valtolina}}]{InterpretationPaperEPTA2023}
{EPTA Collaboration}, {InPTA Collaboration}, {Antoniadis}, J., {et~al.} 2024,
  \aap, 685, A94

\bibitem[{{EPTA Collaboration} {et~al.}(2023{\natexlab{b}}){EPTA
  Collaboration}, {InPTA Collaboration}, {Antoniadis}, {Arumugam}, {Arumugam},
  {Babak}, {Bagchi}, {Nielsen}, {Bassa}, {Bathula}, {Berthereau}, {Bonetti},
  {Bortolas}, {Brook}, {Burgay}, {Caballero}, {Chalumeau}, {Champion},
  {Chanlaridis}, {Chen}, {Cognard}, {Dandapat}, {Deb}, {Desai}, {Desvignes},
  {Dhanda-Batra}, {Dwivedi}, {Falxa}, {Ferdman}, {Franchini}, {Gair},
  {Goncharov}, {Gopakumar}, {Graikou}, {Grie{\ss}meier}, {Guillemot}, {Guo},
  {Gupta}, {Hisano}, {Hu}, {Iraci}, {Izquierdo-Villalba}, {Jang}, {Jawor},
  {Janssen}, {Jessner}, {Joshi}, {Kareem}, {Karuppusamy}, {Keane}, {Keith},
  {Kharbanda}, {Kikunaga}, {Kolhe}, {Kramer}, {Krishnakumar}, {Lackeos}, {Lee},
  {Liu}, {Liu}, {Lyne}, {McKee}, {Maan}, {Main}, {Mickaliger}, {Ni{\c{t}}u},
  {Nobleson}, {Paladi}, {Parthasarathy}, {Perera}, {Perrodin}, {Petiteau},
  {Porayko}, {Possenti}, {Prabu}, {Leclere}, {Rana}, {Samajdar}, {Sanidas},
  {Sesana}, {Shaifullah}, {Singha}, {Speri}, {Spiewak}, {Srivastava},
  {Stappers}, {Surnis}, {Susarla}, {Susobhanan}, {Takahashi}, {Tarafdar},
  {Theureau}, {Tiburzi}, {van der Wateren}, {Vecchio}, {Krishnan}, {Verbiest},
  {Wang}, {Wang}, \& {Wu}}]{Custom_noise_EPTA}
{EPTA Collaboration}, {InPTA Collaboration}, {Antoniadis}, J., {et~al.}
  2023{\natexlab{b}}, \aap, 678, A49

\bibitem[{{Escala} {et~al.}(2004){Escala}, {Larson}, {Coppi}, \&
  {Mardones}}]{Escala2004}
{Escala}, A., {Larson}, R.~B., {Coppi}, P.~S., \& {Mardones}, D. 2004, \apj,
  607, 765

\bibitem[{{Falxa} {et~al.}(2023){Falxa}, {Babak}, {Baker}, {B{\'e}csy},
  {Chalumeau}, {Chen}, {Chen}, {Cornish}, {Guillemot}, {Hazboun}, {Mingarelli},
  {Parthasarathy}, {Petiteau}, {Pol}, {Sesana}, {Spolaor}, {Taylor},
  {Theureau}, {Vallisneri}, {Vigeland}, {Witt}, {Zhu}, {Antoniadis},
  {Arzoumanian}, {Bailes}, {Bhat}, {Blecha}, {Brazier}, {Brook}, {Caballero},
  {Cameron}, {Casey-Clyde}, {Champion}, {Charisi}, {Chatterjee}, {Cognard},
  {Cordes}, {Crawford}, {Cromartie}, {Crowter}, {Dai}, {DeCesar}, {Demorest},
  {Desvignes}, {Dolch}, {Drachler}, {Feng}, {Ferrara}, {Fiore}, {Fonseca},
  {Garver-Daniels}, {Glaser}, {Goncharov}, {Good}, {Griessmeier}, {Guo},
  {G{\"u}ltekin}, {Hobbs}, {Hu}, {Islo}, {Jang}, {Jennings}, {Johnson},
  {Jones}, {Kaczmarek}, {Kaiser}, {Kaplan}, {Keith}, {Kelley}, {Kerr}, {Key},
  {Laal}, {Lam}, {Lamb}, {Lazio}, {Liu}, {Liu}, {Luo}, {Lynch}, {Madison},
  {Main}, {Manchester}, {McEwen}, {McKee}, {McLaughlin}, {Ng}, {Nice}, {Ocker},
  {Olum}, {Os{\l}owski}, {Pennucci}, {Perera}, {Perrodin}, {Porayko},
  {Possenti}, {Quelquejay-Leclere}, {Ransom}, {Ray}, {Reardon}, {Russell},
  {Samajdar}, {Sarkissian}, {Schult}, {Shaifullah}, {Shannon},
  {Shapiro-Albert}, {Siemens}, {Simon}, {Siwek}, {Smith}, {Speri}, {Spiewak},
  {Stairs}, {Stappers}, {Stinebring}, {Swiggum}, {Tiburzi}, {Turner},
  {Vecchio}, {Verbiest}, {Wahl}, {Wang}, {Wang}, {Wang}, {Wu}, {Zhang},
  {Zhang}, \& {IPTA Collaboration}}]{Falxa2023}
{Falxa}, M., {Babak}, S., {Baker}, P.~T., {et~al.} 2023, \mnras, 521, 5077

\bibitem[{{Fiacconi} {et~al.}(2013){Fiacconi}, {Mayer}, {Ro{\v{s}}kar}, \&
  {Colpi}}]{Fiacconi2013}
{Fiacconi}, D., {Mayer}, L., {Ro{\v{s}}kar}, R., \& {Colpi}, M. 2013, \apjl,
  777, L14

\bibitem[{{Fontecilla} {et~al.}(2017){Fontecilla}, {Chen}, \&
  {Cuadra}}]{Fontecilla2017}
{Fontecilla}, C., {Chen}, X., \& {Cuadra}, J. 2017, \mnras, 468, L50

\bibitem[{{Foster} \& {Backer}(1990)}]{Foster1990}
{Foster}, R.~S. \& {Backer}, D.~C. 1990, \apj, 361, 300

\bibitem[{{Franchini} {et~al.}(2022){Franchini}, {Lupi}, \&
  {Sesana}}]{Franchini2022}
{Franchini}, A., {Lupi}, A., \& {Sesana}, A. 2022, \apjl, 929, L13

\bibitem[{{Franchini} {et~al.}(2023){Franchini}, {Lupi}, {Sesana}, \&
  {Haiman}}]{Franchini2023}
{Franchini}, A., {Lupi}, A., {Sesana}, A., \& {Haiman}, Z. 2023, \mnras, 522,
  1569

\bibitem[{{Franchini} {et~al.}(2021){Franchini}, {Sesana}, \&
  {Dotti}}]{Franchini2021}
{Franchini}, A., {Sesana}, A., \& {Dotti}, M. 2021, \mnras, 507, 1458

\bibitem[{{Gardiner} {et~al.}(2023){Gardiner}, {Kelley}, {Lemke}, \&
  {Mitridate}}]{Gardiner2023}
{Gardiner}, E.~C., {Kelley}, L.~Z., {Lemke}, A.-M., \& {Mitridate}, A. 2023,
  arXiv e-prints, arXiv:2309.07227

\bibitem[{{Goldstein} {et~al.}(2019){Goldstein}, {Sesana}, {Holgado}, \&
  {Veitch}}]{Goldstein2019}
{Goldstein}, J.~M., {Sesana}, A., {Holgado}, A.~M., \& {Veitch}, J. 2019,
  \mnras, 485, 248

\bibitem[{{Guo} {et~al.}(2011){Guo}, {White}, {Boylan-Kolchin}, {De Lucia},
  {Kauffmann}, {Lemson}, {Li}, {Springel}, \& {Weinmann}}]{Guo2011}
{Guo}, Q., {White}, S., {Boylan-Kolchin}, M., {et~al.} 2011, \mnras, 413, 101

\bibitem[{{Haiman} {et~al.}(2009){Haiman}, {Kocsis}, \& {Menou}}]{Haiman2009}
{Haiman}, Z., {Kocsis}, B., \& {Menou}, K. 2009, \apj, 700, 1952

\bibitem[{{Henriques} {et~al.}(2015){Henriques}, {White}, {Thomas}, {Angulo},
  {Guo}, {Lemson}, {Springel}, \& {Overzier}}]{Henriques2015}
{Henriques}, B. M.~B., {White}, S. D.~M., {Thomas}, P.~A., {et~al.} 2015,
  \mnras, 451, 2663

\bibitem[{{Henriques} {et~al.}(2020){Henriques}, {Yates}, {Fu}, {Guo},
  {Kauffmann}, {Srisawat}, {Thomas}, \& {White}}]{Henriques2020}
{Henriques}, B. M.~B., {Yates}, R.~M., {Fu}, J., {et~al.} 2020, \mnras, 491,
  5795

\bibitem[{{Hopkins} \& {Quataert}(2010)}]{HopkinsQuataert2010}
{Hopkins}, P.~F. \& {Quataert}, E. 2010, \mnras, 407, 1529

\bibitem[{{Husa}(2009)}]{Husa2009}
{Husa}, S. 2009, General Relativity and Gravitation, 41, 1667

\bibitem[{{Ivanov} {et~al.}(1999){Ivanov}, {Papaloizou}, \&
  {Polnarev}}]{Ivanov1999}
{Ivanov}, P.~B., {Papaloizou}, J.~C.~B., \& {Polnarev}, A.~G. 1999, \mnras,
  307, 79

\bibitem[{{Ivezi{\'c}} {et~al.}(2019){Ivezi{\'c}}, {Kahn}, {Tyson}, {Abel},
  {Acosta}, {Allsman}, {Alonso}, {AlSayyad}, {Anderson}, {Andrew}, {Angel},
  {Angeli}, {Ansari}, {Antilogus}, {Araujo}, {Armstrong}, {Arndt}, {Astier},
  {Aubourg}, {Auza}, {Axelrod}, {Bard}, {Barr}, {Barrau}, {Bartlett}, {Bauer},
  {Bauman}, {Baumont}, {Bechtol}, {Bechtol}, {Becker}, {Becla}, {Beldica},
  {Bellavia}, {Bianco}, {Biswas}, {Blanc}, {Blazek}, {Blandford}, {Bloom},
  {Bogart}, {Bond}, {Booth}, {Borgland}, {Borne}, {Bosch}, {Boutigny},
  {Brackett}, {Bradshaw}, {Brandt}, {Brown}, {Bullock}, {Burchat}, {Burke},
  {Cagnoli}, {Calabrese}, {Callahan}, {Callen}, {Carlin}, {Carlson},
  {Chandrasekharan}, {Charles-Emerson}, {Chesley}, {Cheu}, {Chiang}, {Chiang},
  {Chirino}, {Chow}, {Ciardi}, {Claver}, {Cohen-Tanugi}, {Cockrum}, {Coles},
  {Connolly}, {Cook}, {Cooray}, {Covey}, {Cribbs}, {Cui}, {Cutri}, {Daly},
  {Daniel}, {Daruich}, {Daubard}, {Daues}, {Dawson}, {Delgado}, {Dellapenna},
  {de Peyster}, {de Val-Borro}, {Digel}, {Doherty}, {Dubois},
  {Dubois-Felsmann}, {Durech}, {Economou}, {Eifler}, {Eracleous}, {Emmons},
  {Fausti Neto}, {Ferguson}, {Figueroa}, {Fisher-Levine}, {Focke}, {Foss},
  {Frank}, {Freemon}, {Gangler}, {Gawiser}, {Geary}, {Gee}, {Geha}, {Gessner},
  {Gibson}, {Gilmore}, {Glanzman}, {Glick}, {Goldina}, {Goldstein}, {Goodenow},
  {Graham}, {Gressler}, {Gris}, {Guy}, {Guyonnet}, {Haller}, {Harris},
  {Hascall}, {Haupt}, {Hernandez}, {Herrmann}, {Hileman}, {Hoblitt}, {Hodgson},
  {Hogan}, {Howard}, {Huang}, {Huffer}, {Ingraham}, {Innes}, {Jacoby}, {Jain},
  {Jammes}, {Jee}, {Jenness}, {Jernigan}, {Jevremovi{\'c}}, {Johns}, {Johnson},
  {Johnson}, {Jones}, {Juramy-Gilles}, {Juri{\'c}}, {Kalirai}, {Kallivayalil},
  {Kalmbach}, {Kantor}, {Karst}, {Kasliwal}, {Kelly}, {Kessler}, {Kinnison},
  {Kirkby}, {Knox}, {Kotov}, {Krabbendam}, {Krughoff}, {Kub{\'a}nek},
  {Kuczewski}, {Kulkarni}, {Ku}, {Kurita}, {Lage}, {Lambert}, {Lange},
  {Langton}, {Le Guillou}, {Levine}, {Liang}, {Lim}, {Lintott}, {Long},
  {Lopez}, {Lotz}, {Lupton}, {Lust}, {MacArthur}, {Mahabal}, {Mandelbaum},
  {Markiewicz}, {Marsh}, {Marshall}, {Marshall}, {May}, {McKercher}, {McQueen},
  {Meyers}, {Migliore}, {Miller}, {Mills}, {Miraval}, {Moeyens}, {Moolekamp},
  {Monet}, {Moniez}, {Monkewitz}, {Montgomery}, {Morrison}, {Mueller},
  {Muller}, {Mu{\~n}oz Arancibia}, {Neill}, {Newbry}, {Nief}, {Nomerotski},
  {Nordby}, {O'Connor}, {Oliver}, {Olivier}, {Olsen}, {O'Mullane}, {Ortiz},
  {Osier}, {Owen}, {Pain}, {Palecek}, {Parejko}, {Parsons}, {Pease},
  {Peterson}, {Peterson}, {Petravick}, {Libby Petrick}, {Petry},
  {Pierfederici}, {Pietrowicz}, {Pike}, {Pinto}, {Plante}, {Plate}, {Plutchak},
  {Price}, {Prouza}, {Radeka}, {Rajagopal}, {Rasmussen}, {Regnault}, {Reil},
  {Reiss}, {Reuter}, {Ridgway}, {Riot}, {Ritz}, {Robinson}, {Roby}, {Roodman},
  {Rosing}, {Roucelle}, {Rumore}, {Russo}, {Saha}, {Sassolas}, {Schalk},
  {Schellart}, {Schindler}, {Schmidt}, {Schneider}, {Schneider}, {Schoening},
  {Schumacher}, {Schwamb}, {Sebag}, {Selvy}, {Sembroski}, {Seppala}, {Serio},
  {Serrano}, {Shaw}, {Shipsey}, {Sick}, {Silvestri}, {Slater}, {Smith},
  {Smith}, {Sobhani}, {Soldahl}, {Storrie-Lombardi}, {Stover}, {Strauss},
  {Street}, {Stubbs}, {Sullivan}, {Sweeney}, {Swinbank}, {Szalay}, {Takacs},
  {Tether}, {Thaler}, {Thayer}, {Thomas}, {Thornton}, {Thukral}, {Tice},
  {Trilling}, {Turri}, {Van Berg}, {Vanden Berk}, {Vetter}, {Virieux},
  {Vucina}, {Wahl}, {Walkowicz}, {Walsh}, {Walter}, {Wang}, {Wang}, {Warner},
  {Wiecha}, {Willman}, {Winters}, {Wittman}, {Wolff}, {Wood-Vasey}, {Wu},
  {Xin}, {Yoachim}, \& {Zhan}}]{Ivezic2019}
{Ivezi{\'c}}, {\v{Z}}., {Kahn}, S.~M., {Tyson}, J.~A., {et~al.} 2019, \apj,
  873, 111

\bibitem[{{Izquierdo-Villalba}
  {et~al.}(2019{\natexlab{a}}){Izquierdo-Villalba}, {Angulo}, {Orsi}, {Hurier},
  {Vilella-Rojo}, {Bonoli}, {L{\'o}pez-Sanjuan}, {Alcaniz}, {Cenarro},
  {Crist{\'o}bal-Hornillos}, {Dupke}, {Ederoclite}, {Hern{\'a}ndez-Monteagudo},
  {Mar{\'\i}n-Franch}, {Moles}, {Mendes de Oliveira}, {Sodr{\'e}}, {Varela}, \&
  {V{\'a}zquez Rami{\'o}}}]{IzquierdoVillalba2019LC}
{Izquierdo-Villalba}, D., {Angulo}, R.~E., {Orsi}, A., {et~al.}
  2019{\natexlab{a}}, \aap, 631, A82

\bibitem[{{Izquierdo-Villalba} {et~al.}(2020){Izquierdo-Villalba}, {Bonoli},
  {Dotti}, {Sesana}, {Rosas-Guevara}, \& {Spinoso}}]{IzquierdoVillalba2020}
{Izquierdo-Villalba}, D., {Bonoli}, S., {Dotti}, M., {et~al.} 2020, \mnras,
  495, 4681

\bibitem[{{Izquierdo-Villalba}
  {et~al.}(2019{\natexlab{b}}){Izquierdo-Villalba}, {Bonoli}, {Spinoso},
  {Rosas-Guevara}, {Henriques}, \&
  {Hern{\'a}ndez-Monteagudo}}]{IzquierdoVillalba2019}
{Izquierdo-Villalba}, D., {Bonoli}, S., {Spinoso}, D., {et~al.}
  2019{\natexlab{b}}, \mnras, 488, 609

\bibitem[{{Izquierdo-Villalba} {et~al.}(2022){Izquierdo-Villalba}, {Sesana},
  {Bonoli}, \& {Colpi}}]{IzquierdoVillalba2022}
{Izquierdo-Villalba}, D., {Sesana}, A., {Bonoli}, S., \& {Colpi}, M. 2022,
  \mnras, 509, 3488

\bibitem[{{Izquierdo-Villalba} {et~al.}(2024){Izquierdo-Villalba}, {Sesana},
  {Colpi}, {Spinoso}, {Bonetti}, {Bonoli}, \&
  {Valiante}}]{IzquierdoVillalba2024}
{Izquierdo-Villalba}, D., {Sesana}, A., {Colpi}, M., {et~al.} 2024, \aap, 686,
  A183

\bibitem[{{Kelley} {et~al.}(2018){Kelley}, {Blecha}, {Hernquist}, {Sesana}, \&
  {Taylor}}]{Kelley2018}
{Kelley}, L.~Z., {Blecha}, L., {Hernquist}, L., {Sesana}, A., \& {Taylor},
  S.~R. 2018, \mnras, 477, 964

\bibitem[{{Kelley} {et~al.}(2019){Kelley}, {Haiman}, {Sesana}, \&
  {Hernquist}}]{Kelley2019}
{Kelley}, L.~Z., {Haiman}, Z., {Sesana}, A., \& {Hernquist}, L. 2019, \mnras,
  485, 1579

\bibitem[{{Kramer} \& {Champion}(2013)}]{Kramer2013}
{Kramer}, M. \& {Champion}, D.~J. 2013, Classical and Quantum Gravity, 30,
  224009

\bibitem[{{Lee}(2016)}]{Lee2016}
{Lee}, K.~J. 2016, in Astronomical Society of the Pacific Conference Series,
  Vol. 502, Frontiers in Radio Astronomy and FAST Early Sciences Symposium
  2015, ed. L.~{Qain} \& D.~{Li}, 19

\bibitem[{{Lentati} {et~al.}(2015){Lentati}, {Taylor}, {Mingarelli}, {Sesana},
  {Sanidas}, {Vecchio}, {Caballero}, {Lee}, {van Haasteren}, {Babak}, {Bassa},
  {Brem}, {Burgay}, {Champion}, {Cognard}, {Desvignes}, {Gair}, {Guillemot},
  {Hessels}, {Janssen}, {Karuppusamy}, {Kramer}, {Lassus}, {Lazarus}, {Liu},
  {Os{\l}owski}, {Perrodin}, {Petiteau}, {Possenti}, {Purver}, {Rosado},
  {Smits}, {Stappers}, {Theureau}, {Tiburzi}, \& {Verbiest}}]{Lentati2015}
{Lentati}, L., {Taylor}, S.~R., {Mingarelli}, C.~M.~F., {et~al.} 2015, \mnras,
  453, 2576

\bibitem[{{Li} {et~al.}(2022){Li}, {Bogdanovi{\'c}}, {Ballantyne}, \&
  {Bonetti}}]{Kunyang2022}
{Li}, K., {Bogdanovi{\'c}}, T., {Ballantyne}, D.~R., \& {Bonetti}, M. 2022,
  \apj, 933, 104

\bibitem[{Lodato {et~al.}(2009)Lodato, Nayakshin, King, \&
  Pringle}]{Lodato2009}
Lodato, G., Nayakshin, S., King, A.~R., \& Pringle, J.~E. 2009, Monthly Notices
  of the Royal Astronomical Society, 398, 1392

\bibitem[{{LSST Science Collaboration} {et~al.}(2009){LSST Science
  Collaboration}, {Abell}, {Allison}, {Anderson}, {Andrew}, {Angel}, {Armus},
  {Arnett}, {Asztalos}, {Axelrod}, {Bailey}, {Ballantyne}, {Bankert},
  {Barkhouse}, {Barr}, {Barrientos}, {Barth}, {Bartlett}, {Becker}, {Becla},
  {Beers}, {Bernstein}, {Biswas}, {Blanton}, {Bloom}, {Bochanski}, {Boeshaar},
  {Borne}, {Bradac}, {Brandt}, {Bridge}, {Brown}, {Brunner}, {Bullock},
  {Burgasser}, {Burge}, {Burke}, {Cargile}, {Chandrasekharan}, {Chartas},
  {Chesley}, {Chu}, {Cinabro}, {Claire}, {Claver}, {Clowe}, {Connolly}, {Cook},
  {Cooke}, {Cooray}, {Covey}, {Culliton}, {de Jong}, {de Vries}, {Debattista},
  {Delgado}, {Dell'Antonio}, {Dhital}, {Di Stefano}, {Dickinson}, {Dilday},
  {Djorgovski}, {Dobler}, {Donalek}, {Dubois-Felsmann}, {Durech},
  {Eliasdottir}, {Eracleous}, {Eyer}, {Falco}, {Fan}, {Fassnacht}, {Ferguson},
  {Fernandez}, {Fields}, {Finkbeiner}, {Figueroa}, {Fox}, {Francke}, {Frank},
  {Frieman}, {Fromenteau}, {Furqan}, {Galaz}, {Gal-Yam}, {Garnavich},
  {Gawiser}, {Geary}, {Gee}, {Gibson}, {Gilmore}, {Grace}, {Green}, {Gressler},
  {Grillmair}, {Habib}, {Haggerty}, {Hamuy}, {Harris}, {Hawley}, {Heavens},
  {Hebb}, {Henry}, {Hileman}, {Hilton}, {Hoadley}, {Holberg}, {Holman},
  {Howell}, {Infante}, {Ivezic}, {Jacoby}, {Jain}, {R}, {Jedicke}, {Jee},
  {Garrett Jernigan}, {Jha}, {Johnston}, {Jones}, {Juric}, {Kaasalainen},
  {Styliani}, {Kafka}, {Kahn}, {Kaib}, {Kalirai}, {Kantor}, {Kasliwal},
  {Keeton}, {Kessler}, {Knezevic}, {Kowalski}, {Krabbendam}, {Krughoff},
  {Kulkarni}, {Kuhlman}, {Lacy}, {Lepine}, {Liang}, {Lien}, {Lira}, {Long},
  {Lorenz}, {Lotz}, {Lupton}, {Lutz}, {Macri}, {Mahabal}, {Mandelbaum},
  {Marshall}, {May}, {McGehee}, {Meadows}, {Meert}, {Milani}, {Miller},
  {Miller}, {Mills}, {Minniti}, {Monet}, {Mukadam}, {Nakar}, {Neill}, {Newman},
  {Nikolaev}, {Nordby}, {O'Connor}, {Oguri}, {Oliver}, {Olivier}, {Olsen},
  {Olsen}, {Olszewski}, {Oluseyi}, {Padilla}, {Parker}, {Pepper}, {Peterson},
  {Petry}, {Pinto}, {Pizagno}, {Popescu}, {Prsa}, {Radcka}, {Raddick},
  {Rasmussen}, {Rau}, {Rho}, {Rhoads}, {Richards}, {Ridgway}, {Robertson},
  {Roskar}, {Saha}, {Sarajedini}, {Scannapieco}, {Schalk}, {Schindler},
  {Schmidt}, {Schmidt}, {Schneider}, {Schumacher}, {Scranton}, {Sebag},
  {Seppala}, {Shemmer}, {Simon}, {Sivertz}, {Smith}, {Allyn Smith}, {Smith},
  {Spitz}, {Stanford}, {Stassun}, {Strader}, {Strauss}, {Stubbs}, {Sweeney},
  {Szalay}, {Szkody}, {Takada}, {Thorman}, {Trilling}, {Trimble}, {Tyson}, {Van
  Berg}, {Vanden Berk}, {VanderPlas}, {Verde}, {Vrsnak}, {Walkowicz},
  {Wandelt}, {Wang}, {Wang}, {Warner}, {Wechsler}, {West}, {Wiecha},
  {Williams}, {Willman}, {Wittman}, {Wolff}, {Wood-Vasey}, {Wozniak}, {Young},
  {Zentner}, \& {Zhan}}]{LSSTScienceBook2009}
{LSST Science Collaboration}, {Abell}, P.~A., {Allison}, J., {et~al.} 2009,
  arXiv e-prints, arXiv:0912.0201

\bibitem[{Mahadevan(1997)}]{Mahadevan1996}
Mahadevan, R. 1997, Astrophys. J., 477, 585

\bibitem[{{Manchester} {et~al.}(2013){Manchester}, {Hobbs}, {Bailes}, {Coles},
  {van Straten}, {Keith}, {Shannon}, {Bhat}, {Brown}, {Burke-Spolaor},
  {Champion}, {Chaudhary}, {Edwards}, {Hampson}, {Hotan}, {Jameson}, {Jenet},
  {Kesteven}, {Khoo}, {Kocz}, {Maciesiak}, {Oslowski}, {Ravi}, {Reynolds},
  {Sarkissian}, {Verbiest}, {Wen}, {Wilson}, {Yardley}, {Yan}, \&
  {You}}]{Manchester2013}
{Manchester}, R.~N., {Hobbs}, G., {Bailes}, M., {et~al.} 2013, \pasa, 30, e017

\bibitem[{{Marshall} {et~al.}(2020){Marshall}, {Mutch}, {Qin}, {Poole}, \&
  {Wyithe}}]{Marshall2020}
{Marshall}, M.~A., {Mutch}, S.~J., {Qin}, Y., {Poole}, G.~B., \& {Wyithe}, J.
  S.~B. 2020, \mnras, 494, 2747

\bibitem[{{Mayer} {et~al.}(2007){Mayer}, {Kazantzidis}, {Madau}, {Colpi},
  {Quinn}, \& {Wadsley}}]{Mayer2007}
{Mayer}, L., {Kazantzidis}, S., {Madau}, P., {et~al.} 2007, Science, 316, 1874

\bibitem[{{McLaughlin}(2013)}]{McLaughlin2013}
{McLaughlin}, M.~A. 2013, Classical and Quantum Gravity, 30, 224008

\bibitem[{{Merloni} {et~al.}(2012){Merloni}, {Predehl}, {Becker},
  {B{\"o}hringer}, {Boller}, {Brunner}, {Brusa}, {Dennerl}, {Freyberg},
  {Friedrich}, {Georgakakis}, {Haberl}, {Hasinger}, {Meidinger}, {Mohr},
  {Nandra}, {Rau}, {Reiprich}, {Robrade}, {Salvato}, {Santangelo}, {Sasaki},
  {Schwope}, {Wilms}, \& {German eROSITA Consortium}}]{Merloni2012}
{Merloni}, A., {Predehl}, P., {Becker}, W., {et~al.} 2012, arXiv e-prints,
  arXiv:1209.3114

\bibitem[{{Miles} {et~al.}(2023){Miles}, {Shannon}, {Bailes}, {Reardon},
  {Keith}, {Cameron}, {Parthasarathy}, {Shamohammadi}, {Spiewak}, {van
  Straten}, {Buchner}, {Camilo}, {Geyer}, {Karastergiou}, {Kramer}, {Serylak},
  {Theureau}, \& {Venkatraman Krishnan}}]{Miles2023}
{Miles}, M.~T., {Shannon}, R.~M., {Bailes}, M., {et~al.} 2023, \mnras, 519,
  3976

\bibitem[{{Miles} {et~al.}(2024){Miles}, {Shannon}, {Reardon}, {Bailes},
  {Champion}, {Geyer}, {Gitika}, {Grunthal}, {Keith}, {Kramer}, {Kulkarni},
  {Nathan}, {Parthasarathy}, {Singha}, {Theureau}, {Thrane}, {Abbate},
  {Buchner}, {Cameron}, {Camilo}, {Moreschi}, {Shaifullah}, {Shamohammadi},
  {Possenti}, \& {Venkatraman Krishnan}}]{Miles2024}
{Miles}, M.~T., {Shannon}, R.~M., {Reardon}, D.~J., {et~al.} 2024, arXiv
  e-prints, arXiv:2412.01153

\bibitem[{{Milosavljevi{\'c}} \& {Merritt}(2001)}]{Milosavljevic2001}
{Milosavljevi{\'c}}, M. \& {Merritt}, D. 2001, \apj, 563, 34

\bibitem[{{Mingarelli} {et~al.}(2017){Mingarelli}, {Lazio}, {Sesana}, {Greene},
  {Ellis}, {Ma}, {Croft}, {Burke-Spolaor}, \& {Taylor}}]{Mingarelli2017}
{Mingarelli}, C. M.~F., {Lazio}, T. J.~W., {Sesana}, A., {et~al.} 2017, Nature
  Astronomy, 1, 886

\bibitem[{Peters \& Mathews(1963)}]{PetersAndMathews1963}
Peters, P.~C. \& Mathews, J. 1963, Phys. Rev., 131, 435

\bibitem[{{Petrov} {et~al.}(2024){Petrov}, {Taylor}, {Charisi}, \&
  {Ma}}]{Petrov2024}
{Petrov}, P., {Taylor}, S.~R., {Charisi}, M., \& {Ma}, C.-P. 2024, \apj, 976,
  129

\bibitem[{{Piro} {et~al.}(2023){Piro}, {Colpi}, {Aird}, {Mangiagli}, {Fabian},
  {Guainazzi}, {Marsat}, {Sesana}, {McNamara}, {Bonetti}, {Rossi}, {Tanvir},
  {Baker}, {Belanger}, {Dal Canton}, {Jennrich}, {Katz}, \&
  {Luetzgendorf}}]{2023MNRAS.521.2577P}
{Piro}, L., {Colpi}, M., {Aird}, J., {et~al.} 2023, \mnras, 521, 2577

\bibitem[{{Planck Collaboration} {et~al.}(2014){Planck Collaboration}, {Ade},
  {Aghanim}, {Armitage-Caplan}, {Arnaud}, {Ashdown}, {Atrio-Barandela},
  {Aumont}, {Baccigalupi}, {Banday}, \& et~al.}]{PlanckCollaboration2014}
{Planck Collaboration}, {Ade}, P.~A.~R., {Aghanim}, N., {et~al.} 2014, \aap,
  571, A16

\bibitem[{{Quinlan} \& {Hernquist}(1997)}]{Quinlan1997}
{Quinlan}, G.~D. \& {Hernquist}, L. 1997, \na, 2, 533

\bibitem[{{Ragusa} {et~al.}(2016){Ragusa}, {Lodato}, \& {Price}}]{Ragusa2016}
{Ragusa}, E., {Lodato}, G., \& {Price}, D.~J. 2016, \mnras, 460, 1243

\bibitem[{{Ravi} {et~al.}(2015){Ravi}, {Wyithe}, {Shannon}, \&
  {Hobbs}}]{Ravi2015}
{Ravi}, V., {Wyithe}, J.~S.~B., {Shannon}, R.~M., \& {Hobbs}, G. 2015, \mnras,
  447, 2772

\bibitem[{{Reardon} {et~al.}(2016){Reardon}, {Hobbs}, {Coles}, {Levin},
  {Keith}, {Bailes}, {Bhat}, {Burke-Spolaor}, {Dai}, {Kerr}, {Lasky},
  {Manchester}, {Os{\l}owski}, {Ravi}, {Shannon}, {van Straten}, {Toomey},
  {Wang}, {Wen}, {You}, \& {Zhu}}]{Reardon2016}
{Reardon}, D.~J., {Hobbs}, G., {Coles}, W., {et~al.} 2016, \mnras, 455, 1751

\bibitem[{{Reardon} {et~al.}(2023){Reardon}, {Zic}, {Shannon}, {Hobbs},
  {Bailes}, {Di Marco}, {Kapur}, {Rogers}, {Thrane}, {Askew}, {Bhat},
  {Cameron}, {Cury{\l}o}, {Coles}, {Dai}, {Goncharov}, {Kerr}, {Kulkarni},
  {Levin}, {Lower}, {Manchester}, {Mandow}, {Miles}, {Nathan}, {Os{\l}owski},
  {Russell}, {Spiewak}, {Zhang}, \& {Zhu}}]{Reardon2023}
{Reardon}, D.~J., {Zic}, A., {Shannon}, R.~M., {et~al.} 2023, \apjl, 951, L6

\bibitem[{{Regan} {et~al.}(2019){Regan}, {Downes}, {Volonteri}, {Beckmann},
  {Lupi}, {Trebitsch}, \& {Dubois}}]{Regan2019}
{Regan}, J.~A., {Downes}, T.~P., {Volonteri}, M., {et~al.} 2019, \mnras, 486,
  3892

\bibitem[{{Rosado} \& {Sesana}(2014)}]{Rosado2014}
{Rosado}, P.~A. \& {Sesana}, A. 2014, \mnras, 439, 3986

\bibitem[{{Rosado} {et~al.}(2015){Rosado}, {Sesana}, \& {Gair}}]{Rosado2015}
{Rosado}, P.~A., {Sesana}, A., \& {Gair}, J. 2015, \mnras, 451, 2417

\bibitem[{{Sersic}(1968)}]{Sersic1968}
{Sersic}, J.~L. 1968, {Atlas de Galaxias Australes}

\bibitem[{{Sesana} {et~al.}(2006){Sesana}, {Haardt}, \& {Madau}}]{Sesana2006}
{Sesana}, A., {Haardt}, F., \& {Madau}, P. 2006, \apj, 651, 392

\bibitem[{{Sesana} \& {Khan}(2015)}]{Sesana2015}
{Sesana}, A. \& {Khan}, F.~M. 2015, \mnras, 454, L66

\bibitem[{{Sesana} \& {Vecchio}(2010)}]{SesanaVecchio2010}
{Sesana}, A. \& {Vecchio}, A. 2010, \prd, 81, 104008

\bibitem[{{Sesana} {et~al.}(2009){Sesana}, {Vecchio}, \&
  {Volonteri}}]{Sesana2009}
{Sesana}, A., {Vecchio}, A., \& {Volonteri}, M. 2009, \mnras, 394, 2255

\bibitem[{{Shakura} \& {Sunyaev}(1973)}]{Shakura&Sunyaev1973}
{Shakura}, N.~I. \& {Sunyaev}, R.~A. 1973, \aap, 24, 337

\bibitem[{{Shen} {et~al.}(2020){Shen}, {Hopkins}, {Faucher-Gigu{\`e}re},
  {Alexander}, {Richards}, {Ross}, \& {Hickox}}]{Shen2020}
{Shen}, X., {Hopkins}, P.~F., {Faucher-Gigu{\`e}re}, C.-A., {et~al.} 2020,
  \mnras, 495, 3252

\bibitem[{{Simon} {et~al.}(2014){Simon}, {Polin}, {Lommen}, {Stappers}, {Finn},
  {Jenet}, \& {Christy}}]{Simon2014}
{Simon}, J., {Polin}, A., {Lommen}, A., {et~al.} 2014, \apj, 784, 60

\bibitem[{{Spinoso} {et~al.}(2023){Spinoso}, {Bonoli}, {Valiante}, {Schneider},
  \& {Izquierdo-Villalba}}]{Spinoso2023}
{Spinoso}, D., {Bonoli}, S., {Valiante}, R., {Schneider}, R., \&
  {Izquierdo-Villalba}, D. 2023, \mnras, 518, 4672

\bibitem[{{Springel} {et~al.}(2005){Springel}, {White}, {Jenkins}, {Frenk},
  {Yoshida}, {Gao}, {Navarro}, {Thacker}, {Croton}, {Helly}, {Peacock}, {Cole},
  {Thomas}, {Couchman}, {Evrard}, {Colberg}, \& {Pearce}}]{Springel2005}
{Springel}, V., {White}, S. D.~M., {Jenkins}, A., {et~al.} 2005, \nat, 435, 629

\bibitem[{{Springel} {et~al.}(2001){Springel}, {White}, {Tormen}, \&
  {Kauffmann}}]{Springel2001}
{Springel}, V., {White}, S.~D.~M., {Tormen}, G., \& {Kauffmann}, G. 2001,
  \mnras, 328, 726

\bibitem[{{Strateva} {et~al.}(2001){Strateva}, {Ivezi{\'c}}, {Knapp},
  {Narayanan}, {Strauss}, {Gunn}, {Lupton}, {Schlegel}, {Bahcall}, {Brinkmann},
  {Brunner}, {Budav{\'a}ri}, {Csabai}, {Castander}, {Doi}, {Fukugita},
  {Gy{\H{o}}ry}, {Hamabe}, {Hennessy}, {Ichikawa}, {Kunszt}, {Lamb}, {McKay},
  {Okamura}, {Racusin}, {Sekiguchi}, {Schneider}, {Shimasaku}, \&
  {York}}]{Strateva2001}
{Strateva}, I., {Ivezi{\'c}}, {\v{Z}}., {Knapp}, G.~R., {et~al.} 2001, \aj,
  122, 1861

\bibitem[{{Susobhanan} {et~al.}(2021){Susobhanan}, {Maan}, {Joshi}, {Prabu},
  {Desai}, {Nobleson}, {Susarla}, {Girgaonkar}, {Dey}, {Batra}, {Gupta},
  {Gopakumar}, {Bagchi}, {Basu}, {Bethapudi}, {Choudhary}, {De},
  {Krishnakumar}, {Manoharan}, {Naidu}, {Pathak}, {Singha}, \&
  {Surnis}}]{Susobhanan2021}
{Susobhanan}, A., {Maan}, Y., {Joshi}, B.~C., {et~al.} 2021, \pasa, 38, e017

\bibitem[{{Tanaka} {et~al.}(2012){Tanaka}, {Menou}, \& {Haiman}}]{Tanaka2012}
{Tanaka}, T., {Menou}, K., \& {Haiman}, Z. 2012, \mnras, 420, 705

\bibitem[{{Truant} {et~al.}(2025){Truant}, {Izquierdo-Villalba}, {Sesana},
  {Shaifullah}, \& {Bonetti}}]{Truant2024}
{Truant}, R.~J., {Izquierdo-Villalba}, D., {Sesana}, A., {Shaifullah}, G.~M.,
  \& {Bonetti}, M. 2025, \aap, 694, A282

\bibitem[{{Vasiliev} {et~al.}(2014){Vasiliev}, {Antonini}, \&
  {Merritt}}]{Vasiliev2014}
{Vasiliev}, E., {Antonini}, F., \& {Merritt}, D. 2014, \apj, 785, 163

\bibitem[{{White} \& {Frenk}(1991)}]{WhiteFrenk1991}
{White}, S.~D.~M. \& {Frenk}, C.~S. 1991, \apj, 379, 52

\bibitem[{{Xu} {et~al.}(2023){Xu}, {Chen}, {Guo}, {Jiang}, {Wang}, {Xu}, {Xue},
  {Nicolas Caballero}, {Yuan}, {Xu}, {Wang}, {Hao}, {Luo}, {Lee}, {Han},
  {Jiang}, {Shen}, {Wang}, {Wang}, {Xu}, {Wu}, {Manchester}, {Qian}, {Guan},
  {Huang}, {Sun}, \& {Zhu}}]{Xu2023}
{Xu}, H., {Chen}, S., {Guo}, Y., {et~al.} 2023, Research in Astronomy and
  Astrophysics, 23, 075024

\bibitem[{{Yates} {et~al.}(2021){Yates}, {Henriques}, {Fu}, {Kauffmann},
  {Thomas}, {Guo}, {White}, \& {Schady}}]{Yates2021}
{Yates}, R.~M., {Henriques}, B. M.~B., {Fu}, J., {et~al.} 2021, \mnras, 503,
  4474

\bibitem[{{Yu}(2002)}]{Yu2002}
{Yu}, Q. 2002, \mnras, 331, 935

\end{thebibliography}

\appendix

\section{Optical magnitude of the $N_{90}$ candidates} \label{appendix:Magnitude_N90}
\begin{figure}
    \centering
    \includegraphics[width=1\columnwidth]{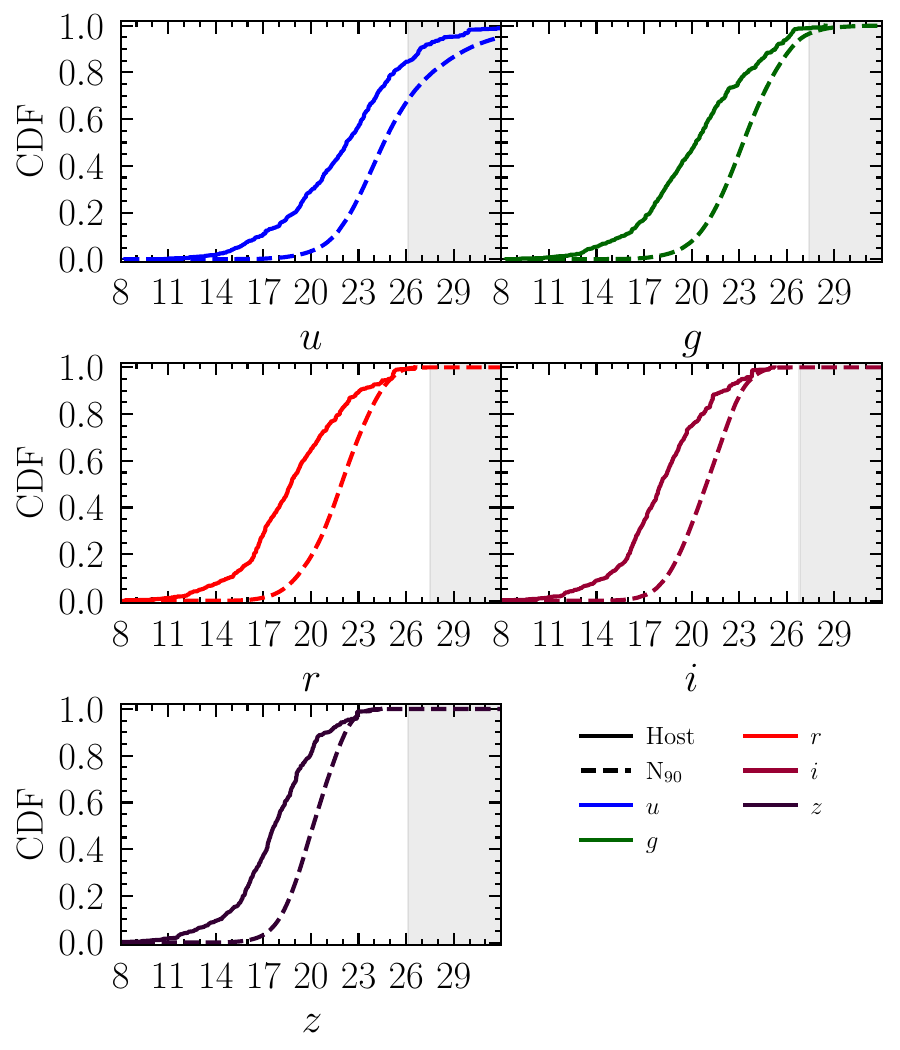}
    \caption{Cumulative distribution function (CDF) of the optical magnitudes ($u,g,r,i,z$) of the galaxies included in the $\rm N_{90}$ sample (dashed lines). For comparison, we have included the CDF of the true hosts (solid lines). Shaded areas correspond to the magnitudes that LSST cannot detect.}
    \label{fig:Magnitude_CDF_N90}
\end{figure}

In this appendix, we study the magnitude distribution of the galaxies composing the $\rm N_{90}$ sample. To this end, in Fig.~\ref{fig:Magnitude_CDF_N90} we present the cumulative distribution of the $u,g,r.i,z$ bands. As we can see, the redder the filter the brighter the galaxies. For instance, only ${\sim}\,20\%$ of the galaxies feature a magnitude of 23 in the $u$-band. This number rises to ${\sim}\,80\%$ in the case of $z$-band. The distribution is compared with the LSST detection limits. As occurred with the true hosts (see Section~\ref{sec:BinaryProperties}) all the $\rm N_{90}$ galaxies are detectable in the $g$, $r$, $i$, $z$-bands.\\

Finally, the distributions are compared with the ones of the true hosts. Interestingly, $\rm N_{90}$ galaxies tend to be slightly dimmer with respect to the true hosts. This is caused by the fact that the latter population is concentrated at $z\,{\lesssim}\,1$ whereas the former can reach larger redshifts ($z\,{>}\,1$) and, thus, intrinsically dimmer. Despite these differences, there is not a clear magnitude cut that can remove a significant number of $ \rm N_{90}$ without losing an important fraction of the true hosts.

\section{Strategies for identifying hosts of CGWs with $\rm S/N\,{>}25$: Galaxy color cuts} \label{appendix:Colors_High_SNR}

\begin{figure}
    \centering
    \includegraphics[width=1\columnwidth]{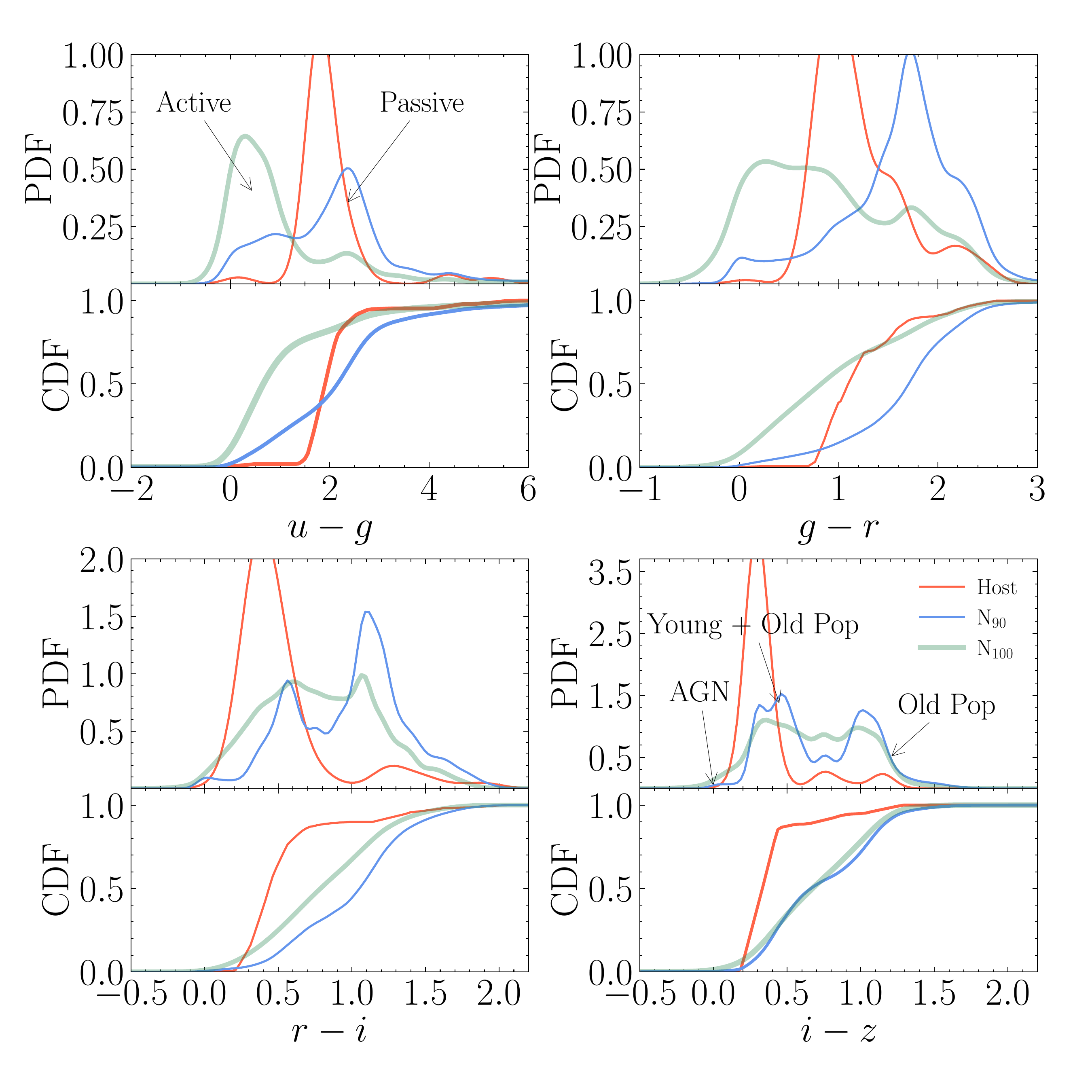}
    \caption{Color distribution of the galaxies within the sky-localization area ($\rm N_{100}$, green), galaxies within the galaxies within the sky-localization area contribution to the 90\% probability of being the true host ($N_{90}$, blue) and the true host (Host, red). \textbf{Upper panel}: Cumulative distribution function (CDF) of the $u\,{-}\,g$, $g\,{-}\,r$, $r\,{-}\,i$ and $i\,{-}\,z$ colors for $\rm N_{100}$, $\rm N_{90}$ and true host. The horizontal black line corresponds to the value where the CDF of the Host reaches 10\%. \textbf{Lower panels}: Probability distribution function (PDF) of the $u\,{-}\,g$, $g\,{-}\,r$, $r\,{-}\,i$ and $i\,{-}\,z$ colors for the three samples studied. In all the plots, the distributions correspond to the galaxies associated with CGWs whose $\rm S/N \,{>}\,25$.}
    \label{fig:Color_distribution_High_SNR}
\end{figure}

In this Appendix, we explore the possibilities of decreasing the number of potential candidates ($\rm N_{90}$) within the sky-localization area restricted for CGWs detected with $\rm S/N\,{>}\,25$. To this end, in Fig.~\ref{fig:Color_distribution_High_SNR} we present the $u\,{-}\,g$, $g\,{-}\,r$ , $r\,{-}\,i$ and $i\,{-}\,z$ color distributions.  As we can see, the distributions of candidates and true hosts are very different. This is particularly evident when we show in the lower panels of Fig.~\ref{fig:Color_distribution_High_SNR} the cumulative distribution functions. These differences can be used to perform a color cut and reduce a large number of candidates. To quantify this, we presented Table~\ref{Table:Fraction_Lost_Hig_SNR}. As shown, with a cut in the $r\,{-}\,i$ or $i\,{-}\,z$ we can diminish the potential host candidates ${\sim}\,70\%$ .  These results suggest that a color-cut is a viable approach for reducing the number of potential host candidates in cases where CGWs are detected with a high $\rm S/N$.

\renewcommand{\arraystretch}{1.05}
\begin{table}[]
\centering
\begin{adjustbox}{width=0.8\columnwidth,center}
\begin{tabular}{lcccc}
\hline 
\multicolumn{1}{c}{Color} & \multicolumn{1}{c}{$u-g$} & \multicolumn{1}{c}{$g-r$} & \multicolumn{1}{c}{$r-i$} & \multicolumn{1}{c}{$i-z$} \\ \hline \hline
 \cellcolor[HTML]{EFEFEF} Hosts lost                & \multicolumn{4}{c}{\cellcolor[HTML]{EFEFEF} \textbf{5\%}} \\
Color cut                 &  ${>}\,1.59$    &   ${>}\,0.80$  &  ${<}\,1.25$    &  ${<}\,0.92$  \\
$\rm \, N_{90}$ reduction  &  19.23\%   &  4.16\%   &   19.69\%  &  27.70\%  \\ \hline \hline
\cellcolor[HTML]{EFEFEF} Hosts lost                & \multicolumn{4}{c}{\cellcolor[HTML]{EFEFEF} \textbf{10\%}} \\
Color cut                 &  ${>}\,1.65$    &  ${>}\,0.87$    &  ${<}\,1.25$    &  ${<}\,0.73$  \\
$\rm \, N_{90}$ reduction &  20.12\%   &  6.25\%    & 19.69\%    &  34.21\%  \\ \hline \hline
\cellcolor[HTML]{EFEFEF} Hosts lost                & \multicolumn{4}{c}{\cellcolor[HTML]{EFEFEF} \textbf{20\%}} \\
Color cut                 &  ${>}\,1.66$    & ${>}\,0.87$     &  ${<}\,0.58$    &  ${<}\,0.33$   \\
$\rm \, N_{90}$ reduction &  24.80\%   &   6.25\%   &   75.00\%  & 77.14\%  \\\hline 
\end{tabular}
\end{adjustbox}
\caption{Fraction of $\rm N_{90}$ galaxies removed after applying different optical color cuts with LSST filters. The host lost refers to the probability that the true host is removed after the color cut. These numbers only consider the galaxies associated with CGWs whose $\rm S/N\,{>}\,25$.}
\label{Table:Fraction_Lost_Hig_SNR}
\end{table}

\end{document}